\documentclass[pra,showpacs,twocolumn]{revtex4}

\usepackage{amssymb}
\usepackage{amsmath}
\usepackage{graphicx}
\usepackage{verbatim}
\usepackage{soul}

\usepackage{color}
\definecolor{blue}{rgb}{0,0,1}
\definecolor{red}{rgb}{1,0,0}
\definecolor{grey}{rgb}{0.6,0.6,0.6}

\def    \bse{\begin{subequations}}
\def    \ese{\end{subequations}}

\def \be{\begin{equation}}
\def \ee{\end{equation}}
\def \bew{\begin{widetext}\begin{equation}}
\def \eew{\end{equation}\end{widetext}}

\def \hH{\hat{H}}
\def \hHL{\hat{H}_{\rm{L}}}
\def \hHNL{\hat{H}_{\rm{NL}}}
\def \hHeff{\hat{H}_{\rm{eff}}}
\def \hc{\hat{c}}
\def \hcs{\hat{c}_{\sigma}}
\def \hcsdag{\hat{c}_{\sigma}^{\dag}}
\def \hcm{\hat{c}_-}
\def \hcmdag{\hat{c}_-^\dagger}
\def \hcp{\hat{c}_+}
\def \hcpdag{\hat{c}_+^\dagger}

\def \ha{\hat{a}}
\def \hadag{\hat{a}^{\dag}}
\def \hb{\hat{b}}
\def \hbdag{\hat{b}^{\dag}}
\def \hd{\hat{d}}
\def \hddag{\hat{d}^{\dag}}
\def \nthM{\bar{n}^{\rm{M}}_{\mathrm{th}}}
\def \bnom{\bar{n}^0_{-}}
\def \bnop{\bar{n}^0_{+}}
\def \bnos{\bar{n}^0_{\sigma}}
\def \bnintm{\bar{n}^{\rm{int}}_{-}}
\def \bnintp{\bar{n}^{\rm{int}}_{+}}
\def \bnints{\bar{n}^{\rm{int}}_{\sigma}}

\def \bneffp{\bar{n}^{\rm{eff}}_{+}[\omega]}
\def \bneffs{\bar{n}^{\rm{eff}}_{\sigma}[\omega]}
\def \bneffd{\bar{n}^{\rm{eff}}_{d}[\omega]}
\def \bndtot{\bar{n}_{d}^{\textrm{tot}}}
\def \Tom{T^0_{-}}

\def \Tos{T^0_{\sigma}}
\def \Teffd{T^{\mathrm{eff}}_{d}[\omega]}

\def \Gammaints{\Gamma^{\rm{int}}_{\sigma}[\omega]}
\def \Gres{G_{\rm{res}}}

\def \GRos{G^R_{\sigma}[\omega]}

\def \GAos{G^A_{\sigma}[\omega]}

\def \GKos{G^K_{\sigma}[\omega]}

\def \GRtots{\mathcal{G}^R_{\sigma}[\omega]}

\def \GAtots{\mathcal{G}^A_{\sigma}[\omega]}

\def \GKtots{\mathcal{G}^K_{\sigma}[\omega]}
\def \SigmaRm{\Sigma^R_-[\omega]}
\def \SigmaRp{\Sigma^R_+[\omega]}
\def \SigmaRs{\Sigma^R_{\sigma}[\omega]}

\def \SigmaAs{\Sigma^A_{\sigma}[\omega]}

\def \SigmaKs{\Sigma^K_{\sigma}[\omega]}
\def \Sd{S_d[\omega]}
\def \Ceffm{\mathcal{C}_{-}^{\rm{eff}}}
\def \Ceffp{\mathcal{C}_{+}^{\rm{eff}}}
\def \Ceffs{\mathcal{C}_{\sigma}^{\rm{eff}}}
\def \abm{\alpha_{b,-}}
\def \abp{\alpha_{b,+}}
\def \abs{\alpha_{b,\sigma}}
\def \babm{\bar{\alpha}_{b,-}}
\def \babp{\bar{\alpha}_{b,+}}
\def \babs{\bar{\alpha}_{b,\sigma}}
\def \adm{\alpha_{d,-}}
\def \adp{\alpha_{d,+}}
\def \ads{\alpha_{d,\sigma}}
\def \badm{\bar{\alpha}_{d,-}}
\def \badp{\bar{\alpha}_{d,+}}
\def \bads{\bar{\alpha}_{d,\sigma}}

\begin{document}

\title{Real photons from vacuum fluctuations in optomechanics: the role of polariton interactions}
\author{Marc-Antoine Lemonde}
\affiliation{Department of Physics, McGill University, 3600 rue
University, Montreal, QC Canada H3A 2T8}
\author{Aashish~A.~Clerk}
\affiliation{Department of Physics, McGill University, 3600 rue
University, Montreal, QC Canada H3A 2T8}

\date{\today}

\begin{abstract}
We study nonlinear interactions in a strongly driven optomechanical cavity, in regimes where the interactions give rise to resonant scattering between optomechanical polaritons and
are thus strongly enhanced.  We use a Keldysh formulation and self-consistent perturbation theory, allowing us to include self energy diagrams at all orders in the interaction.  Our main focus is understanding how non-equilibrium effects are modified by the polariton interactions, in particular the generation of non-zero effective polariton temperatures from vacuum fluctuations (both in the incident cavity drive and in the mechanical dissipation).  We discuss how these effects could be observed in the output spectrum of the cavity.  Our work also provides a technical toolkit that will be useful for studies of more complex optomechanical systems.
\end{abstract}


\maketitle

%%%%%%%%%%%%%%%%%%%%%%%%%%%%%%%%%%%%%%%%%%%%%%%%%%%%%%%%%%%%%%%%%%%%%%%%%%%%%%%%%%%%%%%%%%%%%%%%%%%%%%%%%%%%%%%%%%%%%%%%%%%%%%%%%%%%%%%%%%

\section{Introduction}

The rapidly growing field of cavity optomechanics seeks to understand the interaction between photons and mechanical motion in driven electromagnetic cavities, hopefully in truly quantum regime \cite{MarquardtRMP,OptomechanicsBook}. The past few years have seen many breakthroughs, including laser cooling of mechanical motion to the ground state \cite{Teufel_Nature_2011,Chan_Nature_2011}, 
and the optomechanical generation of squeezed light leaving the cavity \cite{Brooks_Nature_2012,Safavi-Naeini_Nature_2013,Regal2013}.
Although the basic radiation pressure interaction between a mechanical resonator and cavity photons is intrinsically nonlinear, almost all the remarkable achievements in the field to date rely on working in strongly driven regimes where the dynamics is essentially linear.  To see truly nonlinear effects in the simplest setting, one needs to achieve a single photon, single phonon optomechanical coupling $g$ which exceeds both the mechanical frequency $\omega_M$ and the cavity damping rate $\kappa$  \cite{Nunnenkamp_PRL_2011,Rabl_PRL_2011,Kronwald_PRA_2013,Kronwald_PRL_2013}. With the exception of experiments using cold atoms \cite{Murch_Nature_2008,Brennecke_Science_2008} this parameter regime remains challenging for experiments.

Nonlinear effects which only require $g \sim \kappa$ can be achieved in slightly more complex settings where the nonlinear interaction becomes resonant.  This can occur in an optomechanical setup with two optical modes \cite{Ludwig_Safavi_PRL_2012, Komar_Bennett_PRA_2013}.  Alternatively, this can occur in a standard single mode optomechanical cavity which is driven, such that the interaction becomes resonant in a basis of dressed states (so-called optomechanical polaritons).  This occurs both in regimes of weak driving \cite{Borkje_PRL_2013,MAL_PRL_2011} and strong driving \cite{MAL_PRL_2011}; similar physics can also occur in membrane-in-the-middle style optomechanical systems \cite{Xu_ArXiv_2014}. Note that  such strong driving regimes, where the drive-enhanced optomechanical coupling exceeds dissipative rates, has been achieved in several experiments \cite{Teufel_Nature_2011_OMIT,Groeblacher_Nature_2009,Verhagen_Nature_2012}.

%%%%%%%%%%%%%%%%%%%%%
\begin{figure}[t]
	\begin{center}
	\includegraphics[width= 0.8\columnwidth]{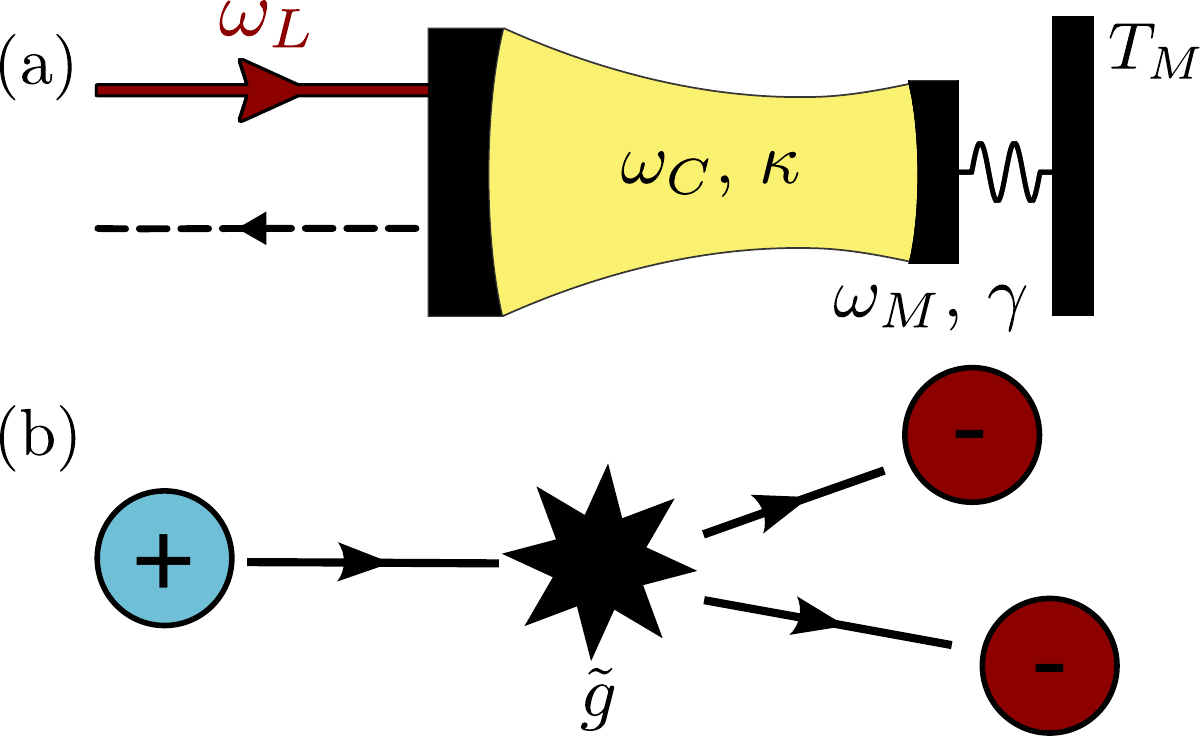}
	\end{center}
	\vspace{-0.5cm}
	\caption {(Color online) (a) A generic optomechanical cavity: 
	a cavity mode of frequency $\omega_C$ and damping rate $\kappa$ is driven by a laser at frequency $\omega_L$, and is coupled via radiation pressure
	to a damped mechanical mode (resonance frequency $\omega_M$, damping rate $\gamma$).  The temperature of the mechanical bath is $T_M$.  
	 (b)  The single-photon optomechanical interaction $g$ can be made resonant 
	 in the basis of optomechanical polaritons; the resulting enhanced interaction shown schematically and described in Eq.~(\ref{Eq:HEff}).}
\label{fig:Schemas}
\end{figure}
%%%%%%%%%%%%%%%%%%%%%

In this paper, we both expand upon and extend the results previously reported in Ref.~\onlinecite{MAL_PRL_2011}.  We consider a standard single-cavity optomechanical system that is driven (possibly strongly), in a regime where the nonlinear interaction is resonantly enhanced.  We again address this system using a Keldysh formulation \cite{Kamenev_AIP_2009, KamenevBook} and perturbation theory; we begin this paper by providing a more complete discussion of this approach to optomechanics, emphasizing the subtleties involved in treating dissipation. 
Our work thus contributes to a growing body of work using the Keldysh technique to address various quantum optics contexts~\cite{Mertens_PRA_1993, Mertens_PRA_1995, Veits_PRA_1997, DallaTorre_PRA_2013, Sieberer_PRB_2014}.
Unlike our previous study, we go beyond a simple lowest-order treatment and introduce a self-consistent perturbation theory.  This corresponds to an infinite partial resummation of self-energy diagrams, and extends the range of couplings and temperatures that we can address. 

Our work also address a new set of physical phenomena.   Ref.~\onlinecite{MAL_PRL_2011} focused on understanding the cavity density of states (DOS), the quantity probed in optomechanically-induced transparency (OMIT) experiments \cite{Agarwal_PRA_2010,Weis_Science_2010,Teufel_Nature_2011_OMIT,Safavi-Naeini_Nature_2011}.  Here, we instead focus on the non-equilibrium state of this system: what are the effective temperatures of the optomechanical polaritons?  Of particular interest is 
how vacuum fluctuations can result in non-zero effective temperatures.  Such effects are often termed ``quantum heating" \cite{Dykman_PRA_2006, Dykman_PRE_2007, Wilhelm_PRL_2007, Dykman_PRA_2011, Ong_PRL_2013}; the simplest example is the amplification of zero-point fluctuations by a parametric amplifier.
Though often described using disparate terms, this physics has been studied in a wide variety of systems ranging from driven nonlinear oscillators \cite{Dykman_PRA_2006, Dykman_PRE_2007, Wilhelm_PRL_2007, Dykman_PRA_2011}, superconducting circuits \cite{Johansson_PRL_2009, Wilson_Nature_2011} and circuit quantum electrodynamics \cite{Peano_PRB_2010}  to phase transitions in driven-dissipative many-body systems \cite{DallaTorre_PRA_2013}.  It also sets a limit to the minimum mechanical temperature achievable using cavity cooling~\cite{Marquartd_PRL_2007}.

Here, we show that quantum heating effects in an optomechanical system lead to observable signatures in the output spectrum of the cavity, i.e. the spectrum of output light that would be measured using a photomultiplier.  Unlike an OMIT experiment, such an experiment probes both the DOS of the cavity (as modified by the optomechanical interactions), as well as the effective temperature of the optomechanical polaritons. This ``quantum heating" is already present at the level of the linearized theory of the optomechanical cavity; here, we investigate how it is modified by the nonlinear interactions. Our work is timely, given that the generation of real photons from mechanical vacuum fluctuations was recently probed experimentally by Lecocq et al.~\cite{Teufel_ArXiV_2014}.

\subsection{Main findings}

Our work has many technical aspects to it that will hopefully be an aid in further studies of nonlinear quantum optomechanics (e.g.~optomechanical lattices
\cite{Chang2011,MarquardtLattice2013,Wei2014,Painter2014,MarquardtLattice2014}).  It also predicts several new physical phenomena, in particular:

\subsubsection{Polariton thermalization}
In the standard linearized theory of a driven optomechanical cavity, quantum heating leads to very different effective temperatures for the two polariton modes (i.e.~normal modes of the linearized theory), 
c.f. Eqs.~(\ref{Eq:ns}).  The nonlinear interaction tends to dilute this effect, as it allows energy exchange between the polaritons and favours their thermalization.  We study this competition in detail for two representative cases:  a laser drive at the red-mechanical sideband (see Fig.~\ref{fig:Sd_D1}), and a laser drive detuned further to the red where nonlinear effects are more important (see Fig.~\ref{fig:Sd_D18}).
The effects of  the nonlinear interaction on this quantum heating physics can be seen experimentally in the cavity output spectrum by tuning the nonlinear interactions into and out of resonance, see Fig.~\ref{fig:NphotonVsG_D10}.

\subsubsection{New instabilities}
For a red-detuned laser, we find that a leading-order treatment of the nonlinear interaction suggests new kind of parametric instability not present in the linearized theory, where one polariton mode acts as an incoherent pump mode for the other, see Sec.~\ref{Sec:ParamInsta}.  Including higher order terms stabilizes the system as expected, see Fig.~\ref{fig:Sd_D065}.

\subsubsection{Two-phonon cavity heating}
We show that nonlinear heating of the cavity (as manifest in the cavity spectrum) are greatly enhanced for red-detuned laser drives near the second mechanical sideband (a detuning $\Delta \approx -2\omega_M$); this is a simple consequence of the lower optomechanical polariton being mostly phononic in this regime, causing effects to be enhanced by the typically large phonon lifetime (see Fig.~\ref{fig:Sd2wm}).

\subsection{Organization of the paper}

The remainder of this paper is organized as follow. We begin in Sec.~\ref{Sec:System} by introducing the basic driven optomechanical system studied in this work, and reviewing the linearized
theory.  We also pay careful attention to the coupling to dissipative baths, and to how quantum heating effects can arise even without nonlinearity.  In Sec.~\ref{Sec:Keldysh}, we introduce the basic aspects of the Keldysh technique as applied to optomechanics, focusing first on the linearized system.  In Sec.~\ref{Sec:PerturbationTheory}, we develop a Keldysh perturbation theory to treat the nonlinear optomechanical interaction, and introduce our self-consistent approach.  In Sec.~\ref{Sec:NoneqDescription}, we discuss how the nonlinear interaction modifies the non-equilibrium physics of the system.
We present a physical picture where interaction effects can be mapped onto a couplings to additional ``self-generated" dissipative baths.  We also investigate in this section the new kind of parametric heating which arises with a red-detuned laser.  In Sec.~\ref{Sec:Observable}, we discuss the observable consequences of our predictions on the cavity output spectrum.  We present our conclusions in Sec.~\ref{Sec:Conclusion}.

%%%%%%%%%%%%%%%%%%%%%%%%%%%%%%%%%%%%%%%%%%%%%%%%%%%%%%%%%%%%%%%%%%%%%%%%%%%%%%%%%%%%%%%%%%%%%%%%%%%%%%%%%%%%%%%%%%%%%%%%%%%%%%%%%%%%%%%%%%

\section{System, treatment of dissipation and linearized theory} \label{Sec:System}

\subsection{Hamiltonian in the polariton basis} \label{Sec:H}

We consider a standard optomechanical system where the frequency of a driven cavity mode is modulated linearly by the position of a mechanical resonator (Fig.~\ref{fig:Schemas}).  The Hamiltonian governing its dynamics is given by \cite{OptomechanicsBook,MarquardtRMP} 
\begin{align}
	\hH = & \omega_C \hadag\ha + \omega_M \hbdag\hb + g \hadag\ha (\hb+\hbdag) \nonumber \\
	& + i( \bar{a}_{\textrm{in}}e^{-i\omega_Lt}\hadag - \textrm{H.c.}) + \hat{H}_{\textrm{diss}}. \label{Eq:StartingHamiltonian}
\end{align}
Here, $\ha$ is the cavity mode, with frequency $\omega_C$ and $\hb$ the mechanical mode with frequency $\omega_M$. The parameter $g$ is the single-photon optomechanical coupling and $\bar{a}_{\textrm{in}}$ is proportional to the amplitude of a classical drive at frequency $\omega_L$. Finally, $\hat{H}_{\rm{diss}}$ describes dissipation due to the coupling to bosonic environments (both for the mechanics and the cavity). Going into a rotating frame at the drive frequency and displacing the cavity field by its classical value, induced by the coherent drive (i.e. $\ha(t) \rightarrow e^{-i \omega_L t}(\bar{a} + \hd(t))$), the Hamiltonian of Eq.~\eqref{Eq:StartingHamiltonian} can be expanded into a quadratic part, known as the linearized optomechanical Hamiltonian, and a nonlinear interaction term such that
\begin{subequations} \label{Eq:Hs}
\begin{align}
	\hH & = \hHL + \hHNL + \hH_{\rm diss}, \\
	\hHL & = -\Delta \hddag\hd + \omega_M \hbdag\hb + G (\hd+\hddag)(\hb+\hbdag), \label{Eq:HL} \\
	\hHNL & = g \hddag\hd (\hbdag + \hb). \label{Eq:HNL}
\end{align}
\end{subequations}
In the above, we have defined the laser detuning $\Delta=\omega_L-\omega_C$ and the many-photon coupling constant $G=\bar{a}g$; we take $\bar{a}, g > 0$ without loss of generality \footnote{The mean value of $\ha$ ($\bar{a}$) is determined by solving the classical equations of motion following from Eq.~(\ref{Eq:StartingHamiltonian}). We also shift the mechanical lowering operator $\hb$ (i.e. $\langle \hb \rangle = 0$ in Eqs.~(\ref{Eq:Hs})) to account for the static radiation pressure force. We also redefine the detuning accordingly.}.

Our general approach is to diagonalize the quadratic part of the Hamiltonian, $\hH_{\textrm{L}}$ (thus treating it exactly), and then treat the terms due to nonlinear interaction ($\hHNL$) as a perturbation.  Unlike treatments based on polaron-transformed Hamiltonians \cite{Rabl_PRL_2011,Nunnenkamp_PRL_2011}, we do not require the coherent driving of the cavity to be so small that it too can be treated perturbatively.  We focus exclusively on a red-detuned laser ($\Delta < 0$), and only require the cavity drive to be weak enough so that $G^2 < -\omega_M\Delta/4 \equiv G_{\rm crit}^2$.  For drives stronger than this critical value (or for a blue-detuned laser), the linearized coherent Hamiltonian $\hHL$ is unstable, and corresponds to a detuned parametric amplifier driven beyond threshold.  This critical value also coincides with the onset of the well-known static optomechanical instability \cite{MarquardtRMP, Dorsel_PRL_1983}. Operating near this instability has been investigated as an alternative promising way to enhance the nonlinear interaction \cite{Xu_ArXiv_2014}. Note that this instability is also closely analogous to the superradiant phase transition in the driven Dicke model studied in \cite{DallaTorre_PRA_2013}.

Focusing on $\Delta < 0$ and $G^2 < -\omega_M\Delta/4$, Eq.~(\ref{Eq:HL}) can be diagonalized via a Bogoliubov transformation to yield:
\begin{equation}
	\hHL = \sum_{\sigma=\pm} E_{\sigma} \hcsdag\hcs.
\end{equation}
Here, $\hat{c}_{\pm}$ destroys an excitation in the eigenmode of $\hHL$ with energy $E_{\pm} > 0$, given by
\begin{equation}
	E_{\pm}=\frac{1}{\sqrt{2}}\left( \omega_M^2 + \Delta^2 \pm \sqrt{(\omega_M^2 - \Delta^2)^2 -16G^2\Delta\omega_M} \right)^{1/2}. \label{Eq:Esigma}
\end{equation}
Note that $E_{-}$ tends to zero as $G$ approaches the critical value at the instability $G_{\rm crit}$. As $\hHL$ does not conserve the total number of photons and phonons, the operators $\hcm, \hcp$ mix photon/phonon annihilation and creation operators:
\begin{subequations} \label{Eq:ChangeOfBasis}
\begin{align}
	\hat{b} & = \abm \hcm + \abp \hcp + \babm \hcmdag + \babp \hcpdag, \label{Eq:PhononToPol} \\
	\hat{d} & = \adm \hcm + \adp \hcp + \badm \hcmdag + \badp \hcpdag. \label{Eq:PhotonToPol}
\end{align}
\end{subequations}
The coefficients $\alpha_{b/d,\pm}$ and $\bar{\alpha}_{b/d,\pm}$ are functions of $\Delta/\omega_M$ and $G/\omega_M$; their explicit form in the case $\Delta = -\omega_M$ is given in Eqs.~(\ref{App:COB1})-(\ref{App:COB2}) of Appendix \ref{Sec:AppendixCOB}. As the excitations described by $\hcm, \hcp$ have both phononic and photonic components, we refer to them as polaritons in what follows.  

Having diagonalized the linearized optomechanical Hamiltonian, we now express the nonlinear Hamiltonian $\hHNL$ in the polariton basis.  We obtain interactions that do not conserve the total number of polaritons,
\begin{align}
	\hHNL \!=\! & \sum_{\sigma,\sigma',\sigma'' = \pm}  \left( g^A_{\sigma \sigma' \sigma''} \hc^\dag_\sigma \hc^\dag_{\sigma'} \hc^\dag_{\sigma''} +
		g^B_{\sigma \sigma' \sigma''} \hc^\dag_\sigma \hc^\dag_{\sigma'} \hc_{\sigma''} + \textrm{H.c.} \right) \nonumber \\
		& +\left( A_- \hcm + A_+ \hcp + \textrm{H.c.} \right).
	\label{Eq:HintPolariton}
\end{align}
Here, the constants $g^{A/B}_{\sigma \sigma' \sigma''} $ and $A_{\sigma}$ are all proportional to $g$ (see Eqs.~\eqref{App:g1}-\eqref{App:g2} of Appendix \ref{Sec:AppendixCOB}).
Note that the linear terms in $\hHNL$ ($\propto A_{\sigma}$) arise from normal ordering $\hHNL$ in the polariton basis; physically, while the zero-polariton state is the vacuum of $\hHL$, this is no longer true when we include the nonlinear interaction.  

%%%%%%%%%%%%%%%%%%%%%%%%%%%%%%%%%%%%%%%%%%%%%%%%%%%%%%%%%%%%%%%%%%%%%%%%%%%%%%%%%%%%%%%%%%%%%%%%%%%%%%%%%%%%%%%%%%%%%%%%%%%%%%%%%%%%%%%%%%

\subsection{Resonant polariton interactions}

As discussed in previous works \cite{MAL_PRL_2011, Borkje_PRL_2013} (and later in \cite{Liu_PRL_2013}), we can enhance the effects of even a weak single-photon coupling $g$ by tuning $G$ and $\Delta$ such that the nonlinear processes in  Eq.~(\ref{Eq:HintPolariton}) that scatters a $+$ polariton into two $-$ polaritons ($\propto g^B_{--+}$) become resonant. This requires $E_+[\Delta, G] = 2E_-[\Delta, G]$; for a given laser detuning in the range $\Delta \in [-2 \omega_M,-\omega_M/2]$, this can always be achieved by tuning $G=\Gres[\Delta]$, where
\begin{equation}
	\Gres[\Delta] \equiv \frac{\sqrt{17\Delta^2\omega_M^2-4(\Delta^4+\omega_M^4)}}{(10\sqrt{-\Delta\omega_M})}. \label{Eq:Gres}
\end{equation}

Once $G$ is tuned to achieve this resonance condition, one can show using standard perturbation theory that all non-resonant nonlinear processes are suppressed by a factor of $\kappa/(E_+-E_-) \propto \kappa/\omega_M$ compared to the resonant process, where $\kappa$ is the cavity damping rate (see \cite{MAL_PRL_2011} for more details). In addition, in this regime, the relative modification of the polariton energies and wavefunctions due to the linear terms in Eq.~(\ref{Eq:HintPolariton}) are strongly suppressed by a factor of $(g / \omega_M)^2$. Thus, if we focus on parameters near this resonant regime, in the resolved sideband regime ($\kappa/\omega_M \ll 1$), and for weak nonlinear interaction ($g/\omega_M \ll 1$), we can both ignore the renormalization of polariton energies and wavefunctions, and make a rotating wave approximation on Eq.~(\ref{Eq:HintPolariton}), keeping only the resonant interaction.  The coherent system Hamiltonian in this regime thus reduces to
\begin{equation}
	\hHeff = \sum_{\sigma=\pm} E_{\sigma} \hcsdag\hcs + \tilde{g}(\hcpdag \hcm \hcm + \textrm{H.c.}), \label{Eq:HEff}
\end{equation} 
where $\tilde{g}=g^{B}_{--+}$ is the effective nonlinear coupling (see Eq.~(\ref{App:gtilde}) in Appendix \ref{Sec:AppendixCOB}). Shown in Fig.~\ref{fig:LinearParams}(a) is the dependence of $\Gres$ as a function of laser detuning $\Delta$, as well as the behaviour of $\tilde{g}$ on $\Delta$, when $G$ is tuned to be $\Gres$. For the rest of this paper, we focus on the dynamics governed by the effective coherent Hamiltonian $\hHeff$.

%%%%%%%%%%%%%%%%%%%%%
\begin{figure}[t]
	\begin{center}
	\includegraphics[width= 0.95\columnwidth]{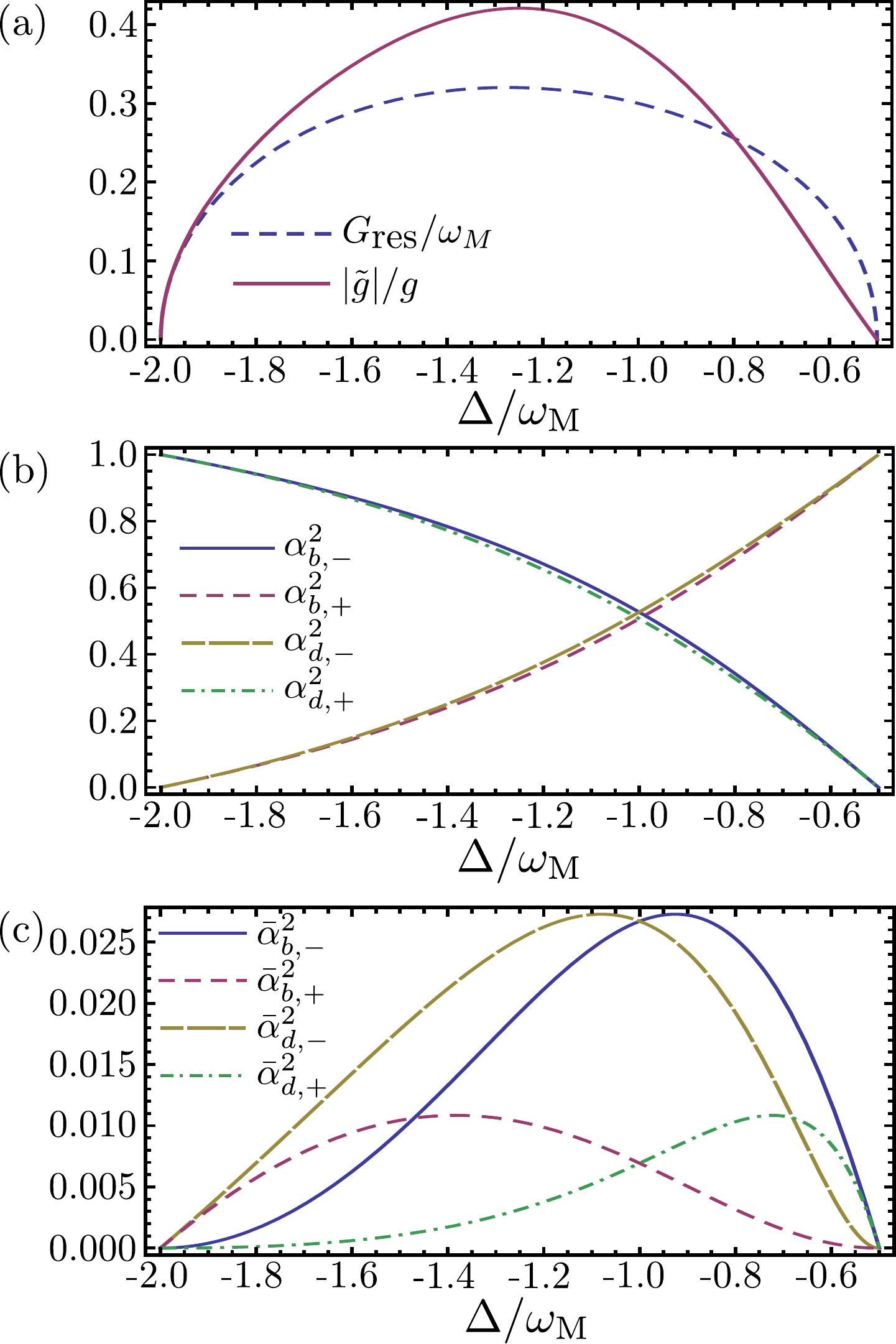}
	\end{center}
	\vspace{-0.5cm}
	\caption {(Color online)
		(a) $\Gres$, the value of the many-photon coupling which leads a resonant nonlinear interaction, as a function of laser detuning 
		$\Delta$ (c.f.~Eq.~\eqref{Eq:Gres}); the corresponding effective nonlinear coupling $\tilde{g}$ (c.f.~Eq.~\eqref{Eq:HEff}) is also plotted. 
		 (b) and (c) Coefficients relating the optomechanical polaritons (eigenstates of the linearized Hamiltonian) to original photon and phonon operators 
		 (c.f.~Eqs.~(\ref{Eq:ChangeOfBasis})) as
		 a function of laser detuning $\Delta$ with $G = \Gres[\Delta]$.
		 In (b), we plot the ``normal" coefficients which relate polariton destruction operators to photon/phonon destruction operators, 
		 i.e.~$\alpha_{b,\sigma} = \langle 0,0 \lvert \hcs\hbdag \rvert 0,0 \rangle$, 
		 with $\lvert 0,0 \rangle$ being the polariton vacuum. In (c) we 
		 plot the ``anomalous" coefficients, i.e.~$\bar{\alpha}_{b,\sigma} = \langle 0,0 \lvert \hcs\hb \rvert 0,0 \rangle$; these coefficients
		 are directly related to the existence of ``quantum heating" effects in the linearized theory (c.f.~Eq.~(\ref{Eq:ncav})).}
\label{fig:LinearParams}
\end{figure}
%%%%%%%%%%%%%%%%%%%%%

%%%%%%%%%%%%%%%%%%%%%%%%%%%%%%%%%%%%%%%%%%%%%%%%%%%%%%%%%%%%%%%%%%%%%%%%%%%%%%%%%%%%%%%%%%%%%%%%%%%%%%%%%%%%%%%%%%%%%%%%%%%%%%%%%%%%%%%%%%

\subsection{Coupling to dissipative reservoirs} \label{Sec:Dissipation}

We now turn to a more careful consideration of the effects of dissipation on our system.  As is standard in optomechanics, both the mechanics and the cavity are taken to be linearly coupled to independent, Markovian bosonic baths (i.e.~baths with constant DOS over the frequency range of interest). This is analogous to the approach taken in input-output treatment of dissipation (see, e.g.~\cite{GardinerZollerBook,Clerk_RMP_2010}).  As we are interested in possibly large many-photon optomechanical couplings $G$, care must still be taken, as the eigenstates of our coherent Hamiltonian are polaritons, not individual photons or phonons; similar issues have recently been addressed in strongly-coupled circuit QED systems \cite{Beaudoin_PRA_2011}.  Moreover, the strong driving of the cavity can also lead directly to ``quantum heating" effects, which manifest themselves directly in the treatment of the cavity dissipation.  We describe these effects more in what follows.

%%%%%%%%%%%%%%%%%%%%%%%%%%%%%%%%%%%%%%%%%%%%%%%%%%%%%%%%%%%%%%%%%%%%%%%%%%%%%%%%%%%%%%%%%%%%%%%%%%%%%%%%%%%%%%%%%%%%%%%%%%%%%%%%%%%%%%%%%%

\subsubsection{Cavity dissipation in the presence of optomechanical coupling and driving}

Consider first the interaction between the cavity and its dissipative reservoir.  In the original lab frame, the cavity-bath interaction will have the following generic form~\cite{GardinerZollerBook}
\begin{align}
	& \hH_{\kappa} \!=\! \sum_{j}\omega_j \hat{f}^{\dag}_j\hat{f}_j, & \hH_{\kappa}^{\rm{int}} \!=\! i\sqrt{\frac{\kappa}{2\pi\rho_C}}\sum_{j}\left( \hat{f}^{\dag}_j-\hat{f}_j\right)(\ha+\hadag). \label{Eq:HCavBathIntIni}
\end{align}
where $\hat{f}_j$ is the anihiliation operator for cavity bath mode $j$ (frequency $\omega_j$), $\kappa$ the damping rate of the photons inside the cavity and $\rho_C$ is the bath DOS.  As we consider a Markovian bath, we take $\kappa$ and $\rho_C$ to be frequency independent.

We next transform to an interaction picture at the drive frequency via the unitary
\begin{equation}
	\hat{U}=\textrm{exp} \left[ -i\omega_L t \left(\hadag\ha+\sum_{j} \hat{f}^{\dag}_j\hat{f}_j \right)\right]. \label{Eq:UnitaryTrans}
\end{equation}
Note that we transform both cavity and bath operators. The result is that in our interaction picture, terms in Eq.~(\ref{Eq:HCavBathIntIni}) which conserve excitation number are time-independent, whereas the remaining excitation non-conserving terms are rapidly oscillation at a frequency 
$\pm 2\omega_L \approx \pm 2\omega_C$ ($\omega_C \gg \vert \Delta \vert$).  
Since $\omega_C$ is much larger than all other energy scales in the rotating frame (i.e.~$\omega_C \gg E_{\sigma}, \lvert\Delta\rvert,\omega_M, g,\kappa$), and since the bath oscillators necessarily have positive energies $\omega_j > 0$, these rapidly oscillating terms can never become resonant.  We can thus safely make a rotating-wave approximation, and drop them. 

With this rotating-wave approximation, the cavity-bath Hamiltonian then becomes time independent. Re-writing the cavity annihilation operator as per Eq.~(\ref{Eq:HL}), one obtains a standard-rotating wave system-bath interaction
\begin{align}
	& \hH_{\kappa} = \sum_{j}(\omega_j-\omega_L) \hat{f}^{\dag}_j\hat{f}_j = \sum_{j}\tilde{\omega}_j \hat{f}^{\dag}_j\hat{f}_j, \label{Eq:HCavBathPol} \\
	& \hH_{\kappa}^{\textrm{int}} = i\sqrt{\frac{\kappa}{2\pi\rho_C}}\sum_{j}\left(\hat{f}^{\dag}_j\hd - \hat{f}_j\hddag\right),  \label{Eq:HCavBathIntPol} 
\end{align}

Note crucially that in this final rotating frame, the transformed bath frequencies $\tilde{\omega}_j$ can be negative (they  extend down to $- \omega_L$). While this may seem innocuous, things become more interesting when we re-write this interaction in terms of polariton operators (c.f.~Eqs.~(\ref{Eq:ChangeOfBasis})):
\begin{equation}
	\hH_{\kappa}^{\textrm{int}} =
		i \sqrt{\frac{\kappa}{2\pi\rho_C}}  \sum_{j,\sigma } \hat{f}^{\dag}_j 
			\left( \alpha_{d,\sigma} \hat{c}_{\sigma} + \bar{\alpha}_{d,\sigma} \hat{c}^\dagger_{\sigma} \right) + \textrm{h.c.}
	\label{eq:HKappaQH}
\end{equation}
The anomalous terms which create or destroy two excitations here {\it should not} be dropped; as the polariton energies $E_{\sigma} \ll \omega_L \sim \omega_C$, these terms can be resonant in our interaction picture, as bath modes having $\tilde{\omega}_j < 0$ can be involved.  Physically, such processes involve the creation of both a polariton and a bath excitation, while at the same time a (classical) drive photon is absorbed.  Such heating processes thus involve the interplay of the system drive and the bath zero-point fluctuations, and are at the heart of quantum activation \cite{Dykman_PRA_2006,Dykman_PRE_2007,Wilhelm_PRL_2007,Dykman_PRA_2011}.

%%%%%%%%%%%%%%%%%%%%%%%%%%%%%%%%%%%%%%%%%%%%%%%%%%%%%%%%%%%%%%%%%%%%%%%%%%%%%%%%%%%%%%%%%%%%%%%%%%%%%%%%%%%%%%%%%%%%%%%%%%%%%%%%%%%%%%%%%%

\subsubsection{Quantum heating and effective temperature}

The anomalous, excitation non-conserving terms in Eq.~(\ref{eq:HKappaQH}) can lead to polariton heating even if the cavity dissipation is at zero temperature.  We can naturally associate an effective temperature to this heating by computing Golden Rule transition rates \cite{Clerk_RMP_2010}.  Considering first the lower-energy $-$ polaritons.  $\hH^{\textrm{int}}_{\kappa}$ will cause transitions uphill in energy between an initial state having $N-1$ polaritons and a final state having $N$ polaritons at a rate $\Gamma_{N,N-1} \propto N |\bar{\alpha}_{d,-}|^2 $.  Similarly, it will cause transitions downhill in energy from the $N$ to $N-1$ polariton state at a rate $\Gamma_{N-1,N} \propto N |\alpha_{d,-}|^2 $.  If these transitions were due to a bath in true thermal equilibrium at temperature $T$, detailed balance dictates that $\Gamma_{N,N-1} / \Gamma_{N-1,N} = \exp(-E_{-} / k_B T )$.  In our case, we can use the ratio of these rates to {\it define} the effective temperature $T_-^{\textrm{cav}}$ of the cavity dissipation as seen by the $-$ polaritons:
\begin{align}
	e^{-E_-/k_BT_-^{\textrm{cav}}} \equiv 
	\frac{\Gamma_{N,N-1} }{ \Gamma_{N-1,N}}
	=\frac{\badm^2}{\adm^2}.
\end{align}
If these transitions were the only dynamics of the $-$ polaritons, they would indeed cause them to reach a thermal state at temperature $T_-^{\textrm{cav}}$, with a thermal occupancy:
\begin{equation}
	n_B[E_-,T_-^{\rm{cav}}] = \frac{1}{e^{E_-/k_BT_-^{\rm{cav}}}-1}=\frac{\badm^2}{\adm^2-\badm^2}. \label{Eq:ncav}
\end{equation}

We thus see the two crucial ingredients needed to obtain a non-zero effective temperature in a bosonic system where the physical bath temperature is zero.  We needed both a coherent drive (yielding effective negative energy bath modes in our interaction picture), and a coherent parametric-amplifier type interactions (i.e.~coherent interactions which do not conserve particle number, and hence yield $\badm \neq 0$).  In a completely analogous fashion, one can also associate an effective temperature $T_+^{\rm{cav}}$ describing the quantum heating of the $+$ polaritons.  We stress
that these effective temperatures have nothing to do with nonlinear interactions.

For a concrete example of this effective temperature physics, consider the special case of a cavity drive at the red mechanical sideband, $\Delta = -\omega_M$.  One finds (c.f.~Eqs.~(\ref{App:COB1})-(\ref{App:COB2})):
\begin{equation}
	\left. n_B[E_\pm,T_{\pm}^{\rm{cav}}] \right\vert_{\Delta=-\omega_M} = \frac{\left(1 - \sqrt{1 \pm 2G/\omega_M}\right)^2}{4\sqrt{1 \pm 2G/\omega_M}}. \label{Eq:Tcav}
\end{equation}
Note that $n_B[E_-,T_{-}^{\rm{cav}}]$ diverges when $G$ approaches the onset of parametric instability, $G \rightarrow \omega_M/2$.  In this limit, $E_- \rightarrow 0$, and the divergence of $n_B[E_-,T_{-}^{\rm{cav}}]$ is equivalent to a fixed effective temperature $T_-^{\rm{cav}} \simeq \omega_M/4 = |\Delta|/4$ near the instability.  Such behaviour is generic for quantum heating near parametric instabilities. 

The type of quantum heating phenomena described here is generic, and is found in a variety of related systems, though the generic nature of the mechanism is often not appreciated. 
Analogous effective temperatures also arise in mean-field treatments of driven dissipative phase transitions.  For example, the study of the driven-dissipative Dicke model in Ref.~\onlinecite{DallaTorre_PRA_2013} finds an effective temperature near the transition identical to that quoted after Eq.~(\ref{Eq:Tcav}).  The physical origin is analogous to that in our system:  it arises from the interplay of coherent parametric-amplifier interactions combined with a coherent linear driving.  
%Finally, we note that this effective temperature physics (which is crucial to strongly-driven optomechanical systems) was neglected in the recent study by Liu \textit{et al.}  \cite{Liu_PRL_2013} of nonlinear effects in driven optomechanical systems.

%%%%%%%%%%%%%%%%%%%%%%%%%%%%%%%%%%%%%%%%%%%%%%%%%%%%%%%%%%%%%%%%%%%%%%%%%%%%%%%%%%%%%%%%%%%%%%%%%%%%%%%%%%%%%%%%%%%%%%%%%%%%%%%%%%%%%%%%%%

\subsubsection{Mechanical dissipation}

We now turn to the interaction between the mechanical resonator and its dissipative bath.  The starting interaction is analogous to Eq.~(\ref{Eq:HCavBathIntIni}) for the cavity dissipation,
\begin{align}
	& \hH_{\gamma} \!=\! \sum_{j}\omega_j \hat{g}^{\dag}_j\hat{g}_j,
	& \hspace{-0.07in} \hH_{\gamma}^{\rm{int}} \!=\! i\sqrt{\frac{\gamma}{2\pi\rho_M}}\sum_{j}\left( \hat{g}^{\dag}_j-\hat{g}_j\right)\left(\hb+\hbdag\right), \label{Eq:HMechBathIntIni}
\end{align}
where $\hat{g}_j$ is a bath annihilation operator, $\gamma$ the mechanical damping rate and $\rho_M$ is the constant DOS of the (Markovian) bath. In what follows, we consider the mechanical bath to be at temperature $T_M$ and define the mean number of excitations inside the bath at $\omega_M$ as
\begin{equation}
	\nthM \equiv n_{B}[\omega_M,T_M], \label{Eq:nMth}
\end{equation}
with $n_{B}[\omega,T]$ being the Bose-Einstein distribution.

As the mechanics is undriven, the bath Hamiltonian and bath-system interaction are unchanged under the transformation of Eq.~(\ref{Eq:UnitaryTrans}) to the rotating frame at the laser frequency.  This time, we re-write the interaction in terms of polariton operators using Eqs.~(\ref{Eq:ChangeOfBasis}) before making further approximations. This difference from the treatment of cavity dissipation stems from the fact that unlike $\omega_{C}$, $\omega_M$ is comparable to $E_{\sigma}$.  It is thus crucial to go to the eigenstates basis of polaritons before assessing which terms may be safely dropped.  In the polariton basis, we have:
\begin{align}
	\hH_{\gamma}^{\rm{int}} = & i\sqrt{\frac{\gamma}{2\pi\rho_M}}\sum_{j, \sigma = \pm} 
		\left( \hat{g}^{\dag}_j-\hat{g}_j\right)  
		\left( \alpha_{b,\sigma} + \bar{\alpha}_{b,\sigma} \right) \left(\hc_\sigma + \hc^\dagger_\sigma \right).
  \label{Eq:HMechBathIntPolFull}
\end{align}

We can now consider the role of terms in Eq.~(\ref{Eq:HMechBathIntPolFull}) that do not conserve excitation number.  In contrast to our treatment of the cavity dissipation, here such anomalous terms can be dropped in a rotating-wave approximation.  As there is no mechanical drive, there are no effective negative energy mechanical bath modes, and hence these terms can never be made resonant.  Thus, we finally obtain a simple rotating-wave interaction between the mechanical bath and the polaritons:
\begin{align}
	\hH_{\gamma}^{\rm{int}} & = i\sqrt{\frac{\gamma}{2\pi\rho_M}}\sum_{j, \sigma=\pm}
		\left[  
			\left(\alpha_\sigma + \bar{\alpha}_\sigma \right)  \hat{g}^{\dag}_j \hc_{\sigma}  + \textrm{H.c.} \right] \label{Eq:HMechBathIntPol}
\end{align}

As there are no excitation non-conserving terms in Eq.~(\ref{Eq:HMechBathIntPol}), it is easy to confirm that the effective temperature of the mechanical dissipation seen by the polaritons is simply equal to the physical temperature of the mechanical dissipation.  As already emphasized, quantum heating requires {\it both} the presence of coherent parametric-amplifier type interactions and the presence of a coherent drive.  Here, while the coherent paramp interactions are present, there is no coherent driving of the mechanics; as such, there is no quantum heating effects involving the mechanical bath.  We see that even if both the cavity and mechanical baths have identical physical temperatures, the polaritons see them as having different effective temperatures.  The driven nature of the system thus gives us interesting non-equilibrium physics even at the level of the linearized (i.e.~quadratic Hamiltonian) theory.

We end this subsection with a caveat on the validity of treating dissipation via Markovian baths.  For the cavity dissipation, this is an excellent approximation, as we are always probing the bath in a narrow interval of width $\sim E_{\sigma}$ around $\omega_C$, an interval over which the bath DOS can be treated as constant (recall that $E_{\sigma} \ll \omega_C$).  In contrast, a similar statement does not hold for the mechanical dissipation:  we will be probing the mechanical bath at frequencies $E_{\sigma}$ which could be significantly different from the mechanical frequency $\omega_M$.  As such, it is not a priori obvious that the bath spectral density can be treated as flat.  For simplicity we will nonetheless use the Markov bath approximation for the mechanics in what follows (consistent with the majority of works in optomechanics).  For the weak dissipation limit of interest, the main effects of a non-flat bath spectral density could be easily incorporated into our calculations.  One would simply make the mechanical contribution $\kappa_M$ to the intrinsic polariton decay rates (c.f.~Eq.~(\ref{Eq:kappas})) proportional to the mechanical bath density of states at the relevant polariton energy.

%%%%%%%%%%%%%%%%%%%%%%%%%%%%%%%%%%%%%%%%%%%%%%%%%%%%%%%%%%%%%%%%%%%%%%%%%%%%%%%%%%%%%%%%%%%%%%%%%%%%%%%%%%%%%%%%%%%%%%%%%%%%%%%%%%%%%%%%%%

\subsection{Lindblad master equation and effective polariton dissipation} \label{Sec:MEQ}

For further insight, it is useful to use the form of the system-bath couplings in Eqs.~(\ref{Eq:HCavBathIntPol}) and (\ref{Eq:HMechBathIntPol}) to derive an approximate Linblad master equation for the dynamics of polaritons in our system.  While we do not use such a master equation in our analysis, it provides a useful comparison point.  Using the standard derivation (see, e.g.,~\cite{GardinerZollerBook}) valid for weakly coupled Markovian baths, we obtain:
\begin{align}
	\dot{\hat{\rho}}(t) =& -i \left[ \hHeff,\hat{\rho}(t) \right] \nonumber \\
	& + \frac{\kappa_-}{2}(\bnom+1)D[\hcm]\hat{\rho} + \frac{\kappa_-}{2}\bnom D[\hcmdag]\hat{\rho}, \nonumber \\
	& + \frac{\kappa_+}{2}(\bnop+1)D[\hcp]\hat{\rho} + \frac{\kappa_+}{2}\bnop D[\hcpdag]\hat{\rho}, \label{Eq:MEQ}
 \end{align}
with the Lindblad super operator $D[\hcs]\hat{\rho}$ defined as
\begin{equation}
	D[\hcs]\hat{\rho} \equiv 2\hcs \hat{\rho}(t)\hcsdag - \hcsdag \hcs \hat{\rho}(t) - \hat{\rho}(t) \hcsdag \hcs.
\end{equation} 

The effective polariton damping rates appearing in Eq.~(\ref{Eq:MEQ}) are
\begin{align}
	\kappa_{\sigma} =\gamma\left( \abs+\babs \right)^2 + \kappa \left( \ads^2-\bads^2 \right) \equiv \kappa_{\sigma}^{\textrm{mech}} + \kappa_{\sigma}^{\textrm{cav}}.
\label{Eq:kappas}
\end{align}
Here, we have introduced $\kappa_{\sigma}^{\textrm{mech}}$ ($\kappa_{\sigma}^{\textrm{cav}}$) as the contribution to the damping rate of the $\sigma$ polariton coming from the interaction with the mechanical resonator (cavity) dissipative bath. The corresponding effective bath thermal occupancies are
\begin{subequations} \label{Eq:ns}
\begin{align}
	\bnos & =\frac{1}{\kappa_{\sigma}}\left[\gamma\left( \abs+\babs \right)^2 n_{B}[E_{\sigma},T_M] + \kappa \bads^2 \right], \\
	& = \frac{\kappa_{\sigma}^{\textrm{mech}} n_{B}[E_{\sigma},T_M] + \kappa_{\sigma}^{\textrm{cav}} n_{B}[E_{\sigma},T_{\sigma}^{\textrm{cav}}]}{\kappa_{\sigma}^{\textrm{mech}} + \kappa_{\sigma}^{\textrm{cav}}} \label{Eq:ns2}
\end{align}
\end{subequations}
Here, $T_M$ is the (physical) temperature of the mechanical bath (cf.~Eq.~(\ref{Eq:nMth})), $T_{\sigma}^{\textrm{cav}}$ is the temperature of the cavity bath as seen by the $\sigma$ polariton (cf.~Eq.~(\ref{Eq:ncav})) and the different coefficients $\alpha$ are given in Eq.~(\ref{Eq:ChangeOfBasis}). As expected, Eq.~(\ref{Eq:ns2}) represents a bosonic mode coupled independently to two disipative baths. An analogous expression holds for the effective mechanical occupancy used to describe cavity-cooling experiments \cite{Marquartd_PRL_2007}. Note that we have made a standard secular approximation, allowing us to drop dissipative terms that do not conserve the number of each polaritons independently in Eq.~\eqref{Eq:MEQ}; this is valid for the regime of interest
$E_{\sigma},\left\lvert E_{+}-E_{-} \right\rvert \gg \kappa,\gamma$. 

\begin{figure}[t]
	\begin{center}
	\includegraphics[width= 0.95\columnwidth]{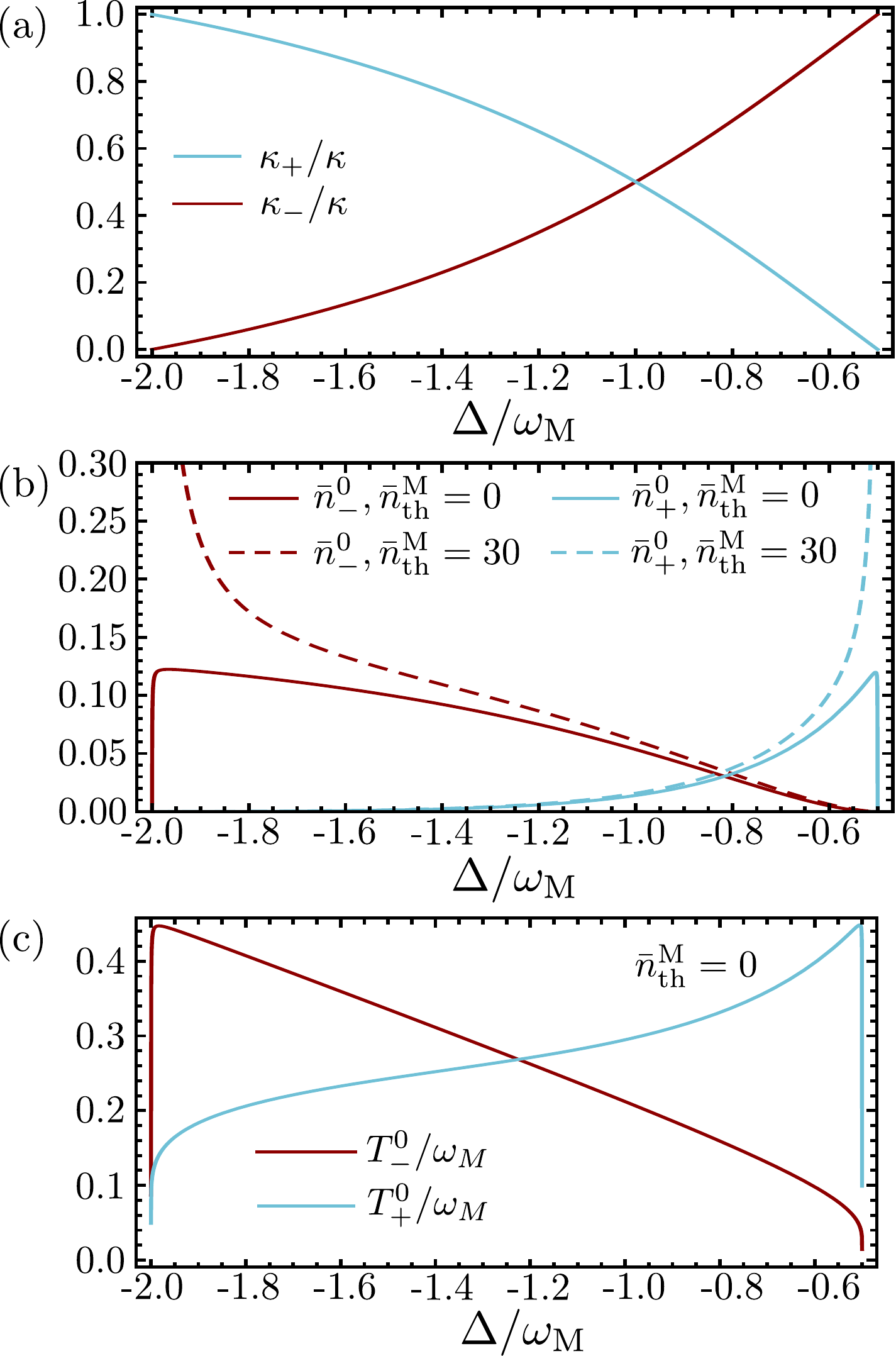}
	\end{center}
	\vspace{-0.5cm}
	\caption {(Color online)
		(a) Damping rates $\kappa_{\sigma}$ of the $\sigma = \pm$ polaritons (cf.~Eq.~(\ref{Eq:kappas})) in the linearized theory (i.e.~$\tilde{g}=0$), as a function of the detuning $\Delta$. 
		For each $\Delta$, we tune the cavity control drive to maintain $G = \Gres(\Delta)$. Near $\Delta = -2\omega_M$, the - (+) polariton damping tends to $\gamma$ ($\kappa$) as it is mostly phonon (photon) like; the converse is true near $\Delta = -\omega_M/2$.
(b) Thermal occupation numbers $\bnos$ of the effective bath coupled to the $\sigma$ polaritons (cf.~Eq.~(\ref{Eq:ns})) also when $\tilde{g}=0$ and $G=\Gres$. Here, $\nthM$ characterises the (physical) temperature of the mechanical bath (cf.~Eq.~\eqref{Eq:nMth}). 
(c) The temperatures corresponding to the thermal occupation $\bnos$ ($k_{B}=1$) in the same regime as panels (a) and (b).
For each curves, $\gamma/\kappa=10^{-4}$ and $\omega_M/\kappa=50$.}
\label{fig:Parambath}
\end{figure}

The thermal occupation number of the effective baths given in Eqs.~(\ref{Eq:ns}) and their corresponding temperatures, defined as $n_{B}[E_{\sigma},\Tos] \equiv \bnos$, are plotted in Fig.~\ref{fig:Parambath} as a function of the detuning $\Delta$.  Anticipating our interest in nonlinear interactions, for each $\Delta$ we adjust the control laser amplitude so that $G = G_{\rm res}$
(i.e.~the value that will make the nonlinear interaction resonant, c.f.~Eq.~(\ref{Eq:Gres})); this can be done for any $\Delta  \in [-2\omega_M,-\omega_M/2]$.  
One sees that even when the physical bath temperature is zero, quantum heating effects can yield effective polariton temperatures as large as $\sim 0.1$ quanta (solid curves).  At non-zero physical temperature (dashed curves), these quantum heating effects persist, but are swamped by the contribution of mechanical noise at the edges of the detuning range considered.  This is simply because near the limits of the detuning range, one polariton species is almost all phononic (see Fig.~\ref{fig:LinearParams}) and becomes very sensitive to thermal fluctuations of the mechanical bath.

Returning to zero physical bath temperature, another striking feature in Fig.~\ref{fig:Parambath}(c) is the sudden drop in effective temperature for the more phonon-like polariton branch at the edges of the detuning interval.  As it will be of interest in what follows, we discuss this behaviour for detunings 
near $\Delta = -2\omega_M$ in more detail; here, one sees a sudden drop in $\bnom$ and $\Tom$. 
As $\Delta \rightarrow - 2 \omega_M$, $\Gres \rightarrow 0$, and the $-$ polariton becomes simply a phonon.  
Expanding $\bnom$ to lowest order in $\Gres/\omega_M$ for $T_M=0$ and $\gamma \ll \kappa$, one gets
\begin{equation}
	\bnom =\frac{\kappa}{\kappa_-}\badm^2 \approx \frac{\frac{1}{9}\frac{G_{\rm{res}}^2}{\omega_M^2}\kappa}{\gamma+\frac{8}{9}\frac{G_{\rm{res}}^2}{\omega_M^2}\kappa}. 
	\label{Eq:nminusNear2omegaM}
\end{equation}
We see there are two competing effects associated with non-zero $\Gres$. The numerator reflects the parametric heating associated with the linearized optomechanical interaction.  The denominator in contrast reflects that the phonon-like polariton has its lifetime decrease as $\Gres$ increases and it becomes more photon like; this is just standard optomechanical optical damping.  
The result is that the net quantum heating is maximized for $[\Gres/\omega_M]^2 \sim \gamma/\kappa$), corresponding to a laser 
detuning $\Delta + 2\omega_M \sim \gamma/\kappa$; the maximum occupancy $\bnos$ that can be obtained is $\frac{1}{8}$ (as seen in Fig.~\ref{fig:Parambath}(b)).

%%%%%%%%%%%%%%%%%%%%%%%%%%%%%%%%%%%%%%%%%%%%%%%%%%%%%%%%%%%%%%%%%%%%%%%%%%%%%%%%%%%%%%%%%%%%%%%%%%%%%%%%%%%%%%%%%%%%%%%%%%%%%%%%%%%%%%%%%%

\section{Keldysh description of the linearized system} \label{Sec:Keldysh}

As shown in the previous section, the driven nature of the system leads to non-equilibrium physics even at the level of the linearized theory (i.e.~the two polariton species see different effective temperatures). Consequently, we need to use the Keldysh formalism \cite{KamenevBook} in order to describe the dynamics and to properly construct a perturbation theory that treats the nonlinear interaction present in $\hHeff$ (Eq.~(\ref{Eq:HEff})).  In this section, we quickly introduce this approach, considering first the linearized Hamiltonian of Eq.~(\ref{Eq:HL}) and the couplings with the environment given by Eqs.~(\ref{Eq:HCavBathIntPol}) and (\ref{Eq:HMechBathIntPol}).  We will also map the resulting free Keldysh theory onto the simple master equation previously derived (Eq.~(\ref{Eq:MEQ})), again
working in the regime where polaritons have a well defined energy (i.e.~$\kappa_{\sigma} \ll E_{\sigma}$).
We stress however that even without interactions, the Keldysh approach is more general than a Lindblad-style master equation, as it is not restricted to Markovian baths. 

In the Keldysh formalism, we represent our linearized optomechanical system by a field theory which is general enough to allow the system to be in an arbitrary, non-equilibrium state. In this field theory, there are two time-dependent fields (classical and quantum) corresponding to each annihilation operator in the original theory.  Consequently, the quadratic action that conserves the number of particles of our two independent bosonic modes (polaritons) will have the following general form:
\begin{align}	
	\hspace{-0.3cm} S_{\rm{L}} =\sum_{\sigma = \pm} & \int\limits_{-\infty}\limits^{\infty}\int\limits_{-\infty}\limits^{\infty}dtdt' (c^*_{\sigma, cl}(t),c^*_{\sigma, q}(t)) \textbf{G}_{\sigma}^{-1}(t-t') \binom{c_{\sigma, cl}(t')}{c_{\sigma, q}(t')}.\label{Eq:SLvsGF}
\end{align}
Here, the $c_{\sigma,q/cl}(t)$ are complex functions of time and $\textbf{G}_{\sigma}^{-1}(t-t')$ is the (operator) inverse of the unperturbed (i.e. $\tilde{g} = 0$) Green function.  The latter is given by a $2\times2$ matrix
\begin{equation}
	\textbf{G}_{\sigma}(t) = \begin{pmatrix}
  	G^K_{\sigma}(t) & G^R_{\sigma}(t) \\
  	G^A_{\sigma}(t) & 0 \\
 	\end{pmatrix}.
\end{equation}
In terms of Heisenberg picture operators, each element is defined as
\begin{subequations}\label{Eq:DefG}
\begin{align}
	G^R_{\sigma}(t) & = \left\lbrace G^A_{\sigma}(t) \right\rbrace^*  \equiv -i \theta(t) \langle [\hc_{\sigma}(t),\hc^{\dag}_{\sigma}(0)] \rangle, \label{Eq:DefGR}\\
	G^K_{\sigma}(t) & \equiv -i \langle \lbrace \hc_{\sigma}(t),\hc^{\dag}_{\sigma}(0)\rbrace \rangle, \label{Eq:DefGK}
\end{align}
\end{subequations}
where the expectations are taken with respect to the initial density matrix without nonlinear interaction ($\tilde{g}=0$). Here, $G^R_{\sigma}(t) $ and $G^A_{\sigma}(t) $ are the standard unperturbed retarded and advanced Green functions, which govern the linear response properties of the unperturbed system. They are also related to the unperturbed DOS of each polariton, given by
\begin{equation}
\rho^{0}_{\sigma}[\omega]=-\frac{1}{\pi}\textrm{Im}[\GRos]. \label{Eq:BareDOS}
\end{equation}
Finally, $G^K_{\sigma}(t)$ is known as the (unperturbed) Keldysh Green function.  It encodes knowledge of the energy distribution function of each polariton (as we will see more clearly below).

With these definitions in hand, one could follow the standard approach used in input-output theory and derive the Heisenberg-Langevin equations from the coherent Hamiltonian of Eq.~(\ref{Eq:HEff}) and the particular form of the system baths coupling given by Eqs.~(\ref{Eq:HCavBathIntPol}) and (\ref{Eq:HMechBathIntPol}). From there, one can directly get the bare Green functions by calculating Eqs.~(\ref{Eq:DefG}). This approach has been used in our previous work \cite{MAL_PRL_2011}.  Here, we instead follow a different but equivalent route to obtain the bare Green functions and Keldysh action.

The goal here is to write the different Green functions such that our description of the linearized theory in the Keldysh formalism is completely equivalent to the master equation of Eq.~(\ref{Eq:MEQ}). As discussed before, Eq.~(\ref{Eq:MEQ}) describes two independent bosonic modes with simple Markovian damping rates $\kappa_{\sigma}$ (cf.~Eq.~(\ref{Eq:kappas})) and energies $E_{\sigma}$ (cf.~Eq.~(\ref{Eq:Esigma})). Consequently, the retarded (advanced) Green functions, which give the response functions of the system, have to adopt the simple following form:
\begin{align}
	\GRos = \left\lbrace \GAos \right\rbrace^* = \frac{1}{\omega - E_{\sigma} + i \kappa_{\sigma}/2}, \label{Eq:GR0}
\end{align}
i.e.~a polariton has an energy $E_{\sigma}$ and a lifetime $1 / \kappa_{\sigma}$.

We now construct the Keldysh Green functions using the same approach. As is standard \cite{KamenevBook}, we define a distribution function $f[\omega]$ that relates the Keldysh and the retarded Green functions, such that
\begin{equation}
	\GKos \equiv -2i (2f[\omega]+1) \textrm{Im}[\GRos]. \label{Eq:FDT}
\end{equation}
This function $f[\omega]$ parameterizes the occupation of different polariton energy eigenstates. As an example, for a system in thermal equilibrium at a certain temperature, $f[\omega]$ would be the corresponding Bose-Einstein distribution and Eq.~(\ref{Eq:FDT}) would be an exact statement of the fluctuation-dissipation theorem \cite{KamenevBook}. In the particular case studied here, the free polaritons have sharply peaked single particle DOS $\rho_{\sigma}[\omega]$ (cf.~Eqs.~(\ref{Eq:BareDOS}) and (\ref{Eq:GR0}) with $\kappa_{\sigma} \ll E_{\sigma}$), such that for the $\sigma$ polariton, the function $f[\omega]$ of Eq.~(\ref{Eq:FDT}) can be 
approximated as $f[\omega] \simeq f[E_{\sigma}]$.  Finally, if we insist that the average occupancy of the polariton matches that in the master equation description, then we must have $f[E_{\sigma}] = \bnos$.  We thus have
\begin{align}
	\GKos &= -2i(2\bnos+1) \textrm{Im}[\GRos]. \label{Eq:GK0}
\end{align}

Describing the linearized theory in the Keldysh formalism using the bare Green functions (\ref{Eq:GR0}) and (\ref{Eq:GK0}) is thus completely equivalent to the Lindblad master equation (\ref{Eq:MEQ}). We recall that the two assumptions underlying these two equivalent descriptions are that the polaritons have sharply peaked DOS (i.e.~dissipation is weak) and that coupling between the $+$ and $-$ polaritons due to dissipation is negligible (i.e.~secular approximation made to derive Eq.~\eqref{Eq:MEQ}, see Sec.~\ref{Sec:MEQ}).

Finally, we note that it is possible to derive exact Langevin equations from the linear Keldysh action given in Eq.~(\ref{Eq:SLvsGF}). Briefly, one first decouples the quadratic quantum-field terms via an exact Hubbard-Stratonovich transormation; this introduces new fields $\xi^{\rm env}_{\sigma}(t)$ which have a Gaussian action.  One can then exactly do the integrals over quantum and classical fields.  The resulting functional delta function corresponds to the Langevin equations:  
\begin{align}
	\partial_t c_{\sigma}(t)=-\left(iE_{\sigma}+\frac{\kappa_{\sigma}}{2}\right)c_{\sigma}(t)-\xi^{\rm{env}}_{\sigma}(t), \label{Eq:LinLangevinEq}
\end{align}
where the noise $\xi^{\rm{env}}_{\sigma}(t)$ is Gaussian with zero mean.  The only non-zero noise correlation functions are given by
\begin{align}
	\langle \xi^{\rm{env}}_{\sigma}(t) [\xi^{\rm{env}}_{\sigma'}(t')]^{*}\rangle = \kappa_{\sigma}(\bar{n}^0_{\sigma}+1/2) \delta(t-t')\delta_{\sigma,\sigma'}. \label{Eq:NoiseCorrPolbath}
\end{align}
As expected, Eqs.~(\ref{Eq:LinLangevinEq}) and (\ref{Eq:NoiseCorrPolbath}) represent two uncoupled damped harmonic oscillators each in contact with their respective finite temperature Markovian baths.

%%%%%%%%%%%%%%%%%%%%%%%%%%%%%%%%%%%%%%%%%%%%%%%%%%%%%%%%%%%%%%%%%%%%%%%%%%%%%%%%%%%%%%%%%%%%%%%%%%%%%%%%%%%%%%%%%%%%%%%%%%%%%%%%%%%%%%%%%%

\section{Keldysh perturbative treatment of polariton interactions} \label{Sec:PerturbationTheory}

\subsection{Self energies and dressed Green functions} 

Have established the Keldysh formulation of the linearized optomechanical theory, we can now address the effects of the nonlinear interaction as a perturbation.
We assume throughout this section that the drive laser has been tuned to make the non-linear interaction resonant, i.e.~$G = \Gres$, and thus consider only the resonant interaction process given in Eq.~(\ref{Eq:HEff}).

\begin{figure}[t]
	\begin{center}
	\includegraphics[width= 0.99\columnwidth]{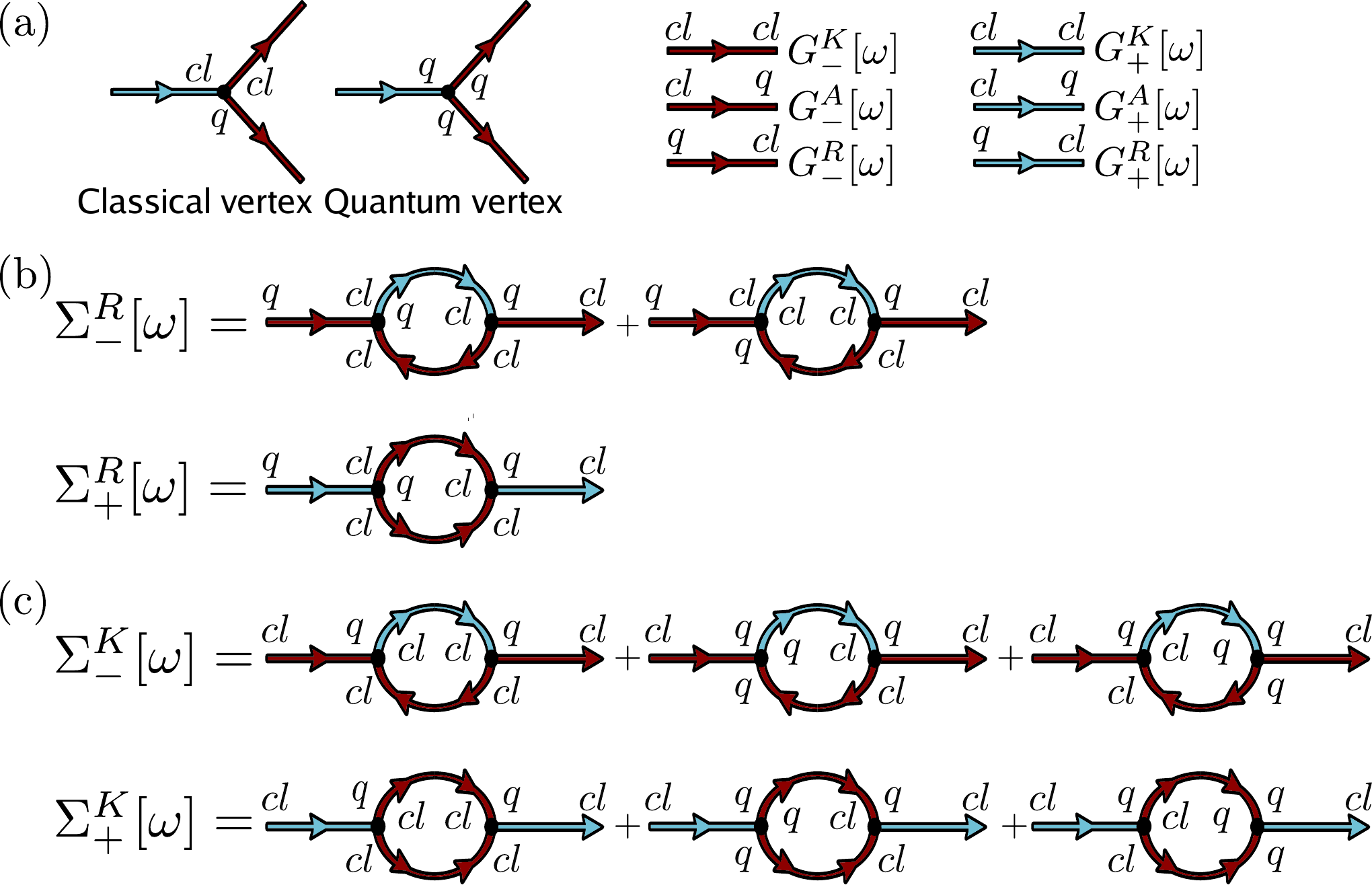}
	\end{center}
	\vspace{-0.5cm}
	\caption {(Color online)
	Diagrams representing the self energies up to second order in $\tilde{g}$ in the Keldysh formalism. The classical vertices are composed of only one quantum component while the quantum vertex is composed of three.}
\label{fig:Diagrams}
\end{figure}

First, the action generated by the nonlinear interaction in $\hat{H}_{\rm{eff}}$ (Eq.~(\ref{Eq:HEff})) in the \textit{cl-q} basis is
\begin{align}
	S_{\rm{NL}} = \frac{\tilde{g}}{\sqrt{2}} \int_{-\infty}^{\infty} & dt \left( c^*_{+,q}c_{-,cl}c_{-,cl} + 2c^*_{+,cl}c_{-,q}c_{-,cl} \right. \nonumber \\
	& \left. + c^*_{+,q}c_{-,q}c_{-,q} + \textrm{C.c.} \right), \label{Eq:SNL}
\end{align}
where the time dependence of the fields are implicit for clarity. Diagrammatically, each terms in the nonlinear action corresponds to a vertex shown in Fig~\ref{fig:Diagrams}(a). The vertices with a single quantum field correct the classical saddle point,
\begin{subequations}
\begin{align}
	\left. \frac{\partial (S_{\textrm{L}} + S_{\textrm{NL}})}{\partial c^*_{+, q}(t)} \right\vert_{c_{\pm, q}=0} = &\int_{-\infty}^{\infty} dt' \left[G^R_{+}(t-t')\right]^{-1}c_{+, cl}(t') \nonumber \\
	& + \frac{\tilde{g}}{\sqrt{2}}c_{-,cl}(t)c_{-,cl}(t), \\
	\left. \frac{\partial (S_{\textrm{L}} + S_{\textrm{NL}})}{\partial c^*_{-, q}(t)} \right\vert_{c_{\pm, q}=0} = &\int_{-\infty}^{\infty} dt' \left[G^R_{-}(t-t')\right]^{-1}c_{-, cl}(t') \nonumber \\
	& + \sqrt{2}\tilde{g}c_{+,cl}(t)c^*_{-,cl}(t).
\end{align}
\end{subequations}
and thus correspond to a classical nonlinear potential. The term with three quantum fields is more of a purely quantum effect.  It could be interpreted as an effective nonlinearity of the quantum noise.

One can still derive Langevin equations from the resulting nonlinear action if one ignores the terms which are cubic in quantum fields.  In this approximation, one obtains modified versions of the Langevin equations in Eq.~(\ref{Eq:LinLangevinEq}):
\begin{subequations}
\begin{align}
	\partial_t c_{-} & =-(iE_{-}+\frac{\kappa_{-}}{2})c_{-} - 2i\tilde{g}c^*_{+}  c_{-} -\xi^{\rm{env}}_{-}, \\
	\partial_t c_{+} & =-(iE_{+}+\frac{\kappa_{+}}{2})c_{+} - i\tilde{g}c_{-} c_{-} -\xi^{\rm{env}}_{+},
\end{align}
\end{subequations}
where the autocorrelation functions of the $\xi^{\rm}_\sigma(t)$ noise are unchanged by the nonlinear interactions.  We have suppressed the explicit time-dependence of fields here for clarity.  

We will not use these approximate quantum Langevin equations further, but proceed in a way that does not neglect terms that are cubic in the quantum fields.  As we are interested in weak nonlinear couplings $\tilde{g}$, we
will compute the self energy ($\Sigma[\omega]$)  of our Keldysh Green functions perturbatively to order $\tilde{g}^2$.  At this order, all relevant scattering processes (see Fig.~\ref{fig:Diagrams}) conserve the number of polaritons independently (i.e.~the self energies are diagonal in the +/- index). Consequently, the Dyson equation that gives the Green functions in presence of interactions can be separately written for each polariton.
\begin{equation}
	\begin{pmatrix}
  	\GKtots & \GRtots \\
  	\GAtots & 0 \\
 	\end{pmatrix}^{-1}
 	= \textbf{G}_{\sigma}[\omega]^{-1}
 	- \begin{pmatrix}
  	\SigmaKs & \SigmaAs \\
  	\SigmaRs & 0 \\
 	\end{pmatrix}.
 	\label{Eq:Dyson}
\end{equation}
Here, we used $\mathcal{G}_{\sigma}^{A,R,K}[\omega]$ to distinguish the full Green functions (i.e.~including the effects of $\tilde{g}$) from the unperturbed ones $G_{\sigma}^{A,R,K}[\omega]$.

%%%%%%%%%%%%%%%%%%%%%%%%%%%%%%%%%%%%%%%%%%%%%%%%%%%%%%%%%%%%%%%%%%%%%%%%%%%%%%%%%%%%%%%%%%%%%%%%%%%%%%%%%%%%%%%%%%%%%%%%%%%%%%%%%%%%%%%%%%

%%%%%%%%%%%%%%%%%%%%%%%%%%%%%%%%%%%%%%%%
%%%%%%%%%%%%%%%%%%%%%%%%%%%%%%%%%%%%%%%%

\subsubsection{Retarded self energies and interaction-induced polariton damping}
 
 The diagrams related to the second order retarded self energies are shown in Fig.~\ref{fig:Diagrams}(b). From these, one straightforwardly calculates
\begin{subequations}\label{Eq:Sigmas}
\begin{align}
	\SigmaRm & = \Ceffm\frac{\kappa_-}{4}\frac{\kappa_-+\kappa_+}{\omega-(E_+-E_-)+i\frac{\kappa_-+\kappa_+}{2}}, \label{Eq:SigmaRm}\\
	\SigmaRp & = \Ceffp\frac{\kappa_+}{2}\frac{\kappa_-}{\omega-2E_-+i\kappa_-}, \label{Eq:SigmaRp}
\end{align}
\end{subequations}
with $\SigmaAs =\lbrace\SigmaRs\rbrace^*$. As discussed in detail in \cite{MAL_PRL_2011}, the self energies $\SigmaRs$ describe the hybridization between the near-resonant $\vert + \rangle$ and $\vert -,- \rangle$ polariton states. The resonant nonlinear processes underlying this hybridization (see Fig.~\ref{fig:Diagrams}(b)) are responsible for the sharply peaked self energies of Eqs.~(\ref{Eq:Sigmas}) ($\kappa_{\sigma} \ll E_{\sigma}$). We have introduced effective cooperativites $\Ceffs$ to parameterize how strong the decay rates resulting from these processes are (i.e.~imaginary part of the self energy) on resonance, compared to the intrinsic polariton linewidth.  Defining
\begin{align}
	\Gammaints & =-2 \textrm{Im}\left[\SigmaRs\right], \label{Eq:Gammaint}
\end{align}
we have
\begin{subequations} \label{Eq:Ceff}
\begin{align}
	\Ceffm & \equiv \frac{\Gamma^{\rm int}_{-}[E_+-E_-]}{\kappa_-}  = \frac{16\tilde{g}^2(\bnom-\bnop)}{\kappa_-(\kappa_-+\kappa_+)}, \label{Eq:Cmeff} \\
	\Ceffp & \equiv \frac{\Gamma^{\rm int}_{+}[2 E_{-}] }{\kappa_+} = \frac{4\tilde{g}^2(2\bnom+1)}{\kappa_-\kappa_+}. \label{Eq:Cpeff}
\end{align}
\end{subequations}

The definitions of $\Ceffm, \Ceffp$ are analogous to the definition of the standard optomechanical cooperativity $\mathcal{C} = 4 G^2 / \kappa \gamma$ as the cavity-induced ``optical damping" of the mechanics to the intrinsic mechanical damping. 
The effective cooperativities are plotted in Fig.~\ref{fig:Ceff} as a function of the detuning (keeping $G = \Gres$ for all detunings).  

A crucial feature of the interaction-induced polariton damping described by Eqs.~(\ref{Eq:Ceff}) is their explicit temperature dependence.  This is a direct consequence of the 
multi-particle nature of the relevant decay process. For the $-$ polariton, we have $\Gamma^{\rm int}_{-} \propto (\bnom-\bnop)$, as expected for a bosonic polarization bubble;
a similar damping rate is found for an oscillator coupled quadratically to an oscillator bath \cite{Dykman_PSS_1975}.  In true thermal equilibrium, the fact that $E_+ > E_-$ guarantees this factor is positive, yielding $\Gamma^{\rm int}_{-} > 0$.   Our system however is not in true thermal equilibrium: as discussed in the previous section, it is possible to have $(\bnom-\bnop) < 0$ by having suitably different
effective temperatures for the two polariton species.  One thus finds that the interactions can lead to negative damping:  $\Gamma^{\rm int}_{-} < 0$.  The physics of this regime and the possibility of true instability are discussed further in Sec.~\ref{Sec:ParamInsta}. 

The retarded polariton self energies presented here directly lead to an interaction-induced modification of the polariton DOS, $\rho_{\sigma}[\omega]$, given by
\begin{equation}
\rho_{\sigma}[\omega] \!\equiv\! -\frac{1}{\pi}\textrm{Im}\left[ \GRtots \right] \!=\! -\frac{1}{\pi}\textrm{Im}\left[ \frac{1}{\omega\!-\!E_{\sigma} \!+\! i \kappa_{\sigma}/2 \!-\! \SigmaRs} \right]\!,\label{Eq:DOS}
\end{equation}
Signatures of $g$ in the DOS (and corresponding changes to OMIT-style experiments) were the focus of our previous work \cite{MAL_PRL_2011}. For strong enough nonlinear coupling $g$, the self energies turn from simply describing an extra broadening of the polaritons, to describing the coherent hybridization of the resonant $|+\rangle$ and  $|-,-\rangle$ polariton states.  The key consequence of this is that the single peak in the $+$ polariton DOS splits.  Further details about this splitting (and how it can be measured via an OMIT-type experiment) can be found in 
Ref.~\onlinecite{MAL_PRL_2011}.

%%%%%%%%%%%%%%%%%%%%%%%%%%%%%%%%%%%%%%%%%%%%%%%%%%%%%%%%%%%%%%%%%%%%%%%%%%%%%%%%%%%%%%%%%%%%%%%%%%%%%%%%%%%%%%%%%%%%%%%%%%%%%%%%%%%%%%%%%%

\subsubsection{Keldysh self energies}

We now turn to the Keldysh self energies, which are also directly calculated from the diagrams of Fig.~\ref{fig:Diagrams}(c). Here, we parametrize each Keldysh self energies via a thermal occupancy factor $\bnints$ associated with the interaction, defined such that
\begin{equation}
	\SigmaKs \equiv  -2i (2\bnints+1) \textrm{Im}\left[\SigmaRs\right]. \label{Eq:SigmaK}
\end{equation}
This parametrization is always possible if we let the $\bnints$ to be frequency dependent. However, 
we find them to be frequency independent and given by
\begin{align}
	\bnintm & = \frac{\bnop(\bnom+1)}{\bnom-\bnop}, 
	& \bnintp = \frac{(\bnom)^2}{2\bnom+1}. \label{Eq:nint}
\end{align}
We stress that these results (as well as the self-energy results above) are based on only keeping the nonlinear polariton interaction in Eq.~\eqref{Eq:HEff}, and thus assume that
$E_+ \simeq 2 E_-$.

%%%%%%%%%%%%%%%%%%%%%%%%%%%%%%%%%%%%%%%%%%%%%%%%%%%%%%%%%%%%%%%%%%%%%%%%%%%%%%%%%%%%%%%%%%%%%%%%%%%%%%%%%%%%%%%%%%%%%%%%%%%%%%%%%%%%%%%%%%

\subsection{Self-consistent calculation} \label{Sec:SelfConsistentApp}

The self-energy results discussed so far (and in Ref.~\onlinecite{MAL_PRL_2011}) only retain diagrams to leading order in $g$.  To capture higher order effects and effectively resum diagrams at all orders in perturbation theory, one can make the diagrams in Fig.~\ref{fig:Diagrams} self-consistent.  One simply replaces all internal propagators in the diagrams by full dressed propagators.  The self energy thus becomes a functional of the full dressed Green function, and the Dyson equation becomes a self-consistent equation for the full Green function. Solving this self-consistent Dyson equation allows to capture a particular ensemble of processes at all orders in $g$.  Note that a related self-consistent Keldysh approach was previously used to study a nonlinear parametric amplifier near threshold \cite{Mertens_PRA_1993, Mertens_PRA_1995, Veits_PRA_1997}.

\begin{figure}[t]
	\begin{center}
	\includegraphics[width= 0.98\columnwidth]{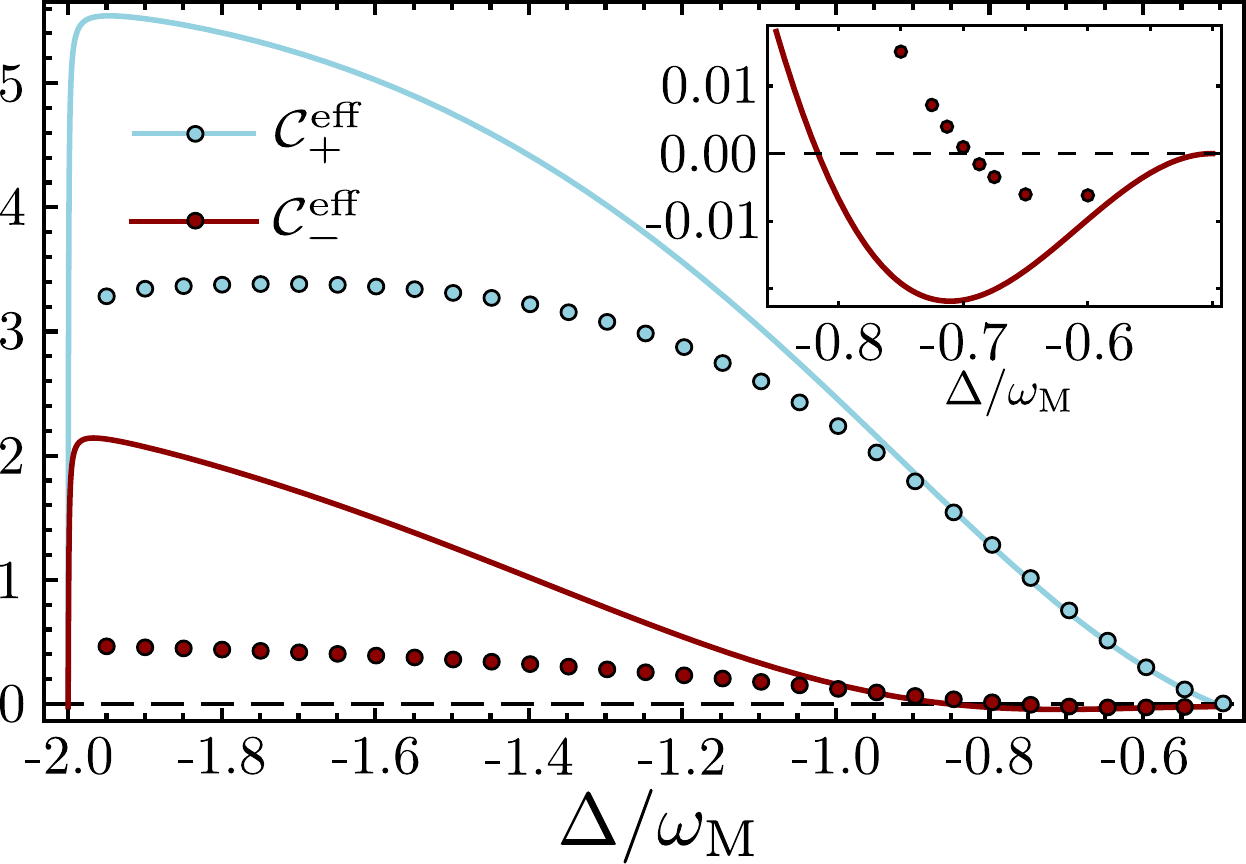}
	\end{center}
	\vspace{-0.5cm}
	\caption{(Color online)
	Solid lines: Leading-order-in-$g$ effective cooperativities $\Ceffs$ associated with the nonlinear interaction (defined in Eq.~(\ref{Eq:Ceff})), 
	as a function of the detuning in the resonant regime ($G=\Gres[\Delta]$).  One sees that the $\Ceffs$ are enhanced near $\Delta = - 2 \omega_M$.  
	The inset zooms on the region where $\Ceffm$ goes below zero, which signals the possibility of a new kind of instability, as discussed explicitly in Sec.~\ref{Sec:ParamInsta}. 
	Circles:  results of the self-consistent perturbation theory described in Sec.~\ref{Sec:SelfConsistentApp}, which includes diagrams at all orders in $g$. 
	For all values of $\Ceffs$ shown here, the approach converges 
	and is in good agreement with numerical simulations of the Lindblad master equation (cf.~Eq.~\eqref{Eq:MEQ}). 
	The parameters used are $\gamma/\kappa=10^{-4}$, $g=\kappa$, $\omega_M=50\kappa$ and $\nthM = 0$.}
\label{fig:Ceff}
\end{figure}

In practice, one solves such self-consistent equations iteratively:  in each step, one calculates the self energy using the current versions of the full Green functions, and then uses these to update the Green functions which will be used for the self-energy calculation in the next iteration. Applying this iterative procedure until convergence solves the self-consistent Dyson equation. To improve the accuracy of our approach, we have applied this iterative strategy for the results shown in Fig.~\ref{fig:Ceff} to Fig.~\ref{fig:Sd2wm}. For each calculation, we have performed 20 iterations, which turns out to be more than enough to get excellence convergence of the Green functions. Note that by using this self-consistent approach, $\bnints$, as defined in Eq.~\eqref{Eq:SigmaK}, becomes frequency dependent.

As expected, the self-consistent approach does not converge if the nonlinear interactions become too strong.  Convergence is not solely controlled by the magnitude of $\tilde{g}$, but is primarily determined by the effective cooperativities $\Ceffs$ defined in Eqs.~(\ref{Eq:Ceff}).  These cooperativities  involve both the magnitude of $\tilde{g}$ and the polariton occupancies, reflecting the fact that large temperature can also enhance the importance of nonlinearity.  As shown in the figures, we find that when the self-consistent approach converges, it also is in excellent agreement with full numerical simulations of the Linblad master equation describing the system, Eq.~(\ref{Eq:MEQ}).

%%%%%%%%%%%%%%%%%%%%%%%%%%%%%%%%%%%%%%%%

%%%%%%%%%%%%%%%%%%%%%%%%%%%%%%%%%%%%%%%%%%%%%%%%%%%%%%%%%%%%%

\section{Influence of polariton interactions on non-equilibrium effects}
\label{Sec:NoneqDescription}

\subsection{Interaction-induced effective environment}

From our previous discussion, we see that the nonlinear interaction gives rise to self energies which modify the single-particle properties of polaritons.  The imaginary part of the retarded self energies describe new interaction-induced polariton damping rates $\Gammaints$ (c.f.~Eq.~(\ref{Eq:Gammaint})), whereas the Keldysh self energies describe new interaction-induced heating effects, with associated thermal occupancy factors $\bar{n}^{\rm int}_\sigma$ (c.f.~Eq.~(\ref{Eq:nint})).  This suggests that in terms of single-particle properties, 
{\it the effects of interactions are equivalent to having coupled the linearized optomechanical system to new dissipative baths}.  In what follows, we make this picture of an ``interaction-induced effective environment" explicit.

First, note that all single particle polariton properties of our system are described by the effective quadratic action
\begin{align}	
	S^{\textrm{eff}}_{\sigma} = \iint\limits_{-\infty}\limits^{\quad\infty}dtdt' (c^*_{\sigma, cl}(t),c^*_{\sigma, q}(t)) \mathcal{G}_{\sigma}^{-1}(t-t') \binom{c_{\sigma, cl}(t')}{c_{\sigma, q}(t')}.\label{Eq:ApproxSNL}
\end{align}
where $\mathcal{G}_{\sigma}^{-1}(t-t')$ is the Fourier transform of the inverse of the $2\times2$ matrix of the full  Green functions given in Eq.~(\ref{Eq:Dyson}).  

Further, as discussed in Sec~\ref{Sec:Keldysh}, this quadratic action is completely equivalent to a set of linear Langevin equations.  Using the standard derivation \cite{KamenevBook}, the effective action in Eq.~(\ref{Eq:ApproxSNL}) is equivalent to the Langevin equations
\begin{align}
	\partial_t c_{\sigma}(t) = & -(iE_{\sigma}+\frac{\kappa_{\sigma}}{2})c_{\sigma}(t) -\int_{-\infty}^{\infty}dt' \Sigma^{R}_{\sigma}(t-t')c_{\sigma}(t') \nonumber \\
	&  - \xi^{\rm{env}}_{\sigma}(t)  - \xi^{\rm{int}}_{\sigma}(t), \label{Eq:AppLangevinEq}
\end{align}
Here, $\Sigma^{R}_{\sigma}(t-t')$ is the Fourier transform of the retarded self energies given in Eqs.~(\ref{Eq:Sigmas}). The noise functions $\xi^{\rm{env}}_{\sigma}(t)$ and $\xi^{\rm{int}}_{\sigma}(t)$ describe complex independent Gaussian noise processes with zero mean.  $\xi^{\rm{env}}_{\sigma}(t)$ describes the intrinsic polariton dissipation as discussed in section~\ref{Sec:Keldysh}; its correlators are given in 
Eq.~(\ref{Eq:NoiseCorrPolbath}).  The only non-zero correlator of the new noise $\xi^{\rm{int}}_{\sigma}(t)$ is
\begin{align}
	& \langle \xi^{\rm{int}}_{\sigma}(t) [\xi^{\rm{int}}_{\sigma'}(t')]^{*}\rangle = \Gamma^{\rm{int}}_{\sigma}(t-t') (\bnints+1/2)\delta_{\sigma,\sigma'}. \label{Eq:NoiseCorrIntbath}
\end{align} 
where $\Gamma^{\rm{int}}_{\sigma}(t-t')$ is the Fourier transform of the frequency-dependent interaction-induced polariton damping given in Eq.~(\ref{Eq:Gammaint}). 

These Langevin equations reproduce the intuitive picture sketched above:  each polariton species is now effectively coupled to two independent dissipative environments, with corresponding damping rates $\kappa_{\sigma}$ and $\Gammaints$, and corresponding thermal occupancies $\bnos$ and $\bnints$.  The first bath corresponds to intrinsic dissipation (i.e.~the intrinsic mechanical and cavity dissipation), whereas the second is due to polariton-polariton interactions.    

It is worth stressing that the induced damping rates $\Gammaints$ are in general sharply peaked functions of frequency, 
due to the resonant nature of the relevant scattering process. For some parameters, the width of $\Gammaints$ can even be much smaller than the width of the density of state $\rho_{\sigma}[\omega]$; e.g.~the + polaritons for $\Delta$ near $-2\omega_M$. In contrast, there are other cases where the density of state is much sharper than the interaction-induced dissipation rate, as is the case for the + polariton for $\Delta$ near $-\omega_M/2$. As a result, the ``interaction-induced" baths cannot always be considered as Markovian.

\begin{figure}[t]
	\begin{center}
	\includegraphics[width= 0.95\columnwidth]{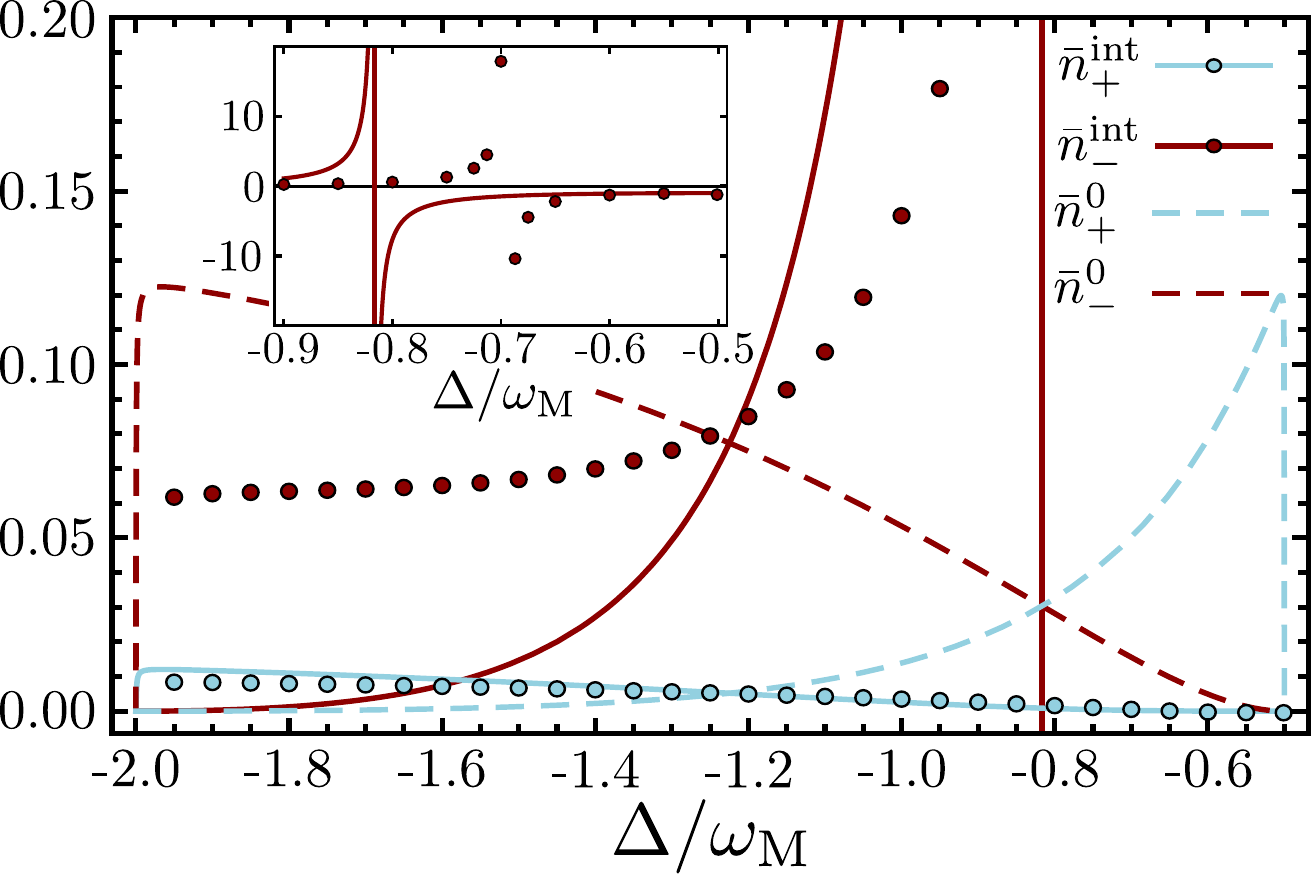}
	\end{center}
	\vspace{-0.5cm}
	\caption {(Color online)
	Effective thermal occupancies for the various baths coupled to the two polariton species, as a function of $\Delta$ with $G = \Gres[\Delta]$.  Solid lines:  occupancies associated with the 
	``interaction-induced" baths to leading order in $g$, c.f.~Eqs.~\eqref{Eq:nint}.  Circles: same, but calculated using the all-orders self-consistent approach of Sec.~\ref{Sec:SelfConsistentApp}.
	Dashed lines:  occupancies of the intrinsic polariton baths, c.f.~Eqs.~\eqref{Eq:ns}.   
	Note the leading-order occupancy for the $-$ polariton interaction bath ($\bnintm$) diverges  when $\bnom = \bnop$ ($\Delta/	
	\omega_M \approx -0.8$) and becomes negative when $\bnop > \bnom$ ($\Delta/\omega_M \gtrsim -0.8$), as shown in the inset.
	This divergence persists in the self-consistent theory, but occurs at smaller-magnitude detunings. 
	This negative occupancy signals the possibility of a new kind of instability, as discussed explicitly in Sec.~\ref{Sec:ParamInsta}.
	All curves are plotted for $\nthM=0$ (cf.~Eq.~(\ref{Eq:nMth})), $\gamma/\kappa=10^{-4}$, $\omega_M/\kappa=50$ and $g=\kappa$.
}
\label{fig:InteractionBaths}
\end{figure}

The thermal occupancies $\bnints$ associated with the interaction-induced environments are plotted in Fig.~\ref{fig:InteractionBaths} as a function of the detuning $\Delta$, in the interesting case where all intrinsic dissipation (i.e.~mechanical bath, cavity bath) are at zero temperature.

%%%%%%%%%%%%%%%%%%%%%%%%%%%%%%%%%%%%%%%%%%%%%%%%%%
%%%%%%%%%%%%%%%%%%%%%%%%%%%%%%%%%%%%%%%%%%%%%%%%%%

%%%%%%%%%%%%%%%%%%%%%%%%%%%%%%%%%%%%%%%%%%%%%%%%%%%%%%%%%%%%%%%%%%%%%%%%%%%%%%%%%%%%%%%%%%%%%%%%%%%%%%%%

\subsection{Interaction-induced quantum heating} \label{Sec:NonlinearQH}

We now discuss in more detail the behaviour of Eq.~(\ref{Eq:nint}) which
gives the thermal occupancies $\bnints$ of the effective ``interaction-induced" dissipative baths introduced in the previous subsection.
For simplicity, we focus on the case of exact resonance, where $G = \Gres$ and hence $E_+ = 2E_-$.
Consider first the case where the linear-theory polariton dissipation is in thermal equilibrium at temperature $T_{\rm eq}$, i.e.~$\bnos = n_B[E_{\sigma},T_{\rm eq}]$.  In this case, it is easy to confirm that for each polariton, $\bnints = \bnos$, i.e.~the ``interaction-induced" dissipation also corresponds to the same temperature $T_{\rm eq}$.  Thus, if without interactions the polaritons start in equilibrium at the same temperature, then the same is true with interactions.

The actual situation is however more complicated: due to quantum heating effects, the effective temperatures of the two polariton species are different even without interactions. 
$\bnom$ and $\bnop$ are thus not related as they would be in thermal equilibrium; this can be parameterized as
\begin{equation}
	\bnop \equiv \frac{\left( \bnom \right)^2}{2 \bnom +1} + \delta \bnop.
\end{equation}
Thermal equilibrium and the condition $E_{+} = 2 E_{-}$ would imply $\delta \bnop = 0$; $\delta \bnop \neq 0$ means that even in the linearized theory, the two polaritons experience different effective temperatures (see Fig.~\ref{fig:Parambath}).

Using this definition, the thermal occupancy of the $-$ polariton interaction-induced bath becomes:
\begin{equation}
	\bnintm = \bnom + \delta \bnop \frac{ (1 + 2 \bnom)^2 }{ \bnom (1 + \bnom) - \delta \bnop (1 + 2 \bnom)}
\end{equation}
Thus, a deviation from true thermal equilibrium in the linear theory (i.e.~without polariton interactions) causes the occupancy of the interaction-induced bath $\bnints$ and the intrinsic bath $\bnos$ (linear-theory dissipation) to deviate from one another. This is not surprising:  in this case, the nonlinear interaction between the two polariton species tends to favour their thermalization, and hence transfers energy from the high-temperature species to the low-temperature species. 

Finally, we also stress that even in the case where the intrinsic mechanical and cavity dissipation is at zero temperature (i.e.~the system only experiences vacuum noise), 
the interaction-bath thermal occupancies $\bnints$ will be non-zero, and are in general different from $\bnos$.
This is shown explicitly in Fig.~\ref{fig:InteractionBaths}.  We thus see that {\it interactions change the effective temperature associated with quantum heating effects.}

%%%%%%%%%%%%%%%%%%%%%%%%%%%%%%%%%%
\begin{figure}[t]
	\begin{center}
	\includegraphics[width= 0.95\columnwidth]{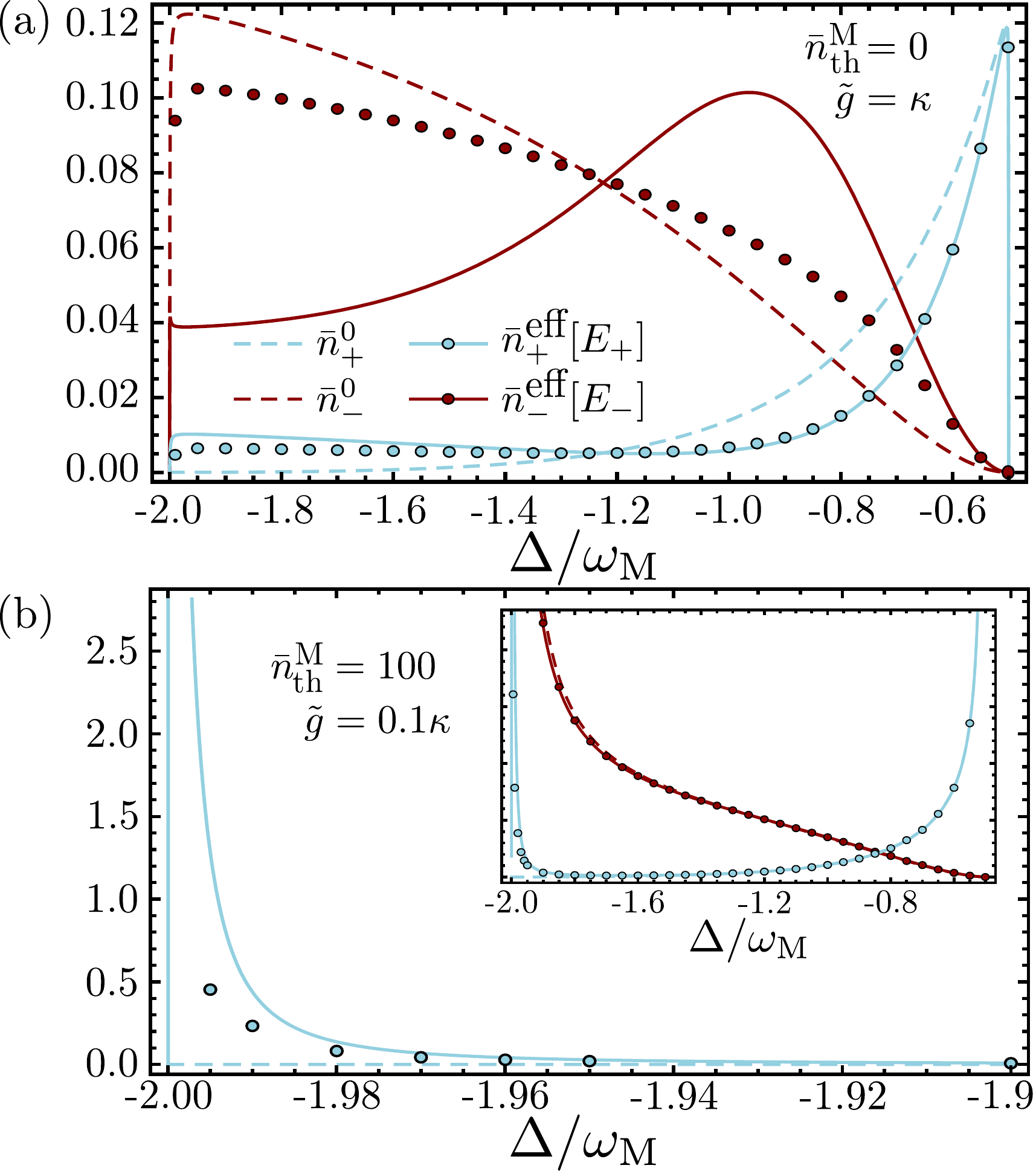}
	\end{center}
	\vspace{-0.5cm}
	\caption {(Color online)
	(a) Solid lines:  net polariton occupancies in the presence of nonlinear interaction, $\bar{n}^{\rm{eff}}_{\sigma}[E_{\sigma}]$, 
	calculated using leading-order self-energies, as a function of $\Delta$ with $G = \Gres[\Delta]$.  Circles: same, but calculated using
	the self-consistent approach (cf.~Sec.~\ref{Sec:SelfConsistentApp}).    Dashed curves:  Occupancies for $g=0$, i.e.~calculated in the linearized theory.  
	In each case, the mechanical temperature is zero ($\nthM = 0$) and $g=\kappa$.  (b) Same as (a), but now with $\nthM = 100$ and $g =0.1\kappa$; main plot and
	inset show different ranges of $\Delta$.  
	Near $-2\omega_M$, one sees an important contribution to the mean number of $+$ polaritons due to nonlinear interaction; this contribution is greatly enhanced by temperatures, as discussed in Sec.~\ref{Sec:TwoPhononsAbs}. 
	All the curves are plotted for $\gamma/\kappa=10^{-4}$ and $\omega_M/\kappa=50$.}
\label{fig:neff}
\end{figure}
%%%%%%%%%%%%%%%%%%%%%%%%%%%%%%%%%%

%%%%%%%%%%%%%%%%%%%%%%%%%%%%%%%%%%%%%%%%%%%%%%%%%%%%%%%%%%%%%%%%%%%%%%%%%%%%%%%%%%%%%%%%%%%%%%%%%%%%%%%%%%%%%%%%%%%%%%%%%%%%%%%%%%%%%%%%%%

\subsection{Polariton energy distribution functions} \label{Sec:PolaritonDist}

The picture established so far is that our optomechanical polaritons are each effectively coupled to two independent effective environments, one of which is self-generated and due
to the nonlinear optomechanical interaction.  In the limit where both the intrinsic cavity and mechanical dissipative baths are at zero temperature ($\nthM = 0$), both these effective environments describe quantum heating physics.  Together, they will determine the total number of polaritons produced by quantum heating, and more specifically, the energy distribution function of the polaritons.  We define this distribution function in the standard manner, as an energy dependent distribution function $\bar{n}_{\sigma}^{\textrm{eff}}[\omega]$.  This quantity is defined via the full Keldysh and retarded polariton Green functions:
\begin{equation}
	\mathcal{G}^K_{\sigma}[\omega] \equiv -2i(2\bar{n}_{\sigma}^{\textrm{eff}}[\omega]+1)\textrm{Im}\left[ \mathcal{G}^R_{\sigma}[\omega] \right] \label{Eq:neff_Def}
\end{equation}
If our polaritons were in thermal equilibrium at temperature $T_{\rm eq}$, then the distribution function  $\bar{n}_{\sigma}^{\textrm{eff}}[\omega]$ would simply be the Bose-Einstein distribution $n_{B}[\omega,T_{\rm{eq}}]$. In contrast, in our system this function will be determined by the thermal occupancies of the two effective baths, and the strength of the couplings (i.e.~damping rates) to each. Using the expression of the dressed Green functions $\mathcal{G}[\omega]$ coming from the Dyson equation (cf. Eq.~(\ref{Eq:Dyson})) and the relation between the self energies given in Eq.~(\ref{Eq:SigmaK}), one finds
\begin{equation}
	\bneffs = \frac{\Gammaints \bnints + \kappa_{\sigma} \bnos}{\Gammaints+\kappa_{\sigma}}. \label{Eq:DetBal}
\end{equation} 
This is exactly the simple expression that would be expected for a free bosonic mode coupled independently to two baths; the same form holds for the linear theory (cf.~Eqs.~(\ref{Eq:ns})). 

In order to focus our attention on the contribution of the nonlinear interaction to $\bneffs$, we rewrite Eq.~(\ref{Eq:DetBal}) using the expressions for $\Gammaints$ (cf.~Eqs.~(\ref{Eq:Sigmas}) and (\ref{Eq:Gammaint})) and $\bnints$ (cf.~Eq.~\ref{Eq:nint}). Doing so, one gets
\begin{align}
	\bneffs =\bnos + I_{\sigma} \frac{\gamma_{\sigma}^2}{(\omega-\omega_{\sigma})^2+\gamma_{\sigma}^2}, \label{Eq:neff}
\end{align}
with
\begin{subequations}
\begin{gather}
	I_{\sigma} = (\bnints-\bnos)  \frac{\Ceffs}{1+\Ceffs}, \\
	\gamma_{-} = \frac{\kappa_-+\kappa_+}{2}\sqrt{1+\Ceffm}, \quad
	\gamma_{+} = \kappa_-\sqrt{1+\Ceffp}, \label{Eq:neffwidth} \\
	\omega_-= E_+-E_-, \quad
	\omega_+=2E_-.
\end{gather}
\end{subequations}
The contribution from the interaction-induced environment appears as a sharp Lorentzian in the polariton distribution functions. This is a direct consequence of the resonant nature of the relevant nonlinear scattering process. 

For exact resonance ($E_+=2E_-$), both $\bneffs$ and the single particle DOS $\rho_{\sigma}[\omega]$ (cf.~Eq.~(\ref{Eq:DOS})) are peaked at $E_{\sigma}$, so that the nonlinear interaction heating effects are maximal. Even in this case though, the frequency dependence of the interaction contribution to $\bneffs$ can be very different than that of the polariton DOS.
In this fully resonant case, the polaritons distribution functions evaluated at $\omega = E_\sigma$ adopt the following simple form
\begin{equation}
	\bar{n}^{\textrm{eff}}_\sigma[E_\sigma] = \frac{\Ceffs \bnints + \bnos}{\Ceffs+ 1}. \label{Eq:DistFuncResonance}
\end{equation}
Eq.~\eqref{Eq:DistFuncResonance} is plotted in Fig.~\ref{fig:neff} as a function of the laser detuning $\Delta$ and is compared to $\bar{n}^{\textrm{eff}}_\sigma[E_\sigma]$ obtained using the self-consistent approach (cf.~Sec.~\ref{Sec:SelfConsistentApp}).

%%%%%%%%%%%%%%%%%%%%%%%%%%%%%%%%%%%%%%%%%%%%%%%%%%%%%%%%%%%%%%%%%%%%%%%%%%%%%%%%%%%%%%%%%%%%%%%%%%%%%%%%%%%%%%%%%%%%%%%%%%%%%%%%%%%%%%%%%%

\subsection{Nonlinear parametric heating} \label{Sec:ParamInsta}

Among the more striking non-equilibrium behaviours possible in the linear theory is the possibility of having $\bnop > \bnom$, i.e.~the thermal occupancy of the higher energy $+$ polariton exceeds that of the $-$ polariton.  We discuss this regime in more detail here, focusing on the exactly resonant case where $E_+=2E_-$.

We start by recalling that the total damping of the $-$ polariton is
\begin{equation}
	\kappa_-^{\rm{tot}}[E_-] = \kappa_- + \Gamma^{\rm{int}}_-[E_-] = \kappa_- \left( 1+ \Ceffm \right). \label{Eq:TotalDampingm}
\end{equation}
From Eqs.~(\ref{Eq:Ceff}), we see that if we have the occupancy inversion $\bnop > \bnom$, then $\Ceffm < 0$, and hence the contribution of the nonlinear interaction to the damping rate of the $-$ polaritons becomes negative.  This is at first glance surprising:  we have opened a new scattering process for the $-$ polariton via the nonlinear interaction, and yet we get an increase in its lifetime.  We also have the possibility of an instability if $\Ceffm < -1$.

From Fig.~\ref{fig:Ceff}, one sees that a negative $\Ceffm$ occurs for detunings near $\Delta = -0.5\omega_M$. In that regime, the $-$ polariton is mostly photonic and the $+$ polariton is mostly phononic. Consequently, by having a high intrinsic mechanical bath temperatures $\nthM \gg 1$, one naturally can achieve the inverted occupancy regime where $\bnop \gg \bnom$. 
In that case and in the limit where $\gamma \ll \kappa$, one can simplify the instability condition, i.e.~$\Ceffm < -1$, to
\begin{equation}
	 \bnop > \bnom + \frac{\kappa_-(\kappa_-+\kappa_+)}{16 \tilde{g}^2} \approx \frac{\kappa_-^2}{16 \tilde{g}^2}, \label{Eq:CondInsta}
\end{equation}
and the final mean number of $-$ polaritons becomes
\begin{equation}
	\bar{n}^{\rm{eff}}_-[E_-] \approx \frac{16\tilde{g}^2\bnop}{\kappa_-^2 - 16\tilde{g}^2\bnop}. \label{Eq:nmparamp}
\end{equation}

The surprising negative damping occurring here can be understood as the result of a parametric instability arising directly from the polariton interaction.  
In fact, Eqs.(\ref{Eq:CondInsta}) and (\ref{Eq:nmparamp}) have exactly the same form as the instability condition and the mean number of 
excitations that one would find for a degenerate parametric amplifier (DPA)
pumped near degeneracy~\cite{GardinerZollerBook, Clerk_RMP_2010}. 
In a DPA, a pump-mode photon scatters into to signal mode photons, and the pump mode is {\it coherently} driven.  In our system, the $-$ polariton plays the role of the signal mode in a DPA, while the $+$ polariton plays the role of a pump mode that has been {\it incoherently} driven by noise.  Despite this
incoherent driving, the form of the above equations is the same as a coherently-driven DPA (see Appendix \ref{Sec:AppendixParamp}).  Note that non-degenerate parametric amplifier instability
can be realized in a linearized optomechanical system driven with a blue-detuned laser, see e.g.~\cite{Marquardt_PRL_2006, Lugwig_NJP_2008}.  In contrast, the instability described here occurs for a red-detuned drive.

Note that our discussion here is based solely on using the leading order results for the polariton self energies.  Including higher-order effects via our self-consistent approach can dramatically change the onset and magnitude of the interaction-induced negative damping.  We discuss this more in Sec.~\ref{Sec:ParampAmpObs}.

%%%%%%%%%%%%%%%%%%%%%%%%%%%%%%%%%%%%%%%%%%%%%%%%%%%%%%%%%%%%%%%%%%%%%%%%%%%%%%%%%%%%%%%%%%%%%%%%%%%%%%%%%%%%%%%%%%%%%%%%%%%%%%%%%%%%%%%%%%

\section{Observable signatures of quantum heating effects} \label{Sec:Observable}

In the previous sections, we have demonstrated how quantum heating effects can lead to a finite density of optomechanical polaritons at zero temperature; we also discussed how these quantum heating effects can be modified by the nonlinear interaction.  In this section, we discuss how these effects lead to observable signatures in the light leaving the optomechanical cavity. We first relate the cavity output spectrum to the polariton distribution functions, and then discuss specific parameter regimes where the heating effects are most prevalent. In addition, we propose a way to effectively control the strength of the nonlinear interaction in experiments by tuning in and out the resonance condition (i.e.~by varying $G$ at fixed detuning $\Delta$). Doing so, one can explicitly isolate and observe the nonlinear 
interaction signatures in the cavity output spectrum.

%%%%%%%%%%%%%%%%%%%%%%%%%%%%%%%%%%%%%%%%%%%%%%%%%%%%%%%%%%%%%%%%%%%%%%%%%%%%%%%%%%%%%%%%%%%%%%%%%%%%%%%%%%%%%%%%%%%%%%%%%%%%%%%%%%%%%%%%%%

\subsection{Polariton energy distribution functions}

To measure polariton occupancies, we consider a measurement of the flux of photons leaving our cavity (assuming a single sided cavity, and that the reflected classical drive tone is filtered away).  The spectrum of this flux is given in the standard manner \cite{GardinerZollerBook} by the normal-ordered cavity spectrum (also known as the ``lesser" Green function within the Keldysh technique),
\begin{equation}
	S_{d}[\omega] \equiv \int_{-\infty}^{\infty} dt e^{i\omega t} \langle \hat{d}^{\dag}(0)\hat{d}(t)\rangle. \label{Eq:DefSd}
\end{equation}

Re-writing Eq.~(\ref{Eq:DefSd}) in terms of polariton Green functions yields
\begin{align}
	\Sd = & 2\pi \sum_{\sigma=\pm} \ads^2 \bneffs  \rho_{\sigma}[\omega] \nonumber \\
	& + 2\pi \sum_{\sigma=\pm} \bads^2 (\bar{n}_{\sigma}^{\textrm{eff}}[-\omega] +1)  \rho_{\sigma}[-\omega].  \label{Eq:SdFull}
\end{align}
Here, the coefficients $\ads$ and $\bads$ are the change-of-basis coefficients introduced in Eq.~\eqref{Eq:PhononToPol} and plotted in Fig.~\ref{fig:LinearParams}, and the polariton DOS, $\rho_{\sigma}[\omega]$, is defined in Eq.~(\ref{Eq:DOS}).  We are still working in a rotating frame with respect to the laser drive frequency $\omega_L$, hence $\omega = 0$ implies output photons leaving at the laser frequency. The negative frequency term means that removing a photon at frequency $\omega$ can involve creating a polariton at frequency $-\omega$, as expected from the presence of ``anomalous" terms in
Eq.~\eqref{Eq:PhotonToPol}. Finally, Eq.~\eqref{Eq:SdFull} reflects the fact that the only non-zero polariton Green functions are those that conserve the number of polaritons independently, as discussed in Secs.~\ref{Sec:Keldysh} and \ref{Sec:PerturbationTheory}.

Using the same assumptions, we also derive the cavity DOS and its energy distribution function:
\begin{align}
	\rho_d[\omega] = & \sum_{\sigma=\pm} \left(\ads^2 \rho_{\sigma}[\omega] - \bads^2 \rho_{\sigma}[-\omega]\right),  \label{Eq:photonDOS} \\
	\bar{n}^{\mathrm{eff}}_d [\omega] = & \frac{1}{2\pi}\frac{\Sd}{\rho_d[\omega]}. \label{Eq:photonNeff}
\end{align}
From the cavity energy distribution, we can use the Bose-Einstein distribution to {\it define} an effective cavity temperature ($k_B = 1$),
\begin{equation}
	\Teffd \equiv \frac{\omega}{\ln\left[ 1 + \frac{1}{\bar{n}^{\mathrm{eff}}_d [\omega]} \right]}, \label{Eq:photonTeff}
\end{equation}
which is always possible if we let the effective temperature to be frequency dependent.

As discussed extensively in \cite{MAL_PRL_2011}, the polariton DOS $\rho_{\sigma}[\omega]$ can be directly measured in an OMIT-style experiment \cite{Agarwal_PRA_2010,Weis_Science_2010,Teufel_Nature_2011_OMIT,Safavi-Naeini_Nature_2011}, where one measures the reflection of a weak additional probe tone incident on the cavity. The cavity spectrum in contrast also yields information on polariton occupancies.  As the polariton energies $E_{\sigma}$ are well separated, the output spectrum will have a series of peaks corresponding to the emission or absorption of a given polariton species.  The magnitude of these peaks is directly proportional to the occupancy of the given polariton.

It is also useful to look at the total number of photons due to a given polariton resonance, which we can obtain by integrating the output spectrum around the corresponding resonance.  We thus introduce:
\begin{equation}
	\bndtot[\omega_0,\delta\omega] \equiv \int_{\omega_0-\delta\omega}^{\omega_0+\delta\omega} \frac{d\omega}{2\pi} \Sd. \label{Eq:ndtot}
\end{equation}
where $\omega_0$ will be taken to be $E_{\pm}$, and $\delta \omega$ will be taken to be larger than the spectral width of the given polariton resonance.

In what follows, we consider signatures of quantum heating in the spectrum for particularly interesting choices of the drive laser detuning $\Delta$.

%%%%%%%%%%%%%%%%%%%%%%%%%%%%%%%%%%%%%%%%%%%%%%%%%%%%%%%%%%%%%%%%%%%%%%%%%%%%%%%%%%%%%%%%%%%%%%%%%%%%%%%%%%%%%%%%%%%%%%%%%%%%%%%%%%%%%%%%%%

\subsection{Limit of zero temperature dissipation: effects of nonlinear interaction on quantum heating}

The first studied limit is for a mechanical bath at zero temperature ($\nthM=0$), where the finite number of polaritons inside the optomechanical cavity exclusively comes from quantum heating. 
We focus on two regimes: the red sideband drive, i.e.~$\Delta = -\omega_M$, where both polaritons are equal mixture of photons and phonons, and the asymmetric polaritons regime, where the polaritons are not equal combinations of photon and phonon. For the latter regime, we chose $\Delta = -1.8\omega_M$ as a representative laser detuning. The results predicted for the cavity driven on the red sideband has the advantage to be robust to temperatures since both polaritons have an important photon part; this implies that even for finite $\nthM$, quantum heating is still the prevalent source of polaritons. In contrast, a laser detuned at $\Delta = -1.8\omega_M$ leads to more striking modifications of the polaritons energy distribution since $\Ceffs[\Delta = -1.8\omega_M] > \Ceffs[\Delta = -\omega_M]$ (cf.~Fig.~\ref{fig:Ceff}). In both cases, we show that the nonlinear interaction modifies the energy distribution of the polaritons as it tends to thermalize the two species.

\subsubsection{Results for symmetric polaritons (red sideband drive)}

\begin{figure}[t]
	\begin{center}
	\includegraphics[width= 0.95\columnwidth]{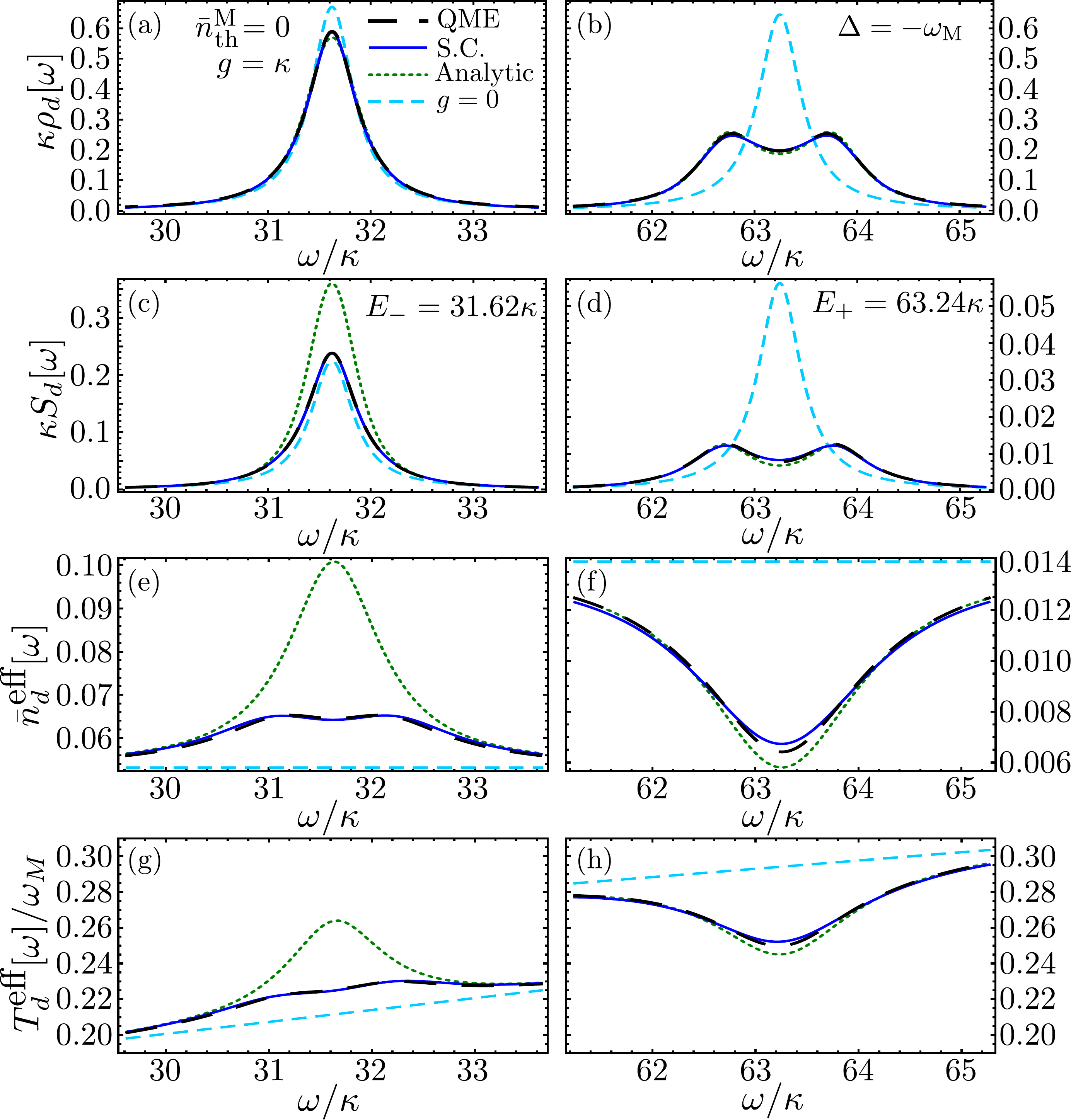}
	\end{center}
	\vspace{-0.5cm}
	\caption {(Color online)
	(a) and (b) Cavity DOS near the $-$ polariton resonance and the $+$ polariton resonance respectively for a cavity driven on the red sideband and when the nonlinear interaction is resonant (i.e.~$G=\Gres$). We work in the frame rotating at the drive frequency so that $\omega=0$ refers to the drive frequency while $E_+ = 2E_- = 63.24\kappa$ in this frame. The light dashed curves represent the linearized theory, the green dotted ones represent the results to leading orders in $\tilde{g}$ (Eq.~\eqref{Eq:DOS} for the DOS), the full blue curves are the results of the self-consistent approach as described in Sec.~\ref{Sec:SelfConsistentApp} and the black curves are for the numerical simulation of the Lindblad master equation shown in Eq.~\eqref{Eq:MEQ}. (c) and (d) Cavity spectrum in the same conditions; the results to leading orders in $\tilde{g}$ are given in Eqs.~\eqref{Eq:neff} and \eqref{Eq:SdFull}. 
 (e) and (f) Cavity energy distribution function (Eq.~\eqref{Eq:photonNeff}) and, (g) and (f) the corresponding (frequency dependent) effective temperatures (Eq.~\eqref{Eq:photonTeff}). The parameters used for all the curves are $\gamma/\kappa=10^{-4}$ and $\omega_M/\kappa=50$, which leads to the leading order effective cooperativities (i.e.~Eq.~\eqref{Eq:Ceff}) $\Ceffm = 0.18$ and $\Ceffp = 2.46$.}
\label{fig:Sd_D1}
\end{figure}

In Fig.~\ref{fig:Sd_D1} 
we plot the cavity DOS (Eq.~\eqref{Eq:photonDOS}), the cavity spectrum (Eq.~\eqref{Eq:SdFull}), its energy distribution function (Eq.~\eqref{Eq:photonNeff}) and the corresponding effective temperature (Eq.~\eqref{Eq:photonTeff}) for $\Delta = -\omega_M$.
The linearized theory ($g=0$) is compared to the case where $g=\kappa$.  For the latter interacting case, we present results obtained from three different methods:  the leading order in $\tilde{g}$ self-energies (Eqs.~\eqref{Eq:DOS} and \eqref{Eq:neff}), the self-consistent approach described in Sec.~\ref{Sec:SelfConsistentApp} and finally,  a numerical simulation of the Lindblad master equation given in Eq.~\eqref{Eq:MEQ}. 
By comparing the three different approaches, one sees that for $g=\kappa$, higher order corrections captured by the self-consistent approach play an important role for the effective distribution functions and the effective temperatures (panel (e)-(h) of Fig.~\ref{fig:Sd_D1}). In contrast, the DOS is already well described at the leading order in $\tilde{g}$, which is in agreement with Ref.~\onlinecite{MAL_PRL_2011}.

The splitting of the $+$ polariton resonance in $\rho_d[\omega]$ and $\Sd$ near $\omega = E_+$ arises from the hybridization between the states $\vert + \rangle$ and $\vert -,- \rangle$; the resulting hybridized states become spectrally resolved for $g \gtrsim \kappa$ (see \cite{MAL_PRL_2011} for more details). The same hybridization phenomena gives rise to a resonance in $\bneffd$ and $\Teffd$
at the $-$ polariton frequency, $\omega = E_-$ (see Fig.~\ref{fig:Sd_D1}(e) and (g)).  

\begin{figure}[t]
	\begin{center}
	\includegraphics[width= 0.95\columnwidth]{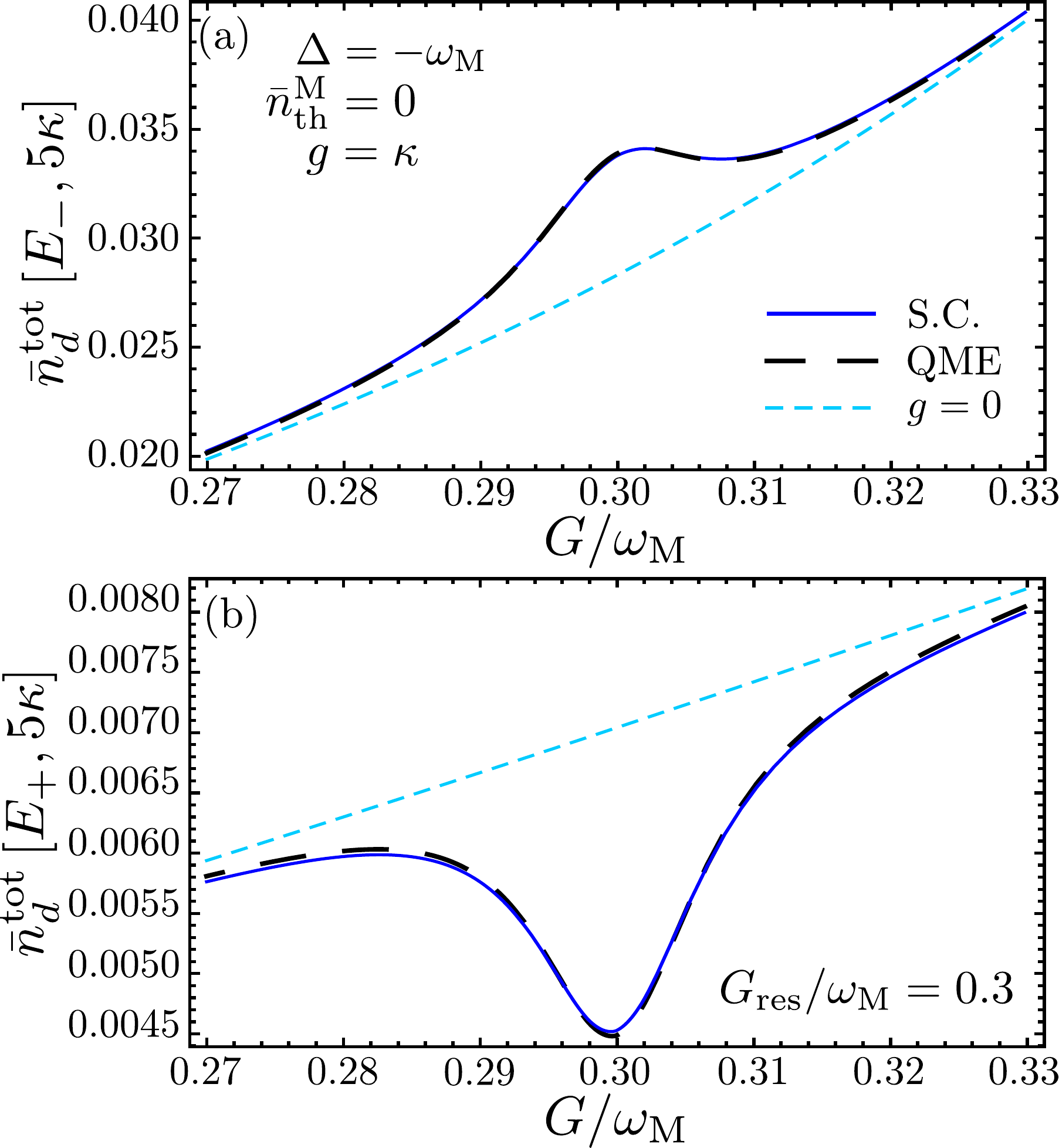}
	\end{center}
	\vspace{-0.5cm}
	\caption {(Color online)
	(a) Output photon flux in a bandwidth of $5\kappa$ near the - polariton resonance $E_-$ (c.f.~Eq.~\eqref{Eq:ndtot}) as a function of the many photon coupling $G$, and for a control laser detuning $\Delta = -\omega_M$. (b) Same as (a) but near the + polariton resonance $E_+$. Note that $G$ can be varied by simply tuning the amplitude of the control laser. On resonance, i.e.~$G=\Gres$, the nonlinear effects are enhanced by a factor of $\omega_M /\kappa= 50$ compared to the off-resonance case, i.e.~$(G-\Gres) \gtrsim \kappa$. 
The light dashed curves represent the linearized theory ($g=0$), the solid blue curves are for the self-consistent approach as described in Sec.~\ref{Sec:SelfConsistentApp} and the black dashed curves are for the numerical simulation of the Lindblad master equation shown (\ref{Eq:MEQ}), but this time, using the full nonlinear part of the Hamiltonian in Eq.~(\ref{Eq:HintPolariton}) that includes all the non-resonant nonlinear processes. For all the curves, we used $\gamma/\kappa = 10^{-4}$ and $\nthM = 0$.}
\label{fig:NphotonVsG_D10}
\end{figure}

In Fig.~\ref{fig:NphotonVsG_D10}, we plot the number of photons leaving the cavity near each polariton resonances, as defined in Eq.~\eqref{Eq:ndtot}, and show that one can effectively isolate the effects of nonlinear interaction.
To do so, one varies $G$ around $\Gres$ such that the nonlinear interaction get amplified by a factor of $\omega_M/\kappa$ when the nonlinear process becomes resonant, i.e.~for $G=\Gres$, compared to the off-resonant case, i.e.~$G-\Gres \gtrsim \kappa$. Away from resonance, the number of photons leaving the cavity is in good approximation given by the linearised theory (dashed lines in Fig.~\ref{fig:NphotonVsG_D10}). For the parameters here ($\nthM=0$ and $\gamma/\kappa \ll 1$), the linearised theory leads to (see Eqs.~\eqref{Eq:Tcav} and \eqref{App:COB2}):
\begin{equation}
	\bndtot[E_\pm, 5\kappa] \approx \alpha_{d,\pm}^2 \bar{n}^0_{\pm} \approx \frac{1}{8}\frac{(G/\omega_M)^2}{1\pm2G/\omega_M}.
\end{equation}

Fig.~\ref{fig:NphotonVsG_D10} clearly shows the thermalization between polaritons brought about the nonlinear interaction.  Without interactions, at $\Delta = -\omega_M$ the $-$ polaritons have a lower effective temperature than the $+$ polaritons, c.f.~Fig.~\ref{fig:Parambath}(c).  When $G$ is near $\Gres$, the nonlinear interaction ``turns on" and allows the two polariton species to exchange energy and partially thermalize (i.e.~interactions heat up the $-$ polaritons while cooling down the $+$ polaritons.)

%%%%%%%%%%%%%%%%%%%%%%%%%%%%%%%%%%%%%%%%%%%%%

%%%%%%%%%%%%%%%%%%%%%%%%%%%%%%%%%%%%%%%%%%%%%
%%%%%%%%%%%%%%%%%%%%%%%%%%%%%%%%%%%%%%%%%%%%%

\subsubsection{Results for asymmetric polaritons}

\begin{figure}[t]
	\begin{center}
	\includegraphics[width= 0.95\columnwidth]{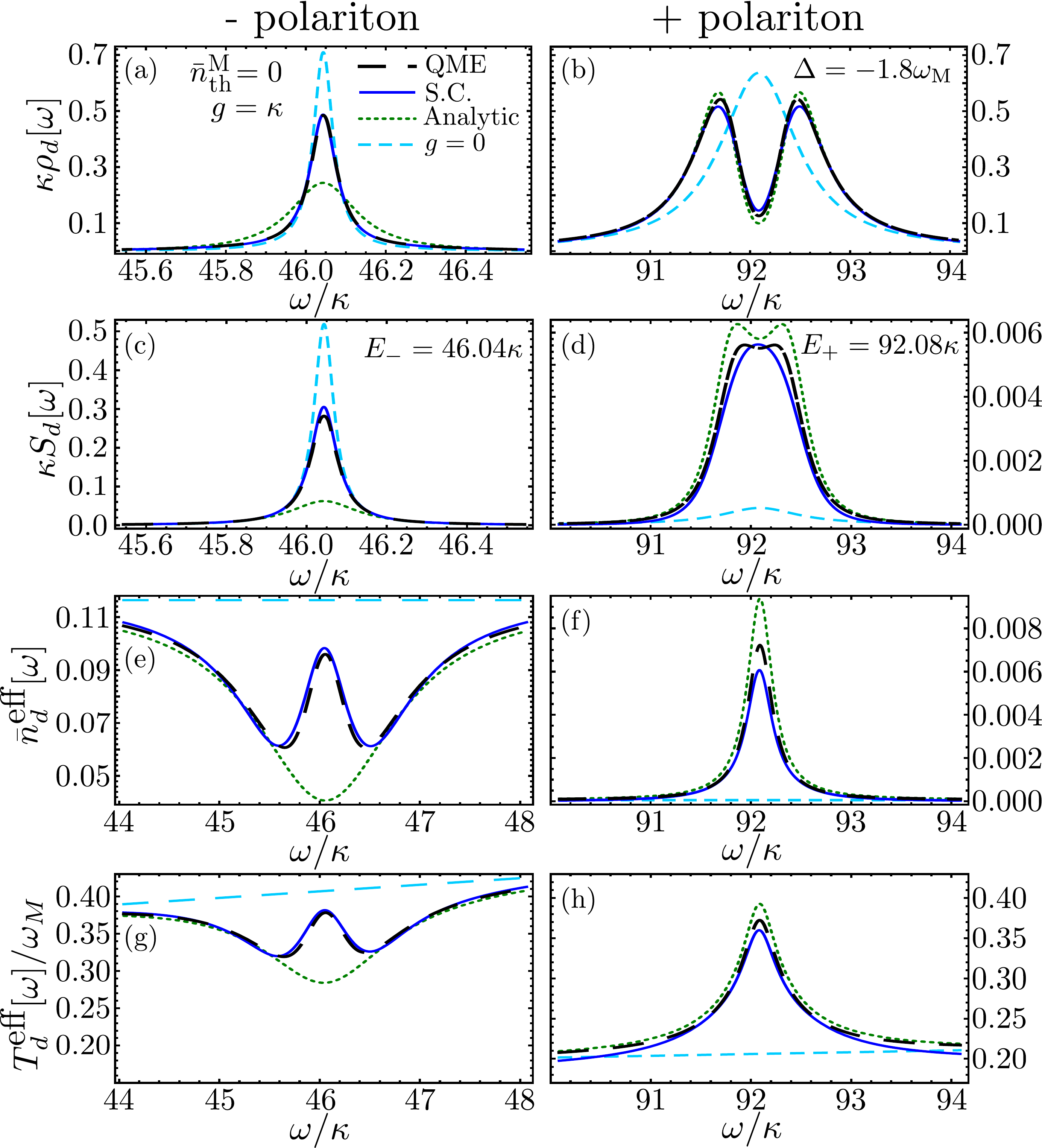}
	\end{center}
	\vspace{-0.5cm}
	\caption {(Color online)
	Same as Fig.~\ref{fig:Sd_D1} but for laser detuning $\Delta = -1.8\omega_M$. In that case, $E_+ = 2E_- = 92.08\kappa$ in the rotating frame and the corresponding leading order (i.e.~Eq.~\eqref{Eq:Ceff}) effective cooperativities $\Ceffm = 1.92$ and $\Ceffp = 5.40$.}
\label{fig:Sd_D18}
\end{figure}

In Fig.~\ref{fig:Sd_D18}, we plot the same functions as in Fig.~\ref{fig:Sd_D1}, but in the case where the laser detuning is $\Delta = -1.8\omega_M$ (and again, $G = \Gres$). For this more negative detuning, the polaritons are no longer an equal mixture of photons and phonons: the - polariton is more phonon-like while the + is more photon-like. This particular asymmetry leads to larger values of $\Ceffs$ than in the red sideband regime, mainly because of the long-lifetime of the phonon like polariton (cf.~Fig.~\ref{fig:Ceff}).
Due to these larger $\Ceffs$, the results obtained to the leading order in $\tilde{g}$ are not sufficient to recover the numerical simulation of the Lindblad master equation (cf.~Eq.~\eqref{Eq:MEQ}) even for the DOS. It is than crucial to use the self-consistent approach to properly describe the effects of nonlinear interaction. Moreover, from panels ($g$) and ($h$), one sees that unlike the case $\Delta = -\omega_M$, the nonlinear interaction cools down the - polaritons and heats up the + polaritons.  This is also consistent with a partial thermalization, as for $\Delta = -1.8 \omega_M$, without interactions the effective temperature of the $-$ polaritons is greater than that of the $+$ polaritons (c.f. Fig.~\ref{fig:Parambath}c).

%%%%%%%%%%%%%%%%%%%%%%%%%%%%%%%%%%%%%%%%%%%%%%%%%%%%%%%%%%%%%%%%%%%%%%%%%%%%%%%%%%%%%%%%%%%%%%%%%%%%%%%%%%%%%%%%%%%%%%%%%%%%%%%%%%%%%%%%%%

\subsection{Parametric amplification of the $-$ polaritons} \label{Sec:ParampAmpObs}

While our emphasis in this work has been on quantum heating effects involving zero temperature dissipation, our approach can also conveniently describe thermal nonlinear phenomena.  Perhaps the most striking example of this occurs for detuning $\Delta \approx -\omega_M/2$, where leading order perturbation theory predicts the presence of a parametric instability at finite temperatures (see Sec.~\ref{Sec:ParamInsta}). In that regime, the nonlinear interaction acts as an incoherently pumped degenerate parametric amplifier with the $+$ polariton (mainly phonon) being the (incoherent) pump and the $-$ polariton (mainly photon) being the signal mode.

\begin{figure}[t]
	\begin{center}
	\includegraphics[width= 0.99\columnwidth]{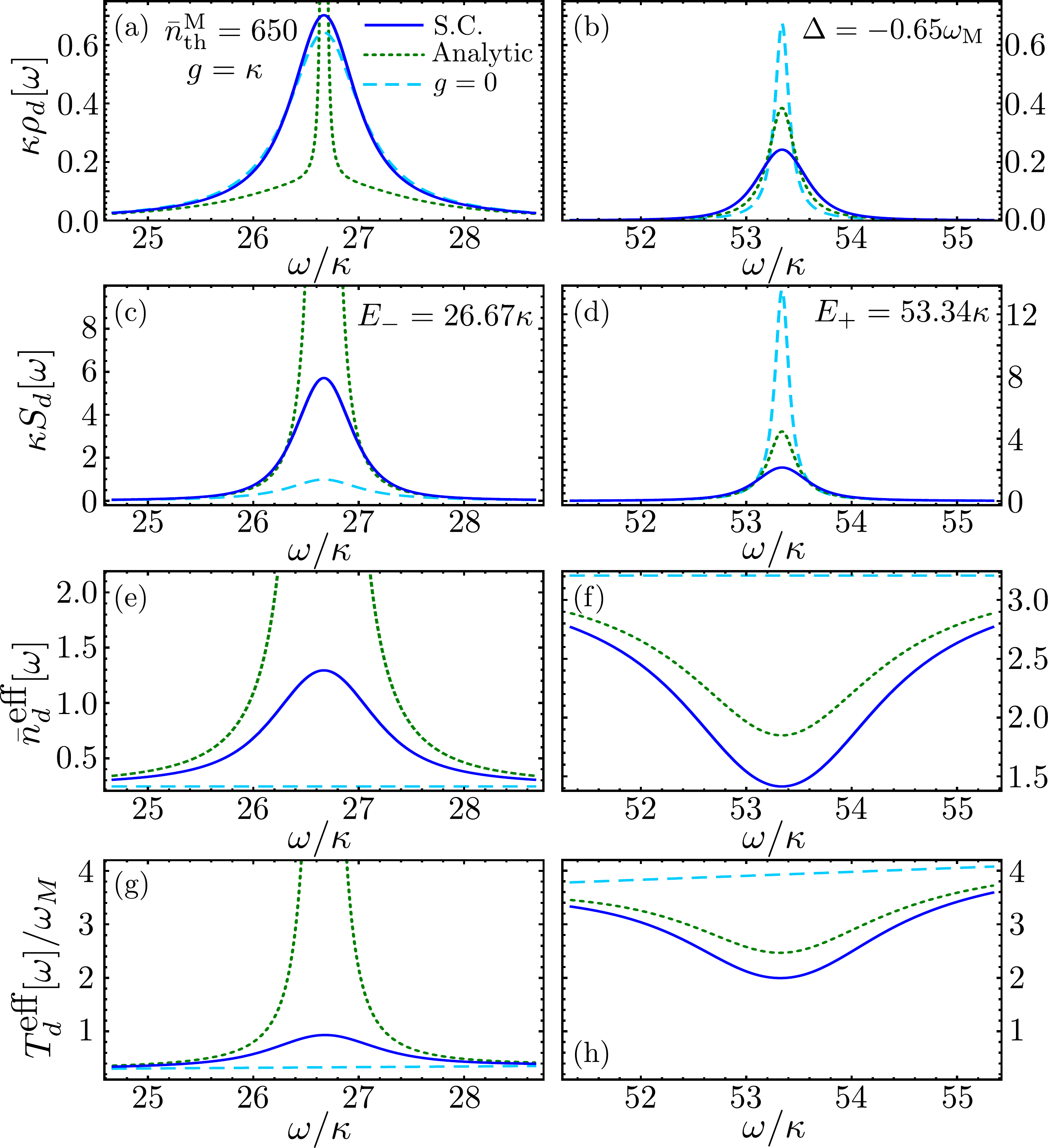}
	\end{center}
	\vspace{-0.5cm}
	\caption {(Color online)
	Same as Fig.~\ref{fig:Sd_D1} but for laser detuning $\Delta = -0.65\omega_M$ and finite temperature $\nthM = 650$. In these circumstances, leading order perturbation theory predicts that the system is close to a parametric instability (see Sec.~\ref{Sec:ParamInsta}). Note the mechanical damping rate here is larger than in previous plots ($\gamma = 10^{-3}\kappa$), as this enhances the pumping effects of mechanical temperature.  For $\omega_M = 50\kappa$ and $g=\kappa$, the leading order effective cooperativies (cf.~Eq.~\eqref{Eq:Ceff}) are $\Ceffm = -0.97$ and $\Ceffp = 0.75$. In the frame that rotates at the drive frequency, $E_+ = 2E_- = 53.34\kappa$. One sees from these plots the crucial importance of higher-order-in-$g$ terms.  }
\label{fig:Sd_D065}
\end{figure}

In Fig.~\ref{fig:Sd_D065}, we show the cavity DOS (cf.~Eq.~\eqref{Eq:photonDOS}), the cavity spectrum $\Sd$ (cf.~Eq.~\eqref{Eq:SdFull}) and the corresponding energy distribution function $\bar{n}^{\mathrm{eff}}_d [\omega]$ (cf.~Eq.~\eqref{Eq:photonNeff}) for a laser detuning $\Delta = -0.65\omega_M$ and $\nthM=650$. Combined with a damping rate of the mechanical resonator $\gamma=10^{-3}\kappa$, its resonant frequency $\omega_M=50\kappa$ and $g=\kappa$, one obtains, to leading order in $g$, an effective cooperativity $\Ceffm = -0.97$ (cf.~Eq.~\eqref{Eq:Ceff}). As discussed in Sec.~\ref{Sec:ParamInsta}, for $\Ceffm = -1$, the leading order perturbation theory predicts a parametric instability caused by the nonlinear interaction. For $\Ceffm \gtrsim -1$, one thus expects an important narrowing of the cavity DOS as well as an important heating of the cavity near the $-$ polariton resonance.  
This predictions from the leading-order self energy are shown in Fig.~\ref{fig:Sd_D065}. 

Not surprisingly, higher-order corrections (as captured by the self-consistent self energy) are especially important in this regime and strongly contribute to prevent the system from going unstable.
More precisely, it is the hybridization between the states $\vert + \rangle$ and $\vert -- \rangle$ that competes with the parametric amplification of the - polaritons; the high number of - polaritons leads to an important modification of the energy of the hybridized states $\tfrac{1}{\sqrt{2}}(\vert + \rangle \pm \vert -- \rangle)$ so that the nonlinear interaction cease to be resonant.
The result of this competition is shown clearly in Fig.~\ref{fig:Sd_D065}.  While the leading order theory predicts a photon occupancy at the $-$ polariton resonance of $\bar{n}^{\mathrm{eff}}_d [E_-]  \approx 50$, in the self-consistent approach, one only obtains a value $\simeq 1.3$.  This is of course still much larger than what would be obtained without interaction; in that case, $\bar{n}^{\mathrm{eff}}_d [E_-] \simeq 0$.

%%%%%%%%%%%%%%%%%%%%%%%%%%%%%%%%%%%%%%%%%%%%%%

%%%%%%%%%%%%%%%%%%%%%%%%%%%%%%%%%%%%%%%%%%%%%%%%%%%%%%%%%%%%%%%%%%%%%%%%%%%%%%%%%%%%%%%%%%%%%%%%%%%%%%%%%%%%%%%%%%%%%%%%%%%%%%%%%%%%%%%%%%

\subsection{Effective two phonons absorption} \label{Sec:TwoPhononsAbs}
Another striking example of temperature-enhanced nonlinear effects occurs for
a laser detuning $\Delta \approx -2\omega_M$. In this regime, the $-$ polaritons are mostly phonons, which leads to a 
thermal ``stimulated emission" enhancement of $\SigmaRm$, leading to a strong modification of the cavity DOS.  This physics follows from Eqs.~(\ref{Eq:Sigmas}), and was discussed extensively in 
Ref.~\onlinecite{MAL_PRL_2011}.  However, as shown in Fig.~\ref{fig:neff}, one also obtains significant nonlinearity-induced heating of the cavity in this regime, as we now describe.

Recall that if $\Delta =-2 \omega_M$, then $\Gres=0$ and the $+$ ($-$) polaritons are exactly photons (phonons). A necessary consequence is that the amplitude for the resonant nonlinear interaction vanishes, $\tilde{g}=0$.  One thus ideally wants a detuning close to, but not exactly equal to $- 2 \omega_M$ such that $0 < \Gres \ll \omega_M$. In the high temperatures limit ($\nthM \gg 1$) and for weak nonlinear interaction $\Ceffp \ll 1$ (i.e.~$g \ll \kappa$), the expression to lowest order in $G/\omega_M$ for the $+$ resonance in the cavity spectrum  simplifies to:
\begin{align}
	S_d[\omega] \approx \frac{\kappa}{\gamma}\left( \frac{G}{\omega_M}\right)^2  \left( \frac{g}{\kappa} \right)^2 \frac{16(\nthM)^2}{\left( 1+\frac{8}{9}\frac{G^2}{\omega_M^2}\frac{\kappa}{\gamma}\right)^3}\frac{\kappa_-^2}{(\omega-E_+)^2+\kappa_-^2}. \label{Eq:Sdnear2wm}
\end{align}
As discussed near Eq.~\eqref{Eq:nminusNear2omegaM}, the competition between  the parametric heating associated with the linearized optomechanical interaction and the standard optomechanical optical damping leads to an optimal value of $G/\omega_M$ where the heating is maximal. In the optimal case, $(G/\omega_M)^2 = \frac{9}{16}\frac{\gamma}{\kappa}$ and $S_d[E_+] \sim \left( \frac{g}{\kappa} \right)^2 (\nthM)^2$. 
Details of the calculations that leads to Eq.~(\ref{Eq:Sdnear2wm}) are presented in Appendix \ref{Sec:AppNear2omega}.

In Fig.~\ref{fig:Sd2wm}, the cavity DOS and the cavity spectrum near the $+$ polariton resonance is plotted for a laser detuning near $-2\omega_M$ and for finite mechanical bath temperatures $\nthM$. In panel (a), we show the sharp dip in the density of state due to nonlinear interaction, also described in Ref.~\onlinecite{MAL_PRL_2011}. Note however that effects of higher order in $g$, captured in the self-consistent approach, considerably modify the predictions made in Ref.~\onlinecite{MAL_PRL_2011}, where only effects to leading order in $g$ were considered. This sharp feature is completely analogous to the optomechanical induced transparency (OMIT) observed in the optomechanical cavity weekly driven on the red sideband~\cite{Weis_Science_2010,Teufel_Nature_2011_OMIT,Safavi-Naeini_Nature_2011,Agarwal_PRA_2010}, except that here, it is the nonlinear interaction that is involved. In panel (b), we show that nonlinear interaction greatly modifies the cavity spectrum; no output light is predicted by the linearized theory while a sharp signal is produced when nonlinear interaction is considered. 

The effective two phonons absorption becomes a very important process in the high temperature limit. In that case, the nonlinear interaction becomes easier to observe, but the phenomenon tends to become purely classical. In order to support this statement, we present a classical treatment of this phenomenon in Appendix \ref{App:ClassicalTwoPhononsAps} and show that it succeeds to recover the right dependence in temperatures and single-photon coupling constant $g$ of the two phonons absorption signature in the cavity spectrum at the lowest order in $g$.

\begin{figure}[t]
	\begin{center}
	\includegraphics[width= 0.95\columnwidth]{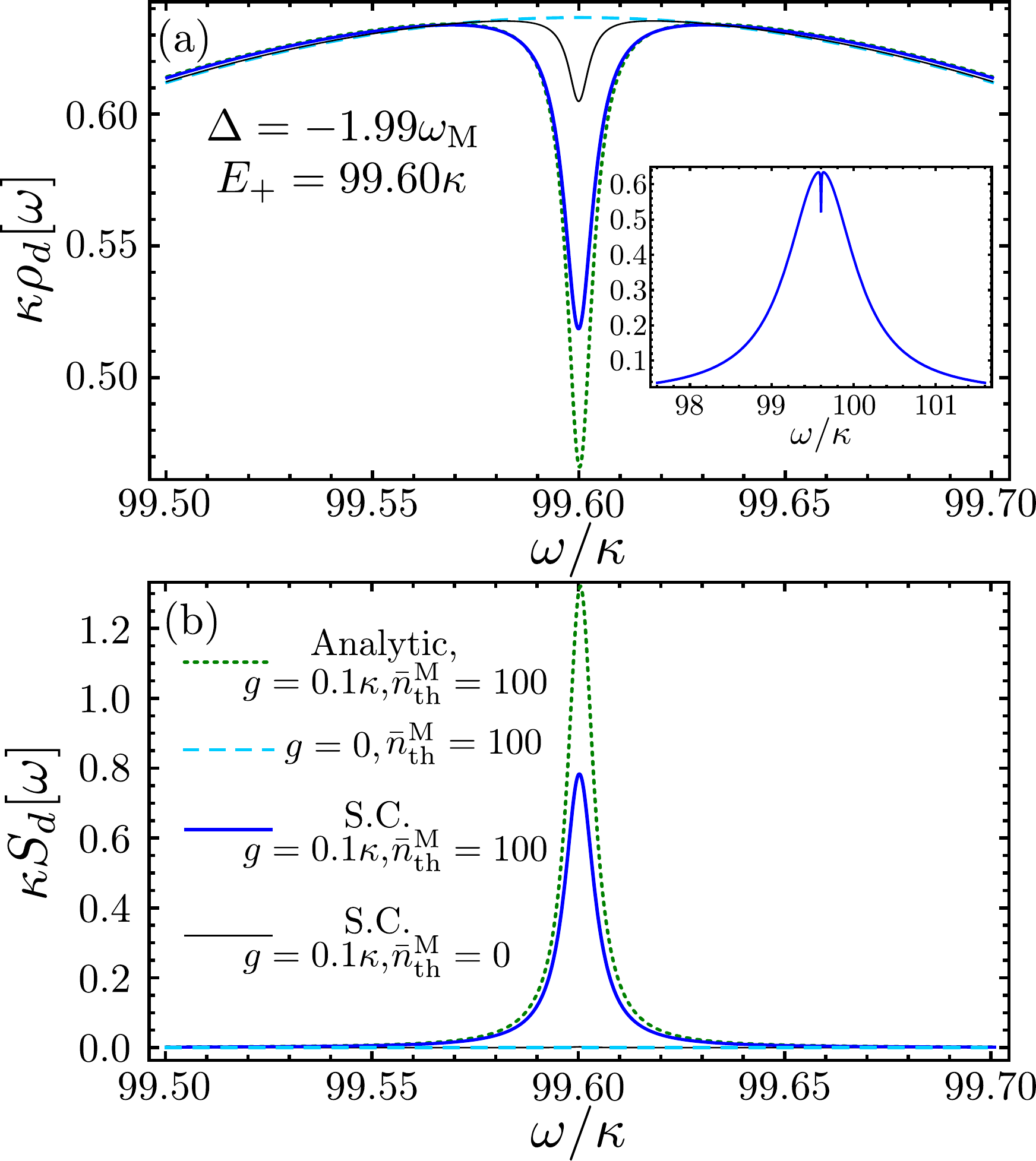}
	\end{center}
	\vspace{-0.5cm}
	\caption {(Color online)
	(a) Signatures of the effective two-phonon absorption in the cavity DOS near the + resonance; the inset shows a larger range of frequency than the main plot. The light-blue dashed curve represents the linearized theory ($g=0$), the green dotted one represents the results to leading order in $g$ (Eqs.~\eqref{Eq:DOS}, \eqref{Eq:neff} and \eqref{Eq:SdFull}) and, the full dark-blue and black curves are the results of the self-consistent approach (S.C.) of Sec.~(\ref{Sec:SelfConsistentApp}) for $\nthM= 100$ and $\nthM = 0$ respectively. (b) Signatures of the effective two-phonon absorption in the cavity spectrum in the same circumstances than (a). It shows the striking temperature-enhanced effect of nonlinear interaction in the + polariton population. The parameters used are $\gamma/\kappa=10^{-4}$ and $\omega_M/\kappa=50$.}
\label{fig:Sd2wm}
\end{figure}

%%%%%%%%%%%%%%%%%%%%%%%%%%%%%%%%%%%%%%%%%%%%%%%%%%%%%%%%%%%%%%%%%%%%%%%%%%%%%%%%%%%%%%%%%%%%%%%%%%%%%%%%%%%%%%%%%%%%%%%%%%%%%%%%%%%%%%%%%%

\section{Conclusions} \label{Sec:Conclusion}

In this work, we have described the effects of nonlinear interaction on the non-equilibrium state of the optomechanical cavity. We have shown the tendency of the nonlinear interaction to thermalize the polaritons, that it can lead to a new parametric instability for a red-detuned laser drive as well as a temperature-enhanced effective two phonons absorption. In addition to these results, we have presented in details many technical aspects with the aim to provide the proper tools to investigate nonlinear effects in more complicated optomechanical systems. This work also opens the path to a more detailed characterization of this new parametric instability and further investigations of its consequences.  

%%%%%%%%%%%%%%%%%%%%%%%%%%%%%%%%%%%%%%%%%%%%%%%%%%%%%%%%%%%%%%%%%%%%%%%%%%%%%%%%%%%%%%%%%%%%%%%%%%%%%%%%%%%%%%%%%%%%%%%%%%%%%%%%%%%%%%%%%%

\section*{Acknowledgements}

We thank Nicolas Didier and F\'elix Beaudoin for helpful discussions.  This work was supported by NSERC and FQRNT.

\appendix

\section{Change of basis for a detuning on the red sideband ($\Delta = -\omega_M$)} \label{Sec:AppendixCOB}

In this Appendix, we show particular examples of the transformation used to go from the photon/phonon basis to the polariton basis for $\Delta = -\omega_M$.

We start with the change-of-basis coefficients $\alpha_{b/d,\pm}$ and $\bar{\alpha}_{b/d,\pm}$ introduced in Eqs.~(\ref{Eq:ChangeOfBasis}),
\begin{align} 
	\alpha_{b,\pm} & = \pm \frac{1}{\sqrt{8\omega_M}} \frac{\omega_M + E_{\pm}}{\sqrt{E_{\pm}}}, \label{App:COB1} \\
	\bar{\alpha}_{b,\pm} & = \pm \frac{1}{\sqrt{8\omega_M}} \frac{\omega_M - E_{\pm}}{\sqrt{E_{\pm}}}, \\
	\alpha_{d,\pm} & = \frac{1}{\sqrt{8\omega_M}} \frac{\omega_M + E_{\pm}}{\sqrt{E_{\pm}}}, \\ 
	\bar{\alpha}_{d,\pm} & = \frac{1}{\sqrt{8\omega_M}} \frac{\omega_M - E_{\pm}}{\sqrt{E_{\pm}}}, \label{App:COB2}
\end{align}
with 
\begin{align}
	& E_{\pm} = \omega_M\sqrt{1 \pm 2G/\omega_M}, \\
	& G_{\rm{res}}/\omega_M = 3/10. \label{App:Gres}
\end{align}
From these coefficients, we can express the different effective nonlinear coupling constants $g^{A/B}_{\sigma \sigma' \sigma''}$ and $A_{\pm}$ introduced in Eq.~(\ref{Eq:HintPolariton}). Here, we only show few examples.
\begin{align}
	g^{A}_{++-} + & g^{A}_{+-+} + g^{A}_{-++} \nonumber \\
	= &  g\left[ \adp\badp(\abm+\babm) \right. \nonumber \\
	& \left. + (\badm\adp + \badp\adm)(\abp + \babp) \right], \label{App:g1} \nonumber \\
	= & \frac{g}{4\sqrt{2}}\left(\frac{\omega_M}{E_-}\right)^{3/4}\left[ \frac{E_--E_++\frac{G}{\omega_M}(E_-+2E_+)}{E_+} \right],
\end{align}
\begin{align}
	\tilde{g}  \equiv g^{B}_{--+} & = g\left[ \adm\badm(\abp+\babp) \right. \nonumber \\
	& \left. + (\badm\badp + \adm\adp)(\abm + \babm) \right], \label{App:gtilde} \nonumber \\
	=  - & \frac{g}{4\sqrt{2}} \left(\frac{\omega_M}{E_+}\right)^{3/4}\left[ \frac{(1+2\frac{G}{\omega_M})E_-+(1-\frac{G}{\omega_M})E_+}{E_-} \right],
\end{align}
\begin{align}
	A_{-} = & g\left[ (2\adm\badm+\badm^2+\badp^2)(\abm+\babm)\right. \nonumber \\
	& + \left. (\badm\adp + \badp\adm)(\abp + \babp) \right], \nonumber \\
	= & \frac{g}{4\sqrt{2}}\left(\frac{\omega_M}{E_-}\right)^{3/4}\left[ \frac{2(\frac{E_-}{\omega_M}-1)E_++\frac{G}{\omega_M}(E_+-E_-)}{E_+} \right], \label{App:g2}
\end{align}

%%%%%%%%%%%%%%%%%%%%%%%%%%%%%%%%%%%%%%%%%%%%%%%%%%%%%%%%%%%%%%%%%%%%%%%%%%%%%%%%%%%%%%%%%%%%%%%%%%%%%%%%%%%%%%%%%%%%%%%%%%%%%%%%%%%%%%%%%%

\section{Parametric amplifier description} \label{Sec:AppendixParamp}

In this appendix, we derive the condition to get an instability from the parametric amplifier Hamiltonian and get the corresponding mean number of signal excitations. These calculations are based on the formalism introduced in \cite{Clerk_RMP_2010}.

Starting with the nonlinear interaction in the effective Hamiltonian of Eq.~\eqref{Eq:HEff} and considering the {\it hypothetical} case where the + mode is {\it coherently} pumped, such that $\langle \hcp \rangle = \bnop$, we get, in the mean field approximation:
\begin{equation}
	\hHeff \approx \tilde{g}\sqrt{\bnop}(\hcm \hcm + \rm{H.c.}). \label{App:Hparamp}
\end{equation}
Using standard input-output theory, one derives the following equation of motion
\begin{align}
	\partial_t \hcm & = i[\hHeff,\hcm]-\frac{\kappa_-}{2}\hcm - \sqrt{\kappa_-}\hat{\xi}_-, \\
	& = -2\tilde{g}\sqrt{\bnop}\hcmdag-\frac{\kappa_-}{2}\hcm - \sqrt{\kappa_-}\hat{\xi}_-.
\end{align}
Here, $\hat{\xi}_-$ represents an incoming field, which includes noise coming from the bath coupled to the $-$ polaritons (as in Eq.~(\ref{Eq:LinLangevinEq})). Since we will focus only on quantities evaluated at $\omega = E_-$ and that $\hHeff$ of Eq.~(\ref{App:Hparamp}) is written in the interaction picture where $E_- = 0$, we can seek for the particular solution given by $\partial_- \hcm = 0$. In that case, we have:
\begin{gather}
		\hcm = -\frac{4i\tilde{g}\sqrt{\bnop}}{\kappa_-}\hcmdag - \frac{2}{\sqrt{\kappa_-}}\hat{\xi}_-, \\
		\Rightarrow \hcm = \frac{2}{\sqrt{\kappa_-}}\left( 1- \lvert Q \rvert ^2 \right)^{-1} \left( Q \hat{\xi}_-^{\dag}- \hat{\xi}_- \right),
\end{gather}
with
\begin{equation}
	Q \equiv \frac{4i\tilde{g}\sqrt{\bnop}}{\kappa_-}.
\end{equation}
Imposing the right commutation relation $[\hcm,\hcmdag] = 1$, one finds
\begin{equation}
	[\hat{\xi}_-, \hat{\xi}^{\dag}_-] =  \frac{\kappa_-\left( 1- \vert Q \vert ^2 \right)}{4},
\end{equation}
 which has the right units since $\langle \hat{\xi}^{\dag}_-\hat{\xi}_- \rangle$ represents a rate at which excitations are coming in. In the case of Gaussian noise with zero mean (i.e. $\langle \hat{\xi}^{\dag}_-\hat{\xi}_- \rangle = 0$), as studied all along this work, we find that the mean number of $-$ polaritons inside the optomechanical cavity is
 \begin{equation}
 	\langle \hcmdag \hcm \rangle = \frac{\vert Q \vert^2}{1 - \vert Q \vert ^2} = \frac{16\tilde{g}^2\bnop}{\kappa_-^2-16\tilde{g}^2\bnop}. \label{App:nmparamps}
 \end{equation}
From this result, one immediately sees that for $\vert Q \vert^2 = 1$, the mean number of polaritons diverges. This condition sets the threshold above which the parametric amplifier goes unstable. More precisely, the system becomes unstable when
\begin{equation}
	\bnop > \frac{\kappa_-^2}{16\tilde{g}^2}. \label{App:InstCond}
\end{equation}
These results have exactly the same form than Eqs.~\eqref{Eq:CondInsta} and \eqref{Eq:nmparamp}, but here, we have explicitly used the fact that the + polaritons are coherently pumped such that $\langle \hcp \rangle = \bnop$.

%%%%%%%%%%%%%%%%%%%%%%%%%%%%%%%%%%%%%%%%%%%%%%%%%%%%%%%%%%%%%%%%%%%%%%%%%%%%%%%%%%%%%%%%%%%%%%%%%%%%%%%%%%%%%%%%%%%%%%%%%%%%%%%%%%%%%%%%%%

\section{Expansion of the photon spectrum function for $\Delta \approx -2\omega_M$} \label{Sec:AppNear2omega}

In this Appendix, we show details of the calculation that leads to Eq.~(\ref{Eq:Sdnear2wm}) starting from Eqs.~(\ref{Eq:neff}) and Eq.~(\ref{Eq:SdFull}). Here, the limit of interest is for $\Delta \approx -2\omega_M$, $G/\omega_M \sim \gamma/\kappa \ll 1$, $\nthM \gg 1$ and $\Ceffp \ll 1$. In these circumstances, photons are mainly $+$ polaritons, such that Eq.~(\ref{Eq:Sdnear2wm}) for frequencies near $E_+$ reduces to
\begin{equation}
	S_d[\omega]\vert_{\omega \approx E_+} \approx 2\pi \bneffp \rho_+[\omega].
\end{equation}
Also, for $\Delta \approx -2\omega_M$ and $G/\omega_M \sim \gamma/\kappa \ll 1$, the mean number of photons inside the cavity without nonlinear interaction, $\bnop$, is negligible compared to the nonlinear contribution. Thus, from Eq.~(\ref{Eq:neff}), we see that $\bneffp$ reduces to a sharply peaked Lorenztian with witdh $\gamma_+ \approx \kappa_- \approx \gamma(1+\Gamma_{\rm{opt}}) \sim \gamma$ with the optical damping rate $\Gamma_{\rm{opt}} \approx \frac{8}{9}\frac{G^2}{\omega_M^2}\kappa$. Consequently, $\rho_+[\omega]$ (Eq.~(\ref{Eq:DOS})), which has a characteristic width of $\kappa_+ \approx \kappa$, can be evaluates at $E_+$. Doing so, one gets
\begin{align}
	S_d[\omega]\vert_{\omega \approx E_+} & \approx 2\pi \bneffp \rho_+[E_+], \\
   \rho_+[E_+] & = \frac{2}{\pi}\frac{1}{\kappa_+(1+\Ceffp)} \approx \frac{2}{\pi\kappa}.
\end{align}
Now, we can write $\bneffp$ as (we have dropped $\bnop$ and explicitely used $E_+=2E_-$)
\begin{align}
	\bneffp \approx \bnintp \Ceffp \frac{\kappa_-^2}{(\omega-E_+)^2+\kappa_-^2},
\end{align}
with
\begin{align}
	\bnintp\Ceffp = \frac{4\tilde{g}^2(\bnom)^2}{\kappa_-\kappa_+}.
\end{align}
Thus
\begin{align}
	S_d[\omega]\vert_{\omega \approx E_+} & \approx \frac{16}{\kappa^2} \frac{\tilde{g}^2(\bnom)^2}{\kappa_-}\frac{\kappa_-^2}{(\omega-E_+)^2+\kappa_-^2},
\end{align}
Again, for this particular limit, we can approximate $\bnom \approx \frac{\gamma \nthM}{\kappa_-}$ and $\tilde{g}^2 \approx (\frac{G}{\omega_M})^2g^2$, such that we get Eq.~(\ref{Eq:Sdnear2wm}), i.e.
\begin{equation}
	\hspace{-0.15cm} \kappa S_d[\omega]\vert_{\omega \approx E_+} \approx \frac{\kappa}{\gamma} \left( \frac{G}{\omega_M} \right)^2 \left( \frac{g}{\kappa} \right)^2 \frac{16(\nthM)^2}{(1+\frac{\Gamma_{\rm{opt}}}{\gamma})}\frac{\kappa_-^2}{(\omega-E_+)^2+\kappa_-^2}
\end{equation}

%%%%%%%%%%%%%%%%%%%%%%%%%%%%%%%%%%%%%%%%%%%%%%%%%%%%%%%%%%%%%%%%%%%%%%%%%%%%%%%%%%%%%%%%%%%%%%%%%%%%%%%%%%%%%%%%%%%%%%%%%%%%%%%%%%%%%%%%%%

\section{Classical treatment of the effective two phonons absorption} \label{App:ClassicalTwoPhononsAps}

As shown in the Sec.~\ref{Sec:TwoPhononsAbs}, the nonlinear process corresponding to the absorption of two phonons by a photon from the classical drive is greatly enhanced by temperature. In particular, even if the nonlinear interaction directly comes from single photon dynamics, one can show that in the high temperatures limit, this phenomenon becomes purely classical. In what follows, we present an accurate classical description of this limit.

The classical Langevin equations for the cavity field coupled to a mechanical resonator via radiation pressure force in the limit of weak displacements are
\begin{gather}
	\dot{a}(t) = \left( -\frac{\kappa}{2}-i\omega_C\left( 1-\frac{x(t)}{L_0}\right)\right)a(t)+i\bar{a}_{in}e^{-i\omega_Lt}, \label{Eq:ClassicalEOMCavity} \\
	\ddot{x}(t)+\frac{\gamma}{2}\dot{x}+\omega_M^2 x = \frac{|a(t)|^2}{m L_0}. \label{Eq:ClassicalEOMResonator}
\end{gather}
Here, $m$ is the mass of the mechanical oscillator, $L_0$ is the length of the cavity when $x(t)=0$ and the damping rate $\kappa$ has been introduced following the standard input-output formalism \cite{Clerk_RMP_2010}. The cavity field is normalized such that the steady state mean energy inside the cavity is $ \bar{U}_{\textrm{cav}}=|\bar{a}|^2 = \kappa|\bar{a}_{\textrm{in}}|^2/(( \frac{\kappa}{2})^2+\Delta^2)$, where $\bar{a}$ is defined as the steady state mean value of the cavity field, $a(t)=(\bar{a}+\delta a(t))e^{-i\omega_Lt}$.

We solve these nonlinear coupled equations of motion via a perturbation approach. Starting with a sinusoidal displacement $x(t)=x_0 \mathrm{sin}(\omega_M t)$, we get the consequent evolution of the optical mode and then, the perturbed displacement of the mechanical resonator. This is a good approximation only in the weak coupling regime ($G \ll \kappa$) where the eigenstates have a well defined number of photons and phonons. This method is thus restricted to the extreme detuning case ($\Delta \approx -2\omega_M$) studied in Sec.~\ref{Sec:TwoPhononsAbs}.

Following this method, one finds that the steady state cavity field is given by
\begin{equation}
	a(t)	= ia_{\textrm{in}}\sum \limits_{n,m=-\infty}^{+\infty}\frac{i^{(n-m)}J_n(\epsilon)J_m(\epsilon)}{\frac{\kappa}{2}+i(n\omega_M-\Delta)}e^{i((n+m)\omega_M-\omega_L)t}. \label{Eq:ClassicalField}
\end{equation}
Here, $J_n(x)$ are the Bessel functions of the first kind and where, in the small displacement limit, $\epsilon \equiv \frac{x_0}{L_0}\frac{\omega_C}{\omega_M}$ plays the role of the perturbation parameter.

Note that by computing the radiation pressure force acting on the mechanical resonator in Eq.~(\ref{Eq:ClassicalEOMResonator}) and by expanding to the lowest order in $\epsilon$, one recovers the standard optical damping and optical spring constant coming from the linear interaction (cf.~Eq.~(\ref{Eq:HL})) in the weak coupling regime~\cite{Marquartd_PRL_2007}.

We focus here on the peak in the cavity spectrum that comes from the nonlinear interaction. According to Eq.~(\ref{Eq:DefSd}), the proper definition of the classical counterpart of the photon spectrum to use is
\begin{equation}
	S_{\textrm{cl}}[\omega] \equiv \frac{1}{\omega_C}\int\limits_{-\infty}^{\infty}dt e^{i\omega t}\langle \delta a^*(t_0) \delta a(t_0+t)\rangle_{t_0}.
	\label{Eq:SdClassical}
\end{equation}
Here, the fluctuations of the cavity field $\delta a(t)$ are given by keeping only the $n \neq -m$ terms in Eq.~(\ref{Eq:ClassicalField}). The mean value $\langle...\rangle_{t_{0}}$ means taking the average over all initial time $t_0$ and the coefficient $1/ \omega_C$ is there to ensure that units match the definition of the photon spectrum used in the quantum treatment (cf.~Eq.~\eqref{Eq:DefSd}).

Using Eq.~(\ref{Eq:ClassicalField}) and focusing only on the frequency range near $2\omega_M$ with $\Delta \approx -2\omega_M$, one finds that in the good cavity limit ($\kappa \ll \omega_M$), the classical cavity spectrum reads
\begin{equation}
	S_{\textrm{cl}}[\omega] \approx 16 \left( \frac{G}{\omega_M} \right)^2 \left( \frac{g}{\kappa} \right)^2 (\nthM)^2 \pi \delta (\omega-2\omega_M).
\end{equation}
Where we have used the definition of the single photon coupling constant, $g \equiv \frac{x_{\textrm{ZPF}}\omega_C}{L_0}$, with $x_{\textrm{ZPF}}$ being the zero point motion of the mechanical resonator and the semi-classical relation $\frac{1}{2}m\omega_M^2x_0^2 = \omega_M \nthM$ valid for $\nthM \gg 1$. 

If one takes $\Gamma_{\textrm{opt}}=0$ and $\gamma \rightarrow 0$ in Eq.~(\ref{Eq:Sdnear2wm}), $S_d[\omega\approx E_+]$ and $S_{\textrm{cl}}[\omega\approx2\omega_M]$ agree. The classical approach fails in getting the width of the peak since the mechanical damping rate is not taken into account while calculating the effects of the mechanical motion on the cavity field. Moreover, the optical damping doesn't come out in this result in the level of approximation we have used.

%%%%%%%%%%%%%%%%%%%%%%%%%%%%%%%%%%%%%%%%%%%%%%%%%%%%%%%%%%%%%%%%%%%%%%%%%%%%%%%%%%%%%%%%%%%%%%%%%%%%%%%%%%%%%%%%%%%%%%%%%%%%%%%%%%%%%%%%%%

\bibliographystyle{apsrev}
\bibliography{MALPapersRefs}

\begin{thebibliography}{54}
\expandafter\ifx\csname natexlab\endcsname\relax\def\natexlab#1{#1}\fi
\expandafter\ifx\csname bibnamefont\endcsname\relax
  \def\bibnamefont#1{#1}\fi
\expandafter\ifx\csname bibfnamefont\endcsname\relax
  \def\bibfnamefont#1{#1}\fi
\expandafter\ifx\csname citenamefont\endcsname\relax
  \def\citenamefont#1{#1}\fi
\expandafter\ifx\csname url\endcsname\relax
  \def\url#1{\texttt{#1}}\fi
\expandafter\ifx\csname urlprefix\endcsname\relax\def\urlprefix{URL }\fi
\providecommand{\bibinfo}[2]{#2}
\providecommand{\eprint}[2][]{\url{#2}}

\bibitem[{\citenamefont{Aspelmeyer et~al.}(2013)\citenamefont{Aspelmeyer,
  Kippenberg, and Marquardt}}]{MarquardtRMP}
\bibinfo{author}{\bibfnamefont{M.}~\bibnamefont{Aspelmeyer}},
  \bibinfo{author}{\bibfnamefont{T.~J.} \bibnamefont{Kippenberg}},
  \bibnamefont{and}
  \bibinfo{author}{\bibfnamefont{F.}~\bibnamefont{Marquardt}},
  \bibinfo{journal}{arXiv:1303.0733v1}  (\bibinfo{year}{2013}).

\bibitem[{\citenamefont{Aspelmeyer et~al.}(2014)\citenamefont{Aspelmeyer,
  Kippenberg, and Marquardt}}]{OptomechanicsBook}
\bibinfo{author}{\bibfnamefont{M.}~\bibnamefont{Aspelmeyer}},
  \bibinfo{author}{\bibfnamefont{T.~J.} \bibnamefont{Kippenberg}},
  \bibnamefont{and}
  \bibinfo{author}{\bibfnamefont{F.}~\bibnamefont{Marquardt}},
  \emph{\bibinfo{title}{Cavity Optomechanics}} (\bibinfo{publisher}{Springer,
  Berlin}, \bibinfo{year}{2014}).

\bibitem[{\citenamefont{Teufel et~al.}(2011{\natexlab{a}})\citenamefont{Teufel,
  Donner, Li, Harlow, Allman, Cicak, Sirois, Whittaker, Lehnert, and
  Simmonds}}]{Teufel_Nature_2011}
\bibinfo{author}{\bibfnamefont{J.}~\bibnamefont{Teufel}},
  \bibinfo{author}{\bibfnamefont{T.}~\bibnamefont{Donner}},
  \bibinfo{author}{\bibfnamefont{D.}~\bibnamefont{Li}},
  \bibinfo{author}{\bibfnamefont{J.}~\bibnamefont{Harlow}},
  \bibinfo{author}{\bibfnamefont{M.}~\bibnamefont{Allman}},
  \bibinfo{author}{\bibfnamefont{K.}~\bibnamefont{Cicak}},
  \bibinfo{author}{\bibfnamefont{A.}~\bibnamefont{Sirois}},
  \bibinfo{author}{\bibfnamefont{J.}~\bibnamefont{Whittaker}},
  \bibinfo{author}{\bibfnamefont{K.}~\bibnamefont{Lehnert}}, \bibnamefont{and}
  \bibinfo{author}{\bibfnamefont{R.}~\bibnamefont{Simmonds}},
  \bibinfo{journal}{Nature} \textbf{\bibinfo{volume}{475}},
  \bibinfo{pages}{359} (\bibinfo{year}{2011}{\natexlab{a}}).

\bibitem[{\citenamefont{Chan et~al.}(2011)\citenamefont{Chan, Alegre,
  Safavi-Naeini, Hill, Krause, Gr{\"o}blacher, Aspelmeyer, and
  Painter}}]{Chan_Nature_2011}
\bibinfo{author}{\bibfnamefont{J.}~\bibnamefont{Chan}},
  \bibinfo{author}{\bibfnamefont{T.~M.} \bibnamefont{Alegre}},
  \bibinfo{author}{\bibfnamefont{A.~H.} \bibnamefont{Safavi-Naeini}},
  \bibinfo{author}{\bibfnamefont{J.~T.} \bibnamefont{Hill}},
  \bibinfo{author}{\bibfnamefont{A.}~\bibnamefont{Krause}},
  \bibinfo{author}{\bibfnamefont{S.}~\bibnamefont{Gr{\"o}blacher}},
  \bibinfo{author}{\bibfnamefont{M.}~\bibnamefont{Aspelmeyer}},
  \bibnamefont{and} \bibinfo{author}{\bibfnamefont{O.}~\bibnamefont{Painter}},
  \bibinfo{journal}{Nature} \textbf{\bibinfo{volume}{478}}, \bibinfo{pages}{89}
  (\bibinfo{year}{2011}).

\bibitem[{\citenamefont{Brooks et~al.}(2012)\citenamefont{Brooks, Botter,
  Schreppler, Purdy, Brahms, and Stamper-Kurn}}]{Brooks_Nature_2012}
\bibinfo{author}{\bibfnamefont{D.~W.} \bibnamefont{Brooks}},
  \bibinfo{author}{\bibfnamefont{T.}~\bibnamefont{Botter}},
  \bibinfo{author}{\bibfnamefont{S.}~\bibnamefont{Schreppler}},
  \bibinfo{author}{\bibfnamefont{T.~P.} \bibnamefont{Purdy}},
  \bibinfo{author}{\bibfnamefont{N.}~\bibnamefont{Brahms}}, \bibnamefont{and}
  \bibinfo{author}{\bibfnamefont{D.~M.} \bibnamefont{Stamper-Kurn}},
  \bibinfo{journal}{Nature} \textbf{\bibinfo{volume}{488}},
  \bibinfo{pages}{476} (\bibinfo{year}{2012}).

\bibitem[{\citenamefont{Safavi-Naeini et~al.}(2013)\citenamefont{Safavi-Naeini,
  Gr{\"o}blacher, Hill, Chan, Aspelmeyer, and
  Painter}}]{Safavi-Naeini_Nature_2013}
\bibinfo{author}{\bibfnamefont{A.~H.} \bibnamefont{Safavi-Naeini}},
  \bibinfo{author}{\bibfnamefont{S.}~\bibnamefont{Gr{\"o}blacher}},
  \bibinfo{author}{\bibfnamefont{J.~T.} \bibnamefont{Hill}},
  \bibinfo{author}{\bibfnamefont{J.}~\bibnamefont{Chan}},
  \bibinfo{author}{\bibfnamefont{M.}~\bibnamefont{Aspelmeyer}},
  \bibnamefont{and} \bibinfo{author}{\bibfnamefont{O.}~\bibnamefont{Painter}},
  \bibinfo{journal}{Nature} \textbf{\bibinfo{volume}{500}},
  \bibinfo{pages}{185} (\bibinfo{year}{2013}).

\bibitem[{\citenamefont{Purdy et~al.}(2013)\citenamefont{Purdy, Yu, Peterson,
  Kampel, and Regal}}]{Regal2013}
\bibinfo{author}{\bibfnamefont{T.~P.} \bibnamefont{Purdy}},
  \bibinfo{author}{\bibfnamefont{P.~L.} \bibnamefont{Yu}},
  \bibinfo{author}{\bibfnamefont{R.~W.} \bibnamefont{Peterson}},
  \bibinfo{author}{\bibfnamefont{N.~S.} \bibnamefont{Kampel}},
  \bibnamefont{and} \bibinfo{author}{\bibfnamefont{C.~A.} \bibnamefont{Regal}},
  \bibinfo{journal}{Phys. Rev. X} \textbf{\bibinfo{volume}{3}},
  \bibinfo{pages}{031012} (\bibinfo{year}{2013}).

\bibitem[{\citenamefont{Nunnenkamp et~al.}(2011)\citenamefont{Nunnenkamp,
  B{\o}rkje, and Girvin}}]{Nunnenkamp_PRL_2011}
\bibinfo{author}{\bibfnamefont{A.}~\bibnamefont{Nunnenkamp}},
  \bibinfo{author}{\bibfnamefont{K.}~\bibnamefont{B{\o}rkje}},
  \bibnamefont{and} \bibinfo{author}{\bibfnamefont{S.}~\bibnamefont{Girvin}},
  \bibinfo{journal}{Physical Review Letters} \textbf{\bibinfo{volume}{107}},
  \bibinfo{pages}{063602} (\bibinfo{year}{2011}).

\bibitem[{\citenamefont{Rabl}(2011)}]{Rabl_PRL_2011}
\bibinfo{author}{\bibfnamefont{P.}~\bibnamefont{Rabl}}, \bibinfo{journal}{Phys.
  Rev. Lett.} \textbf{\bibinfo{volume}{107}}, \bibinfo{pages}{063601}
  (\bibinfo{year}{2011}).

\bibitem[{\citenamefont{Kronwald et~al.}(2013)\citenamefont{Kronwald, Ludwig,
  and Marquardt}}]{Kronwald_PRA_2013}
\bibinfo{author}{\bibfnamefont{A.}~\bibnamefont{Kronwald}},
  \bibinfo{author}{\bibfnamefont{M.}~\bibnamefont{Ludwig}}, \bibnamefont{and}
  \bibinfo{author}{\bibfnamefont{F.}~\bibnamefont{Marquardt}},
  \bibinfo{journal}{Phys. Rev. A} \textbf{\bibinfo{volume}{87}},
  \bibinfo{pages}{013847} (\bibinfo{year}{2013}).

\bibitem[{\citenamefont{Kronwald and Marquardt}(2013)}]{Kronwald_PRL_2013}
\bibinfo{author}{\bibfnamefont{A.}~\bibnamefont{Kronwald}} \bibnamefont{and}
  \bibinfo{author}{\bibfnamefont{F.}~\bibnamefont{Marquardt}},
  \bibinfo{journal}{Phys. Rev. Lett.} \textbf{\bibinfo{volume}{111}},
  \bibinfo{pages}{133601} (\bibinfo{year}{2013}).

\bibitem[{\citenamefont{Murch et~al.}(2008)\citenamefont{Murch, Moore, Gupta,
  and Stamper-Kurn}}]{Murch_Nature_2008}
\bibinfo{author}{\bibfnamefont{K.~W.} \bibnamefont{Murch}},
  \bibinfo{author}{\bibfnamefont{K.~L.} \bibnamefont{Moore}},
  \bibinfo{author}{\bibfnamefont{S.}~\bibnamefont{Gupta}}, \bibnamefont{and}
  \bibinfo{author}{\bibfnamefont{D.~M.} \bibnamefont{Stamper-Kurn}},
  \bibinfo{journal}{Nature} \textbf{\bibinfo{volume}{4}}, \bibinfo{pages}{561}
  (\bibinfo{year}{2008}).

\bibitem[{\citenamefont{Brennecke et~al.}(2008)\citenamefont{Brennecke, Ritter,
  Donner, and Esslinger}}]{Brennecke_Science_2008}
\bibinfo{author}{\bibfnamefont{F.}~\bibnamefont{Brennecke}},
  \bibinfo{author}{\bibfnamefont{S.}~\bibnamefont{Ritter}},
  \bibinfo{author}{\bibfnamefont{T.}~\bibnamefont{Donner}}, \bibnamefont{and}
  \bibinfo{author}{\bibfnamefont{T.}~\bibnamefont{Esslinger}},
  \bibinfo{journal}{Science} \textbf{\bibinfo{volume}{322}},
  \bibinfo{pages}{235} (\bibinfo{year}{2008}).

\bibitem[{\citenamefont{Ludwig et~al.}(2012)\citenamefont{Ludwig,
  Safavi-Naeini, Painter, and Marquardt}}]{Ludwig_Safavi_PRL_2012}
\bibinfo{author}{\bibfnamefont{M.}~\bibnamefont{Ludwig}},
  \bibinfo{author}{\bibfnamefont{A.~H.} \bibnamefont{Safavi-Naeini}},
  \bibinfo{author}{\bibfnamefont{O.}~\bibnamefont{Painter}}, \bibnamefont{and}
  \bibinfo{author}{\bibfnamefont{F.}~\bibnamefont{Marquardt}},
  \bibinfo{journal}{Phys. Rev. Lett.} \textbf{\bibinfo{volume}{109}},
  \bibinfo{pages}{063601} (\bibinfo{year}{2012}).

\bibitem[{\citenamefont{Komar et~al.}(2013)\citenamefont{Komar, Bennett,
  Stannigel, Habraken, Rabl, Zoller, and Lukin}}]{Komar_Bennett_PRA_2013}
\bibinfo{author}{\bibfnamefont{P.}~\bibnamefont{Komar}},
  \bibinfo{author}{\bibfnamefont{S.~D.} \bibnamefont{Bennett}},
  \bibinfo{author}{\bibfnamefont{K.}~\bibnamefont{Stannigel}},
  \bibinfo{author}{\bibfnamefont{S.~J.~M.} \bibnamefont{Habraken}},
  \bibinfo{author}{\bibfnamefont{P.}~\bibnamefont{Rabl}},
  \bibinfo{author}{\bibfnamefont{P.}~\bibnamefont{Zoller}}, \bibnamefont{and}
  \bibinfo{author}{\bibfnamefont{M.~D.} \bibnamefont{Lukin}},
  \bibinfo{journal}{Phys. Rev. A} \textbf{\bibinfo{volume}{87}},
  \bibinfo{pages}{013839} (\bibinfo{year}{2013}).

\bibitem[{\citenamefont{Borkje et~al.}(2013)\citenamefont{Borkje, Nunnenkamp,
  Teufel, and Girvin}}]{Borkje_PRL_2013}
\bibinfo{author}{\bibfnamefont{K.}~\bibnamefont{Borkje}},
  \bibinfo{author}{\bibfnamefont{A.}~\bibnamefont{Nunnenkamp}},
  \bibinfo{author}{\bibfnamefont{J.~D.} \bibnamefont{Teufel}},
  \bibnamefont{and} \bibinfo{author}{\bibfnamefont{S.~M.}
  \bibnamefont{Girvin}}, \bibinfo{journal}{Phys. Rev. Lett.}
  \textbf{\bibinfo{volume}{111}}, \bibinfo{pages}{053603}
  (\bibinfo{year}{2013}).

\bibitem[{\citenamefont{Lemonde et~al.}(2013)\citenamefont{Lemonde, Didier, and
  Clerk}}]{MAL_PRL_2011}
\bibinfo{author}{\bibfnamefont{M.-A.} \bibnamefont{Lemonde}},
  \bibinfo{author}{\bibfnamefont{N.}~\bibnamefont{Didier}}, \bibnamefont{and}
  \bibinfo{author}{\bibfnamefont{A.~A.} \bibnamefont{Clerk}},
  \bibinfo{journal}{Phys. Rev. Lett.} \textbf{\bibinfo{volume}{111}},
  \bibinfo{pages}{053602} (\bibinfo{year}{2013}).

\bibitem[{\citenamefont{Xu et~al.}(2014)\citenamefont{Xu, Gullans, and
  Taylor}}]{Xu_ArXiv_2014}
\bibinfo{author}{\bibfnamefont{X.}~\bibnamefont{Xu}},
  \bibinfo{author}{\bibfnamefont{M.}~\bibnamefont{Gullans}}, \bibnamefont{and}
  \bibinfo{author}{\bibfnamefont{J.~M.} \bibnamefont{Taylor}},
  \bibinfo{journal}{arXiv:1404.3726}  (\bibinfo{year}{2014}).

\bibitem[{\citenamefont{Teufel et~al.}(2011{\natexlab{b}})\citenamefont{Teufel,
  Li, Allman, Cicak, Sirois, Whittaker, and
  Simmonds}}]{Teufel_Nature_2011_OMIT}
\bibinfo{author}{\bibfnamefont{J.}~\bibnamefont{Teufel}},
  \bibinfo{author}{\bibfnamefont{D.}~\bibnamefont{Li}},
  \bibinfo{author}{\bibfnamefont{M.}~\bibnamefont{Allman}},
  \bibinfo{author}{\bibfnamefont{K.}~\bibnamefont{Cicak}},
  \bibinfo{author}{\bibfnamefont{A.}~\bibnamefont{Sirois}},
  \bibinfo{author}{\bibfnamefont{J.}~\bibnamefont{Whittaker}},
  \bibnamefont{and} \bibinfo{author}{\bibfnamefont{R.}~\bibnamefont{Simmonds}},
  \bibinfo{journal}{Nature} \textbf{\bibinfo{volume}{471}},
  \bibinfo{pages}{204} (\bibinfo{year}{2011}{\natexlab{b}}).

\bibitem[{\citenamefont{Groblacher et~al.}(2009)\citenamefont{Groblacher,
  Hammerer, Vanner, and Aspelmeyer}}]{Groeblacher_Nature_2009}
\bibinfo{author}{\bibfnamefont{S.}~\bibnamefont{Groblacher}},
  \bibinfo{author}{\bibfnamefont{K.}~\bibnamefont{Hammerer}},
  \bibinfo{author}{\bibfnamefont{M.~R.} \bibnamefont{Vanner}},
  \bibnamefont{and}
  \bibinfo{author}{\bibfnamefont{M.}~\bibnamefont{Aspelmeyer}},
  \bibinfo{journal}{Nature} \textbf{\bibinfo{volume}{460}},
  \bibinfo{pages}{724} (\bibinfo{year}{2009}).

\bibitem[{\citenamefont{Verhagen et~al.}(2010)\citenamefont{Verhagen,
  Deleglise, Weis, Schliesser, and Kippenderg}}]{Verhagen_Nature_2012}
\bibinfo{author}{\bibfnamefont{E.}~\bibnamefont{Verhagen}},
  \bibinfo{author}{\bibfnamefont{S.}~\bibnamefont{Deleglise}},
  \bibinfo{author}{\bibfnamefont{S.}~\bibnamefont{Weis}},
  \bibinfo{author}{\bibfnamefont{A.}~\bibnamefont{Schliesser}},
  \bibnamefont{and}
  \bibinfo{author}{\bibfnamefont{T.}~\bibnamefont{Kippenderg}},
  \bibinfo{journal}{Nature} \textbf{\bibinfo{volume}{482}}, \bibinfo{pages}{63}
  (\bibinfo{year}{2010}).

\bibitem[{\citenamefont{Kamenev and Levchenko}(2009)}]{Kamenev_AIP_2009}
\bibinfo{author}{\bibfnamefont{A.}~\bibnamefont{Kamenev}} \bibnamefont{and}
  \bibinfo{author}{\bibfnamefont{A.}~\bibnamefont{Levchenko}},
  \bibinfo{journal}{Advances in Physics} \textbf{\bibinfo{volume}{58:3}},
  \bibinfo{pages}{197} (\bibinfo{year}{2009}).

\bibitem[{\citenamefont{Kamenev}(2011)}]{KamenevBook}
\bibinfo{author}{\bibfnamefont{A.}~\bibnamefont{Kamenev}},
  \emph{\bibinfo{title}{Field Theory of Non-Equilibrium Systems}}
  (\bibinfo{publisher}{Cambridge University Press}, \bibinfo{year}{2011}), ISBN
  \bibinfo{isbn}{9781139500296}.

\bibitem[{\citenamefont{Mertens et~al.}(1993)\citenamefont{Mertens, Kennedy,
  and Swain}}]{Mertens_PRA_1993}
\bibinfo{author}{\bibfnamefont{C.~J.} \bibnamefont{Mertens}},
  \bibinfo{author}{\bibfnamefont{T.~A.~B.} \bibnamefont{Kennedy}},
  \bibnamefont{and} \bibinfo{author}{\bibfnamefont{S.}~\bibnamefont{Swain}},
  \bibinfo{journal}{Phys. Rev. A} \textbf{\bibinfo{volume}{48}},
  \bibinfo{pages}{2374} (\bibinfo{year}{1993}).

\bibitem[{\citenamefont{Mertens et~al.}(1995)\citenamefont{Mertens, Hasty,
  Roark, Nowakowski, and Kennedy}}]{Mertens_PRA_1995}
\bibinfo{author}{\bibfnamefont{C.~J.} \bibnamefont{Mertens}},
  \bibinfo{author}{\bibfnamefont{J.~M.} \bibnamefont{Hasty}},
  \bibinfo{author}{\bibfnamefont{H.~H.} \bibnamefont{Roark}},
  \bibinfo{author}{\bibfnamefont{D.}~\bibnamefont{Nowakowski}},
  \bibnamefont{and} \bibinfo{author}{\bibfnamefont{T.~A.~B.}
  \bibnamefont{Kennedy}}, \bibinfo{journal}{Phys. Rev. A}
  \textbf{\bibinfo{volume}{52}}, \bibinfo{pages}{742} (\bibinfo{year}{1995}).

\bibitem[{\citenamefont{Veits and Fleischhauer}(1997)}]{Veits_PRA_1997}
\bibinfo{author}{\bibfnamefont{O.}~\bibnamefont{Veits}} \bibnamefont{and}
  \bibinfo{author}{\bibfnamefont{M.}~\bibnamefont{Fleischhauer}},
  \bibinfo{journal}{Phys. Rev. A} \textbf{\bibinfo{volume}{55}},
  \bibinfo{pages}{3059} (\bibinfo{year}{1997}).

\bibitem[{\citenamefont{Torre et~al.}(2013)\citenamefont{Torre, Diehl, Lukin,
  Sachdev, and Strack}}]{DallaTorre_PRA_2013}
\bibinfo{author}{\bibfnamefont{E.~G.~D.} \bibnamefont{Torre}},
  \bibinfo{author}{\bibfnamefont{S.}~\bibnamefont{Diehl}},
  \bibinfo{author}{\bibfnamefont{M.~D.} \bibnamefont{Lukin}},
  \bibinfo{author}{\bibfnamefont{S.}~\bibnamefont{Sachdev}}, \bibnamefont{and}
  \bibinfo{author}{\bibfnamefont{P.}~\bibnamefont{Strack}},
  \bibinfo{journal}{Phys. Rev. A} \textbf{\bibinfo{volume}{87}},
  \bibinfo{pages}{023831} (\bibinfo{year}{2013}).

\bibitem[{\citenamefont{Sieberer et~al.}(2014)\citenamefont{Sieberer, Huber,
  Altman, and Diehl}}]{Sieberer_PRB_2014}
\bibinfo{author}{\bibfnamefont{L.~M.} \bibnamefont{Sieberer}},
  \bibinfo{author}{\bibfnamefont{S.~D.} \bibnamefont{Huber}},
  \bibinfo{author}{\bibfnamefont{E.}~\bibnamefont{Altman}}, \bibnamefont{and}
  \bibinfo{author}{\bibfnamefont{S.}~\bibnamefont{Diehl}},
  \bibinfo{journal}{Phys. Rev. B} \textbf{\bibinfo{volume}{89}},
  \bibinfo{pages}{134310} (\bibinfo{year}{2014}).

\bibitem[{\citenamefont{Agarwal and Huang}(2010)}]{Agarwal_PRA_2010}
\bibinfo{author}{\bibfnamefont{G.~S.} \bibnamefont{Agarwal}} \bibnamefont{and}
  \bibinfo{author}{\bibfnamefont{S.}~\bibnamefont{Huang}},
  \bibinfo{journal}{Phys. Rev. A} \textbf{\bibinfo{volume}{81}},
  \bibinfo{pages}{041803} (\bibinfo{year}{2010}).

\bibitem[{\citenamefont{Weis et~al.}(2010)\citenamefont{Weis, Rivi{\`e}re,
  Del{\'e}glise, Gavartin, Arcizet, Schliesser, and
  Kippenberg}}]{Weis_Science_2010}
\bibinfo{author}{\bibfnamefont{S.}~\bibnamefont{Weis}},
  \bibinfo{author}{\bibfnamefont{R.}~\bibnamefont{Rivi{\`e}re}},
  \bibinfo{author}{\bibfnamefont{S.}~\bibnamefont{Del{\'e}glise}},
  \bibinfo{author}{\bibfnamefont{E.}~\bibnamefont{Gavartin}},
  \bibinfo{author}{\bibfnamefont{O.}~\bibnamefont{Arcizet}},
  \bibinfo{author}{\bibfnamefont{A.}~\bibnamefont{Schliesser}},
  \bibnamefont{and} \bibinfo{author}{\bibfnamefont{T.~J.}
  \bibnamefont{Kippenberg}}, \bibinfo{journal}{Science}
  \textbf{\bibinfo{volume}{330}}, \bibinfo{pages}{1520} (\bibinfo{year}{2010}).

\bibitem[{\citenamefont{Safavi-Naeini et~al.}(2011)\citenamefont{Safavi-Naeini,
  Alegre, Chan, Eichenfield, Winger, Lin, Hill, Chang, and
  Painter}}]{Safavi-Naeini_Nature_2011}
\bibinfo{author}{\bibfnamefont{A.~H.} \bibnamefont{Safavi-Naeini}},
  \bibinfo{author}{\bibfnamefont{T.~M.} \bibnamefont{Alegre}},
  \bibinfo{author}{\bibfnamefont{J.}~\bibnamefont{Chan}},
  \bibinfo{author}{\bibfnamefont{M.}~\bibnamefont{Eichenfield}},
  \bibinfo{author}{\bibfnamefont{M.}~\bibnamefont{Winger}},
  \bibinfo{author}{\bibfnamefont{Q.}~\bibnamefont{Lin}},
  \bibinfo{author}{\bibfnamefont{J.~T.} \bibnamefont{Hill}},
  \bibinfo{author}{\bibfnamefont{D.}~\bibnamefont{Chang}}, \bibnamefont{and}
  \bibinfo{author}{\bibfnamefont{O.}~\bibnamefont{Painter}},
  \bibinfo{journal}{Nature} \textbf{\bibinfo{volume}{472}}, \bibinfo{pages}{69}
  (\bibinfo{year}{2011}).

\bibitem[{\citenamefont{Marthaler and Dykman}(2006)}]{Dykman_PRA_2006}
\bibinfo{author}{\bibfnamefont{M.}~\bibnamefont{Marthaler}} \bibnamefont{and}
  \bibinfo{author}{\bibfnamefont{M.~I.} \bibnamefont{Dykman}},
  \bibinfo{journal}{Phys. Rev. A} \textbf{\bibinfo{volume}{73}},
  \bibinfo{pages}{042108} (\bibinfo{year}{2006}).

\bibitem[{\citenamefont{Dykman}(2007)}]{Dykman_PRE_2007}
\bibinfo{author}{\bibfnamefont{M.~I.} \bibnamefont{Dykman}},
  \bibinfo{journal}{Phys. Rev. E} \textbf{\bibinfo{volume}{75}},
  \bibinfo{pages}{011101} (\bibinfo{year}{2007}).

\bibitem[{\citenamefont{Serban and Wilhelm}(2007)}]{Wilhelm_PRL_2007}
\bibinfo{author}{\bibfnamefont{I.}~\bibnamefont{Serban}} \bibnamefont{and}
  \bibinfo{author}{\bibfnamefont{F.~K.} \bibnamefont{Wilhelm}},
  \bibinfo{journal}{Phys. Rev. Lett.} \textbf{\bibinfo{volume}{99}},
  \bibinfo{pages}{137001} (\bibinfo{year}{2007}).

\bibitem[{\citenamefont{Dykman et~al.}(2011)\citenamefont{Dykman, Marthaler,
  and Peano}}]{Dykman_PRA_2011}
\bibinfo{author}{\bibfnamefont{M.~I.} \bibnamefont{Dykman}},
  \bibinfo{author}{\bibfnamefont{M.}~\bibnamefont{Marthaler}},
  \bibnamefont{and} \bibinfo{author}{\bibfnamefont{V.}~\bibnamefont{Peano}},
  \bibinfo{journal}{Phys. Rev. A} \textbf{\bibinfo{volume}{83}},
  \bibinfo{pages}{052115} (\bibinfo{year}{2011}).

\bibitem[{\citenamefont{Ong et~al.}(2013)\citenamefont{Ong, Boissonneault,
  Mallet, Doherty, Blais, Vion, Esteve, and Bertet}}]{Ong_PRL_2013}
\bibinfo{author}{\bibfnamefont{F.~R.} \bibnamefont{Ong}},
  \bibinfo{author}{\bibfnamefont{M.}~\bibnamefont{Boissonneault}},
  \bibinfo{author}{\bibfnamefont{F.}~\bibnamefont{Mallet}},
  \bibinfo{author}{\bibfnamefont{A.~C.} \bibnamefont{Doherty}},
  \bibinfo{author}{\bibfnamefont{A.}~\bibnamefont{Blais}},
  \bibinfo{author}{\bibfnamefont{D.}~\bibnamefont{Vion}},
  \bibinfo{author}{\bibfnamefont{D.}~\bibnamefont{Esteve}}, \bibnamefont{and}
  \bibinfo{author}{\bibfnamefont{P.}~\bibnamefont{Bertet}},
  \bibinfo{journal}{Phys. Rev. Lett.} \textbf{\bibinfo{volume}{110}},
  \bibinfo{pages}{047001} (\bibinfo{year}{2013}).

\bibitem[{\citenamefont{Johansson et~al.}(2009)\citenamefont{Johansson,
  Johansson, Wilson, and Nori}}]{Johansson_PRL_2009}
\bibinfo{author}{\bibfnamefont{J.~R.} \bibnamefont{Johansson}},
  \bibinfo{author}{\bibfnamefont{G.}~\bibnamefont{Johansson}},
  \bibinfo{author}{\bibfnamefont{C.~M.} \bibnamefont{Wilson}},
  \bibnamefont{and} \bibinfo{author}{\bibfnamefont{F.}~\bibnamefont{Nori}},
  \bibinfo{journal}{Phys. Rev. Lett.} \textbf{\bibinfo{volume}{103}},
  \bibinfo{pages}{147003} (\bibinfo{year}{2009}).

\bibitem[{\citenamefont{Wilson et~al.}(2011)\citenamefont{Wilson, Johansson,
  Pourkabirian, Simoen, Johansson, Duty, Nori, and
  Delsing}}]{Wilson_Nature_2011}
\bibinfo{author}{\bibfnamefont{C.~M.} \bibnamefont{Wilson}},
  \bibinfo{author}{\bibfnamefont{G.}~\bibnamefont{Johansson}},
  \bibinfo{author}{\bibfnamefont{A.}~\bibnamefont{Pourkabirian}},
  \bibinfo{author}{\bibfnamefont{M.}~\bibnamefont{Simoen}},
  \bibinfo{author}{\bibfnamefont{J.~R.} \bibnamefont{Johansson}},
  \bibinfo{author}{\bibfnamefont{T.}~\bibnamefont{Duty}},
  \bibinfo{author}{\bibfnamefont{F.}~\bibnamefont{Nori}}, \bibnamefont{and}
  \bibinfo{author}{\bibfnamefont{P.}~\bibnamefont{Delsing}},
  \bibinfo{journal}{Nature} \textbf{\bibinfo{volume}{479}},
  \bibinfo{pages}{373} (\bibinfo{year}{2011}).

\bibitem[{\citenamefont{Peano and Thorwart}(2010)}]{Peano_PRB_2010}
\bibinfo{author}{\bibfnamefont{V.}~\bibnamefont{Peano}} \bibnamefont{and}
  \bibinfo{author}{\bibfnamefont{M.}~\bibnamefont{Thorwart}},
  \bibinfo{journal}{Phys. Rev. B} \textbf{\bibinfo{volume}{82}},
  \bibinfo{pages}{155129} (\bibinfo{year}{2010}).

\bibitem[{\citenamefont{Marquardt et~al.}(2007)\citenamefont{Marquardt, Chen,
  Clerk, and Girvin}}]{Marquartd_PRL_2007}
\bibinfo{author}{\bibfnamefont{F.}~\bibnamefont{Marquardt}},
  \bibinfo{author}{\bibfnamefont{J.~P.} \bibnamefont{Chen}},
  \bibinfo{author}{\bibfnamefont{A.~A.} \bibnamefont{Clerk}}, \bibnamefont{and}
  \bibinfo{author}{\bibfnamefont{S.~M.} \bibnamefont{Girvin}},
  \bibinfo{journal}{Physical Review Letters} \textbf{\bibinfo{volume}{99}}
  (\bibinfo{year}{2007}).

\bibitem[{\citenamefont{Lecocq et~al.}(2014)\citenamefont{Lecocq, Teufel,
  Aumentado, and Simmonds}}]{Teufel_ArXiV_2014}
\bibinfo{author}{\bibfnamefont{F.}~\bibnamefont{Lecocq}},
  \bibinfo{author}{\bibfnamefont{J.~D.} \bibnamefont{Teufel}},
  \bibinfo{author}{\bibfnamefont{J.}~\bibnamefont{Aumentado}},
  \bibnamefont{and} \bibinfo{author}{\bibfnamefont{R.~W.}
  \bibnamefont{Simmonds}}, \bibinfo{journal}{arXiv:1409.0872}
  (\bibinfo{year}{2014}).

\bibitem[{\citenamefont{Chang et~al.}(2011)\citenamefont{Chang, Safavi-Naeini,
  Hafezi, and Painter}}]{Chang2011}
\bibinfo{author}{\bibfnamefont{D.~E.} \bibnamefont{Chang}},
  \bibinfo{author}{\bibfnamefont{A.~H.} \bibnamefont{Safavi-Naeini}},
  \bibinfo{author}{\bibfnamefont{M.}~\bibnamefont{Hafezi}}, \bibnamefont{and}
  \bibinfo{author}{\bibfnamefont{O.}~\bibnamefont{Painter}},
  \bibinfo{journal}{New J. Phys.} \textbf{\bibinfo{volume}{13}},
  \bibinfo{pages}{023003} (\bibinfo{year}{2011}).

\bibitem[{\citenamefont{Schmidt et~al.}(2013)\citenamefont{Schmidt, Peano, and
  Marquardt}}]{MarquardtLattice2013}
\bibinfo{author}{\bibfnamefont{M.}~\bibnamefont{Schmidt}},
  \bibinfo{author}{\bibfnamefont{V.}~\bibnamefont{Peano}}, \bibnamefont{and}
  \bibinfo{author}{\bibfnamefont{F.}~\bibnamefont{Marquardt}},
  \bibinfo{journal}{arXiv:1311.7095}  (\bibinfo{year}{2013}).

\bibitem[{\citenamefont{Chen and Clerk}(2014)}]{Wei2014}
\bibinfo{author}{\bibfnamefont{W.}~\bibnamefont{Chen}} \bibnamefont{and}
  \bibinfo{author}{\bibfnamefont{A.~A.} \bibnamefont{Clerk}},
  \bibinfo{journal}{Phys. Rev. A.} \textbf{\bibinfo{volume}{89}},
  \bibinfo{pages}{033854} (\bibinfo{year}{2014}).

\bibitem[{\citenamefont{Safavi-Naeini et~al.}(2014)\citenamefont{Safavi-Naeini,
  Hill, Meenehan, Chan, Gr{\"o}blacher, and Painter}}]{Painter2014}
\bibinfo{author}{\bibfnamefont{A.~H.} \bibnamefont{Safavi-Naeini}},
  \bibinfo{author}{\bibfnamefont{J.~T.} \bibnamefont{Hill}},
  \bibinfo{author}{\bibfnamefont{S.}~\bibnamefont{Meenehan}},
  \bibinfo{author}{\bibfnamefont{J.}~\bibnamefont{Chan}},
  \bibinfo{author}{\bibfnamefont{S.}~\bibnamefont{Gr{\"o}blacher}},
  \bibnamefont{and} \bibinfo{author}{\bibfnamefont{O.}~\bibnamefont{Painter}},
  \bibinfo{journal}{Phys. Rev. Lett.} \textbf{\bibinfo{volume}{112}},
  \bibinfo{pages}{153603} (\bibinfo{year}{2014}).

\bibitem[{\citenamefont{Peano et~al.}(2014)\citenamefont{Peano, Brendel,
  Schmidt, and Marquardt}}]{MarquardtLattice2014}
\bibinfo{author}{\bibfnamefont{V.}~\bibnamefont{Peano}},
  \bibinfo{author}{\bibfnamefont{C.}~\bibnamefont{Brendel}},
  \bibinfo{author}{\bibfnamefont{M.}~\bibnamefont{Schmidt}}, \bibnamefont{and}
  \bibinfo{author}{\bibfnamefont{F.}~\bibnamefont{Marquardt}},
  \bibinfo{journal}{arXiv:1409.5375}  (\bibinfo{year}{2014}).

\bibitem[{\citenamefont{Dorsel et~al.}(1983)\citenamefont{Dorsel, McCullen,
  Meystre, Vignes, and Walther}}]{Dorsel_PRL_1983}
\bibinfo{author}{\bibfnamefont{A.}~\bibnamefont{Dorsel}},
  \bibinfo{author}{\bibfnamefont{J.~D.} \bibnamefont{McCullen}},
  \bibinfo{author}{\bibfnamefont{P.}~\bibnamefont{Meystre}},
  \bibinfo{author}{\bibfnamefont{E.}~\bibnamefont{Vignes}}, \bibnamefont{and}
  \bibinfo{author}{\bibfnamefont{H.}~\bibnamefont{Walther}},
  \bibinfo{journal}{Phys. Rev. Lett.} \textbf{\bibinfo{volume}{51}},
  \bibinfo{pages}{1550} (\bibinfo{year}{1983}).

\bibitem[{\citenamefont{Liu et~al.}(2013)\citenamefont{Liu, Xiao, Chen, Yu, and
  Gong}}]{Liu_PRL_2013}
\bibinfo{author}{\bibfnamefont{Y.-C.} \bibnamefont{Liu}},
  \bibinfo{author}{\bibfnamefont{Y.-F.} \bibnamefont{Xiao}},
  \bibinfo{author}{\bibfnamefont{Y.-L.} \bibnamefont{Chen}},
  \bibinfo{author}{\bibfnamefont{X.-C.} \bibnamefont{Yu}}, \bibnamefont{and}
  \bibinfo{author}{\bibfnamefont{Q.}~\bibnamefont{Gong}},
  \bibinfo{journal}{Phys. Rev. Lett.} \textbf{\bibinfo{volume}{111}},
  \bibinfo{pages}{083601} (\bibinfo{year}{2013}).

\bibitem[{\citenamefont{Gardiner and Zoller}(2004)}]{GardinerZollerBook}
\bibinfo{author}{\bibfnamefont{C.}~\bibnamefont{Gardiner}} \bibnamefont{and}
  \bibinfo{author}{\bibfnamefont{P.}~\bibnamefont{Zoller}},
  \emph{\bibinfo{title}{Quantum Noise: A Handbook of Markovian and
  Non-Markovian Quantum Stochastic Methods with Applications to Quantum
  Optics}}, Springer Series in Synergetics (\bibinfo{publisher}{Springer},
  \bibinfo{year}{2004}), ISBN \bibinfo{isbn}{9783540223016}.

\bibitem[{\citenamefont{Clerk et~al.}(2010)\citenamefont{Clerk, Devoret,
  Girvin, Marquardt, and Schoelkopf}}]{Clerk_RMP_2010}
\bibinfo{author}{\bibfnamefont{A.~A.} \bibnamefont{Clerk}},
  \bibinfo{author}{\bibfnamefont{M.~H.} \bibnamefont{Devoret}},
  \bibinfo{author}{\bibfnamefont{S.~M.} \bibnamefont{Girvin}},
  \bibinfo{author}{\bibfnamefont{F.}~\bibnamefont{Marquardt}},
  \bibnamefont{and} \bibinfo{author}{\bibfnamefont{R.~J.}
  \bibnamefont{Schoelkopf}}, \bibinfo{journal}{Rev. Mod. Phys.}
  \textbf{\bibinfo{volume}{82}}, \bibinfo{pages}{1155} (\bibinfo{year}{2010}).

\bibitem[{\citenamefont{Beaudoin et~al.}(2011)\citenamefont{Beaudoin, Gambetta,
  and Blais}}]{Beaudoin_PRA_2011}
\bibinfo{author}{\bibfnamefont{F.}~\bibnamefont{Beaudoin}},
  \bibinfo{author}{\bibfnamefont{J.~M.} \bibnamefont{Gambetta}},
  \bibnamefont{and} \bibinfo{author}{\bibfnamefont{A.}~\bibnamefont{Blais}},
  \bibinfo{journal}{Phys. Rev. A} \textbf{\bibinfo{volume}{84}},
  \bibinfo{pages}{043832} (\bibinfo{year}{2011}).

\bibitem[{\citenamefont{Dykman and Krivoglaz}(1975)}]{Dykman_PSS_1975}
\bibinfo{author}{\bibfnamefont{M.~I.} \bibnamefont{Dykman}} \bibnamefont{and}
  \bibinfo{author}{\bibfnamefont{M.~A.} \bibnamefont{Krivoglaz}},
  \bibinfo{journal}{Phys. stat. sol.} \textbf{\bibinfo{volume}{68}},
  \bibinfo{pages}{111} (\bibinfo{year}{1975}).

\bibitem[{\citenamefont{Marquardt et~al.}(2006)\citenamefont{Marquardt, Harris,
  and Girvin}}]{Marquardt_PRL_2006}
\bibinfo{author}{\bibfnamefont{F.}~\bibnamefont{Marquardt}},
  \bibinfo{author}{\bibfnamefont{J.}~\bibnamefont{Harris}}, \bibnamefont{and}
  \bibinfo{author}{\bibfnamefont{S.}~\bibnamefont{Girvin}},
  \bibinfo{journal}{Phys. Rev. Lett.} \textbf{\bibinfo{volume}{96}},
  \bibinfo{pages}{103901} (\bibinfo{year}{2006}).

\bibitem[{\citenamefont{Ludwig et~al.}(2008)\citenamefont{Ludwig, Kubala, and
  Marquardt}}]{Lugwig_NJP_2008}
\bibinfo{author}{\bibfnamefont{M.}~\bibnamefont{Ludwig}},
  \bibinfo{author}{\bibfnamefont{B.}~\bibnamefont{Kubala}}, \bibnamefont{and}
  \bibinfo{author}{\bibfnamefont{F.}~\bibnamefont{Marquardt}},
  \bibinfo{journal}{New Journal of Physics} \textbf{\bibinfo{volume}{10}},
  \bibinfo{pages}{095013+} (\bibinfo{year}{2008}), ISSN
  \bibinfo{issn}{1367-2630}.

\end{thebibliography}

\end{document}